\documentclass[a4paper,11pt]{article}
\pdfoutput=1 

\usepackage{jheppub} 

\usepackage[T1]{fontenc} 
\usepackage{xcolor} 
\usepackage{hyperref} 

\usepackage{ulem}
\usepackage{subcaption}

\usepackage{physics} 
\usepackage{bm}

\usepackage{booktabs}

\usepackage{multirow} 

\usepackage{tabularx} 

\usepackage{makecell}

\title{\boldmath Analytic Gravitational Wave Spectrum in  Next‑to‑Minimal Bouncing Cosmology}

\author[a]{Changhong Li\footnote{Corresponding Author}}

\affiliation[a]{Department of Astronomy, Key Laboratory of Astroparticle Physics of Yunnan Province,\\  School of Physics and Astronomy,  Yunnan University,\\ No.2 Cuihu North Road, Kunming,  650091 China}
\emailAdd{changhongli@ynu.edu.cn}

\abstract{Bouncing cosmology offers a singularity-free
alternative to inflation. However, its minimal
realization, known as minimal bouncing cosmology
(MBC), comprises only four cosmic phases and
produces a stochastic gravitational-wave background
(SGWB) with a single power-law spectrum. Interpreting the pulsar timing
array (PTA) signal within this framework faces
tension with the $\Delta N_{\rm eff}$ bound and
has required trans-Planckian bounce energies,
restricting its observational viability.
In this work, we propose next-to-minimal bouncing
cosmology (NMBC) by adding an earlier phase of
cosmic evolution to MBC.
Using our previously developed primordial
gravitational-wave propagation-matrix method,
with an improved matching prescription,
we derive explicit analytic expressions for
the resulting SGWB spectrum.
The spectrum takes a broken power-law form:
the low- and high-frequency tilts are determined
by the equations of state of the additional
and original contraction phases, respectively,
while the spectral amplitudes depend jointly
on these equations of state and the characteristic
transition-time scales.
This additional freedom allows a PTA-band signal
to coexist with a scale-invariant or red-tilted
high-frequency branch, providing a way to satisfy
the $\Delta N_{\rm eff}$ bound.
For the scale-invariant high-frequency branch,
a pointwise benchmark using the adopted
$\Delta N_{\rm eff}$ radiation budget gives
$\rho_{s\downarrow}^{1/4}\lesssim2.65\,m_{\rm pl}$,
where $m_{\rm pl}$ is the reduced Planck mass.
This provides a conservative comparison estimate
for the setup considered here, since integrating
the spectrum over the adopted illustrative
frequency interval gives a tighter bound.
The resulting upper limit is therefore of
order the reduced Planck scale.
Finally, we use the analytic expressions to
reconstruct illustrative NMBC backgrounds with
different spectral tilts and break frequencies,
producing potentially detectable signals in
different frequency bands.
These examples demonstrate how GW observations
can probe the pre-bounce evolution within NMBC.}

\begin{document}
\maketitle
\flushbottom


\section{Introduction}
\label{sec:introduction}

Bouncing cosmologies offer a compelling resolution
to the Universe's initial singularity through a
nonsingular transition from contraction to
expansion at a finite minimum scale factor
\cite{Novello:2008ra,Brandenberger:2016vhg,
Nojiri:2017ncd,Odintsov:2023weg}.
Primordial gravitational waves (PGWs) offer a
way to test this picture: their generation and
subsequent propagation through the bounce can
leave signatures in the present-day stochastic
gravitational-wave background (SGWB).
Searches for this background across multiple
frequency bands therefore provide complementary
probes of this nonsingular onset~\cite{Caprini:2018mtu}.
These searches include CMB/BICEP, pulsar timing
arrays (PTAs), and ground-based interferometers
(LVK), while proposed laboratory detectors,
such as superconducting circuits and cavities,
could extend the searches to much higher
frequencies
\cite{Li:2024oru,Li:2025ilc,Lai:2025efh,Li:2025mxj}.

Substantial progress has been made in constructing
bouncing models
\cite{Khoury:2001wf,Gasperini:2002bn,Creminelli:2006xe,
Peter:2006hx,Cai:2007qw,Cai:2008qw,Saidov:2010wx,
Li:2011nj,Cai:2011tc,Easson:2011zy,
Bhattacharya:2013ut,Qiu:2015nha,Barrow:2017yqt,
deHaro:2017yll,Ijjas:2018qbo,Boruah:2018pvq,
Nojiri:2019yzg,Silva:2015qna,Silva:2020bnn,
Silva:2023ieb,Nayeri:2005ck,Brandenberger:2006xi,
Fischler:1998st,Cai:2009rd,Li:2014era,Cheung:2014nxi,
Li:2014cba,Li:2015egy,Li:2020nah}
and studying their SGWB signatures
\cite{Boyle:2004gv,Piao:2004jg,Cai:2014bea,
Chowdhury:2015cma,Cai:2016hea,Zhang:2019tct,
Zhu:2023lbf,Papanikolaou:2024fzf,
Ben-Dayan:2024aec,Qiu:2024sdd}.
To turn GW observations into quantitative
constraints on the bounce, we need an explicit
relation between the cosmological background
and both the spectral shape and amplitude
of the SGWB.
Establishing this relation requires tracking
tensor perturbations through contraction,
the bounce, and the subsequent expansion.

\begin{figure}[htbp]
    \centering
    \includegraphics[width=\textwidth]{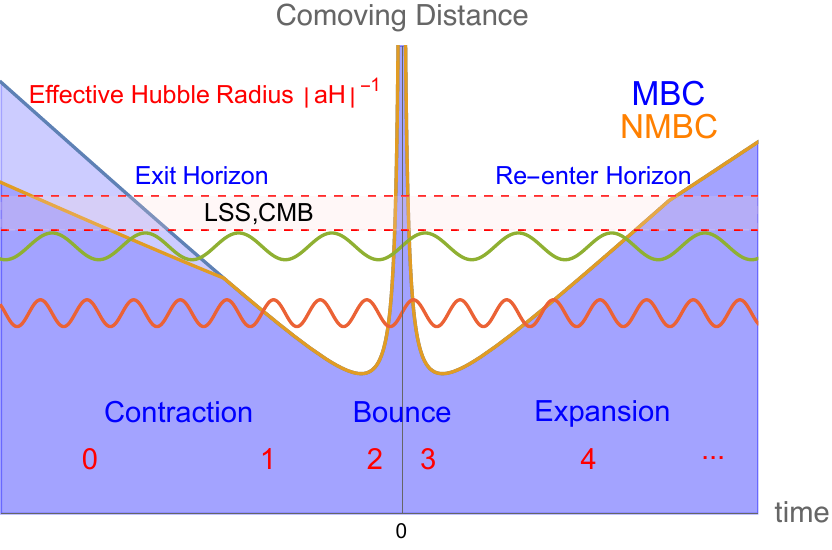}
    \caption{
        Schematic of the effective Hubble radius
        $|aH|^{-1}$ in minimal bouncing cosmology
        (MBC, blue) and next-to-minimal bouncing
        cosmology (NMBC, orange).
        In NMBC, the low-frequency modes
        (green wavy lines) first exit the horizon
        during Phase~0, while the high-frequency
        modes (red wavy lines) first exit during
        Phase~I.
        The additional phase thus allows the two
        frequency branches to probe different
        stages of contraction.
    }
    \label{fig:NMBC}
\end{figure}

In Ref.~\cite{Li:2024dce}, we addressed this
problem by describing minimal bouncing
cosmology (MBC) in terms of four essential
phases: collapsing contraction, bouncing
contraction, bouncing expansion, and
post-bounce expansion
($i=1,2,3,4$; see Fig.~\ref{fig:NMBC}).
Taking a constant effective equation of state
in each phase, we solved the tensor mode
equation analytically and connected the
solutions at the phase boundaries.
On this basis, we developed a matrix-representation
method supported by an inequality-based algebra.
The transformation and boundary-matching
matrices propagate the initial vacuum
fluctuations through these phases, yielding
an analytic PGW spectrum.
This construction provides a systematic route
from a specified bounce history to its
primordial tensor spectrum.
In Appendix~\ref{sec:pgwpropagationmatrix},
we summarize this formalism and improve the
boundary-matching prescription used in the
present work.

In Ref.~\cite{Li:2025ilc}, we used the analytic
PGW spectrum derived in Ref.~\cite{Li:2024dce}
to establish an explicit analytic relation
between the SGWB spectrum and the bounce
energy scale. Under the adopted assumptions, the leading-order,
phase-averaged MBC spectrum follows a single
power law, apart from the standard late-time
transfer effects.
A single spectral tilt links the amplitudes
across widely separated frequency bands,
limiting the freedom to produce an observable
signal in one band while satisfying constraints
in others.
For example, a blue-tilted spectrum normalized
at PTA frequencies can overproduce GW radiation
at higher frequencies and violate the
$\Delta N_{\rm eff}$ bound.
Earlier PTA fits using the MBC spectrum also
inferred bounce energies many orders of
magnitude above the Planck scale
\cite{Lai:2025efh}.
These observational and theoretical tensions
motivate an extension with greater flexibility
in the spectral shape, for example a broken
power-law spectrum arising from a contraction
history in which the low- and high-frequency
branches have different spectral tilts.

We therefore introduce next-to-minimal
bouncing cosmology (NMBC) by adding an earlier
contraction phase, Phase~0, to the four MBC
phases ($i=0,1,2,3,4$; see Fig.~\ref{fig:NMBC}).
Low-frequency modes first exit the horizon
during Phase~0, whereas sufficiently
high-frequency modes first exit during Phase~I.
The additional phase therefore provides a
physical mechanism for modifying the
low-frequency spectrum while preserving
the MBC behavior at high frequencies.
We determine analytically how this extension
changes both the spectral tilt and normalization,
allowing its observational potential to be
assessed in terms of the underlying bounce
parameters.

To establish this connection, we extend our
matrix method and improve its boundary-matching
prescription.%
\footnote{
    Assuming a continuous scale factor and no
    impulsive tensor source, we impose continuity
    of the tensor perturbation and its derivative
    with respect to the future-directed conformal
    time $\tau$, defined by $d\tau=dt/a$.
    This prescription accounts for the differing
    orientations of the positive local conformal-time
    arguments used in the mode functions.
    We apply it consistently to both MBC and NMBC;
    details are given in
    Appendix~\ref{sec:pgwpropagationmatrix}.
}
For the swift and symmetric bounce considered
here, we obtain explicit analytic expressions
for the leading-order, phase-averaged NMBC
SGWB spectrum.
The spectrum has a broken power-law form:
the high-frequency branch recovers the MBC
result, while the additional phase modifies
the low-frequency branch.
The low- and high-frequency spectral tilts
are determined by the equations of state
during Phases~0 and~I, respectively.
The amplitude depends on both equations of
state and the transition-time scales.
These expressions establish an explicit
analytic relation between the observable
SGWB spectrum and the underlying bounce
parameters.
We also show that NMBC recovers the MBC
results in two distinct limits.
This agreement supports the correctness of
our analytic calculation and establishes
a consistent connection between the two
cosmological models.
We further quantify the mismatch between
the asymptotic branches extrapolated to
the break frequency.

Applying the $\Delta N_{\rm eff}$ constraint
to our analytic spectra yields an upper bound
of order the reduced Planck scale on the NMBC
bounce energy $\rho_{s\downarrow}^{1/4}$.
We obtain this result for the scale-invariant
or red-tilted high-frequency branch and the
illustrative integration interval adopted here
(Fig.~\ref{fig:intMBC_Tau_DeltaNeff_Contours_Completeessential}).
The bound follows from the constraint on the
frequency-integrated GW energy density,
which, for the adopted benchmark
$\Delta N_{\rm eff}^{\max}=0.30$, reads
$\int d\ln f\,
\overline{\Omega}_{\rm GW}(f)h^2
\lesssim1.7\times10^{-6}$
\cite{Smith:2006nka,Pagano:2015hma,
Caprini:2018mtu,LISACosmologyWorkingGroup:2022jok}.
For comparison, the illustrative pointwise
benchmarks give
$\rho_{s\downarrow}^{1/4}\lesssim2.65\,m_{\rm pl}$,
as shown in Eq.~\eqref{eq:rhosdlerhosipointmpl}. At the scale-invariant point, the illustrative
integrated benchmark gives
$\rho_{s\downarrow}^{1/4}\lesssim1.03\,m_{\rm pl}$
[Eq.~\eqref{eq:rhosdvalueintessential}],
provided that the adopted interval lies within
the regime of validity of the high-frequency
spectrum. This result establishes a direct connection
between the observational limit on additional
radiation and the high-energy regime of NMBC:
within the adopted setup, bounce energies far
above the Planck scale are excluded.
It therefore provides a quantitative criterion
for assessing the energy scales required by
bouncing models.
We hope that this NMBC example will inspire
further efforts to address trans-Planckian
issues in early-universe model building
\cite{Martin:2000xs,Brandenberger:2000wr,
Brandenberger:2012aj,Kaloper:2002cs,
Easther:2002xe,Bedroya:2019snp,
Bedroya:2019tba,Benetti:2021uea,
Bedroya:2022tbh,Bedroya:2024zta,
Bedroya:2025ris}.

To demonstrate the potential of NMBC for
multi-frequency SGWB searches, we construct
four illustrative spectra with distinct
broken power-law features.
Starting from prescribed spectral tilts,
break frequencies, and pivot amplitudes,
we reconstruct the NMBC background parameters
and their associated physical scales.
Substituting these parameters into the analytic
expressions recovers the prescribed spectra,
confirming the internal consistency of the
reconstruction and spectral calculation.
This demonstrates how phenomenological SGWB
features can be traced back to the underlying
contraction history.

The resulting spectra lie near observational
benchmarks in different frequency bands
(Fig.~\ref{fig:FourExamples_Tau}),
highlighting the opportunities for probing
NMBC with complementary GW observations
\cite{Schmitz:2020syl,Annis:2022xgg,
Bi:2023tib,Lai:2025efh,Chen:2023ryb,Li:2025mxj}.
From a model-building perspective, these
examples demonstrate how the additional phase
expands the range of accessible spectral
shapes and amplitudes.
The higher-dimensional parameter space of NMBC
therefore offers greater freedom to accommodate
observational constraints while retaining
potentially detectable signals.
This flexibility is a central motivation for
introducing the additional phase and highlights
the potential of NMBC for future GW searches.

The paper is organized as follows.
Section~\ref{sec:effparamceBC} specifies the
effective backgrounds and their physical scales.
Section~\ref{sec:SGWBPGWkernel} relates the
propagation kernel to the observable SGWB spectrum.
Sections~\ref{sec:mbcessential}
and \ref{sec:NMBCessential} present the MBC
and NMBC spectra and their associated constraints,
followed by illustrative parameter reconstructions.
Section \ref{sec:conclusions} discuss the implications
and summarize the results.
The appendices collect the detailed matrix
calculations, the comparison of integrated
and pointwise bounds, the phase-node solutions,
and the asymptotic mismatch factors.

\section{Minimal and Next-to-Minimal Bouncing Cosmologies}
\label{sec:effparamceBC}

The distinction between MBC and NMBC lies in
the contraction history preceding the bounce.
We describe cosmic evolution as a sequence
of phases with constant effective
equation-of-state parameters and idealize
the transitions between them as instantaneous.
This piecewise description allows us to solve
the tensor mode equation analytically within
each phase and connect the solutions through
explicit boundary-matching matrices.

\subsection{Cosmic evolution of MBC and NMBC}

For comparison, the simplest inflationary
history before matter--radiation equality
can be represented by two main phases,
connected by an effectively instantaneous
reheating transition:
\begin{enumerate}
    \item
    \textbf{Phase A: Inflationary expansion.}
    The universe undergoes accelerated expansion
    ($\dot a>0$, $\ddot a>0$) with $w<-\tfrac13$.
    The comoving Hubble radius $|aH|^{-1}$ shrinks,
    and the modes of interest exit the horizon
    and evolve toward $k\eta\ll1$.

    \item
    \textbf{Phase B: Post-inflationary expansion.}
    After reheating, the universe enters
    radiation domination with $w=\tfrac13$
    and decelerated expansion
    ($\dot a>0$, $\ddot a<0$).
    The comoving Hubble radius grows,
    allowing modes to re-enter and evolve
    toward $k\eta\gg1$.
\end{enumerate}
Here, a dot denotes differentiation with
respect to cosmic time $t$.
Throughout, ``horizon'' refers to the
comoving Hubble radius $|aH|^{-1}$,
compared with the comoving mode scale $k^{-1}$.

We use the future-directed conformal time
$d\tau=dt/a$, which increases with $t$.
Within each phase $i$, the background and
tensor mode solutions are expressed using
the positive local variable
$\eta^{[i]}=|\tau-\tau_i(0)|>0$.%
\footnote{
    Here $\tau_i(0)$ is the conformal-time
    origin for the phase-$i$ solution.
    We omit the phase label $[i]$ and write
    $\eta$ for simplicity.
    This local variable can increase or
    decrease as $\tau$ advances.
    All bounce interfaces occur at finite,
    positive $\eta$; the asymptotic matching
    conditions do not require $\eta=0$
    or $\eta\to\infty$.
    The derivative relations and matching
    prescription are given in
    Appendix~\ref{sec:pgwpropagationmatrix}.
}
The conditions $k\eta\ll1$ and $k\eta\gg1$
specify the asymptotic regimes used in
the tensor-mode calculation.

A nonsingular bouncing background connects
contraction to expansion at a finite minimum
scale factor.
MBC describes this evolution through four
successive macroscopic phases
($i=1,2,3,4$), illustrated schematically
in Fig.~\ref{fig:NMBC}.
Their background evolution and associated
horizon crossings are as follows
\cite{Li:2024dce}:
\begin{enumerate}
    \item
    \textbf{Phase I: Collapsing contraction.}
    The universe contracts with increasing
    speed ($\dot a<0$, $\ddot a<0$),
    with $w_1>-\tfrac13$.
    The comoving Hubble radius shrinks,
    and the modes of interest first exit
    the horizon and evolve toward
    $k\eta\ll1$.

    \item
    \textbf{Phase II: Bouncing contraction.}
    Contraction slows as the universe
    approaches the bounce
    ($\dot a<0$, $\ddot a>0$),
    with $w_2<-\tfrac13$.
    The comoving Hubble radius grows,
    allowing the relevant modes to re-enter
    and evolve toward $k\eta\gg1$.

    \item
    \textbf{Phase III: Bouncing expansion.}
    After the bounce, the universe undergoes
    accelerated expansion
    ($\dot a>0$, $\ddot a>0$),
    with $w_3<-\tfrac13$.
    The comoving Hubble radius shrinks again,
    and the relevant modes undergo a second
    horizon exit, evolving toward
    $k\eta\ll1$.

    \item
    \textbf{Phase IV: Post-bounce expansion.}
    The universe enters radiation domination
    with $w_4=\tfrac13$ and decelerated expansion
    ($\dot a>0$, $\ddot a<0$).
    The comoving Hubble radius grows again,
    allowing the relevant modes to re-enter
    and evolve toward $k\eta\gg1$.
    The subsequent matter- and dark-energy-dominated
    eras follow the standard cosmological
    history~\cite{Caprini:2018mtu}.
\end{enumerate}
The PGW modes considered here therefore undergo
two horizon exits and two re-entries in MBC.
Their final spectrum depends on propagation
through this entire sequence.

NMBC adds an earlier contraction phase,
labeled $i=0$:
\begin{itemize}
    \item
    \textbf{Phase 0: Early collapsing contraction.}
    The universe contracts with increasing
    speed ($\dot a<0$, $\ddot a<0$),
    with $w_0>-\tfrac13$.
    The modes of interest initially lie
    deep inside the horizon.
    As the comoving Hubble radius shrinks,
    long-wavelength modes first exit during
    this phase, while shorter-wavelength
    modes remain inside until Phase~I.
\end{itemize}
The additional phase separates the initial
horizon-exit histories of low- and high-frequency
modes, providing the physical origin of the
broken power-law SGWB spectrum.
Sufficiently high-frequency modes recover
the leading MBC prediction, while the
low-frequency spectral tilt reflects
the equation of state during Phase~0.

The MBC framework and its earlier GW spectra
were developed in
Refs.~\cite{Li:2024dce,Li:2025ilc}.
In the present calculation, we impose continuity
of $h_k$ and $dh_k/d\tau$ across the interfaces.
For a continuous scale factor, the derivative
condition is equivalent to continuity of
$dh_k/dt$.
This matching prescription yields the MBC
spectrum derived in
Sec.~\ref{sec:mbcessential}, which provides
the baseline for our NMBC extension.%
\footnote{
    In Refs.~\cite{Li:2024dce,Li:2025ilc},
    we imposed continuity of the derivative
    with respect to the positive local
    conformal-time argument $\eta$.
    The resulting leading asymptotic MBC
    kernel is recovered at the phase nodes
    of the present solution, within the
    regime specified in
    Appendix~\ref{sec:phasenodeall}.
}

\subsection{Effective parametrization of MBC and NMBC}

The effective background is specified by
the equation-of-state parameters $\{w_i\}$
and the positive local conformal-time
arguments at the phase boundaries,
$\{\eta_{i\uparrow/\downarrow}\}$.
These quantities provide the inputs for
the GW propagation matrices.
A gradual transition can be approximated
by inserting additional phases.

We take Phase~IV to be radiation dominated,
with $w_4=\tfrac13$, and specialize to a swift
and symmetric bounce~\cite{Li:2025ilc}
(see also the review in Ref.~\cite{Cai:2009zp}).
The swift-bounce limit is
\begin{equation}\label{eq:swiftbouncecon}
    w_2=w_3\to-\infty,
\end{equation}
while symmetry is imposed through
\begin{equation}\label{eq:symmetricbouncecon}
    \eta_{s\downarrow}
    \equiv
    \eta_{1\downarrow}
    =
    \eta_{3\downarrow}.
\end{equation}

With these assumptions, the MBC parameters
are~\cite{Li:2024dce}
\begin{align}
    \text{MBC:}\quad
    &\{w_i\}^{(1)}
    =
    \bigl(w_1,w_2,w_3,w_4\bigr)
    =
    \bigl(w_1,-\infty,-\infty,\tfrac13\bigr),
    \label{eq:MBCsetw}\\
    &\{\eta_{i\uparrow/\downarrow}\}^{(1)}
    =
    \bigl(\eta_1,\eta_2,\eta_3\bigr)
    =
    \bigl(
        \eta_{s\downarrow},
        \eta_{2\uparrow},
        \eta_{s\downarrow}
    \bigr).
    \label{eq:MBCseteta}
\end{align}
NMBC adds the early contraction phase,
characterized by $w_0$ and
$\eta_{0\downarrow}$:
\begin{align}
    \text{NMBC:}\quad
    &\{w_i\}^{(0)}
    =
    \bigl(w_0,w_1,w_2,w_3,w_4\bigr)
    =
    \bigl(w_0,w_1,-\infty,-\infty,\tfrac13\bigr),
    \label{eq:NMBCsetw}\\
    &\{\eta_{i\uparrow/\downarrow}\}^{(0)}
    =
    \bigl(\eta_0,\eta_1,\eta_2,\eta_3\bigr)
    =
    \bigl(
        \eta_{0\downarrow},
        \eta_{s\downarrow},
        \eta_{2\uparrow},
        \eta_{s\downarrow}
    \bigr).
    \label{eq:NMBCseteta}
\end{align}
The superscripts $(1)$ and $(0)$ denote MBC
and NMBC, respectively.

For a mode with comoving wavenumber $k$,
the arrows $\uparrow$ and $\downarrow$
label the large- and small-argument matching
regimes,
$k\eta_{i\uparrow}\gg1$ and
$k\eta_{i\downarrow}\ll1$, respectively.
The modes considered here satisfy
$k\eta_1\ll1$, $k\eta_2\gg1$, and
$k\eta_3\ll1$, so we write
$\eta_1=\eta_{1\downarrow}$,
$\eta_2=\eta_{2\uparrow}$, and
$\eta_3=\eta_{3\downarrow}$.
At the additional Phase~0--I boundary,
$\eta_0=\eta_{0\downarrow}$ denotes the
same background transition scale for both
frequency branches.
The $\downarrow$ label identifies the
low-frequency regime
$k\eta_{0\downarrow}\ll1$, in which this
transition modifies the leading spectrum.
High-frequency modes instead satisfy
$k\eta_{0\downarrow}\gg1$ at the same boundary
and recover the leading MBC result.

Within each constant-$w_i$ phase,
$a^{[i]}\propto(\eta^{[i]})^{\nu_i}$.
The background exponent $\nu_i$ and the
corresponding index $\tilde{\nu}_i$ entering
the tensor mode solutions are
\begin{equation}\label{eq:nutnuw}
    \nu_i
    =
    \frac{2}{3w_i+1},
    \qquad
    \tilde{\nu}_i
    \equiv
    \left|\nu_i-\frac12\right|
    =
    \left|
        \frac{3(1-w_i)}{2(3w_i+1)}
    \right|,
    \qquad w_i\ne-\frac13.
\end{equation}
The MBC equation-of-state parameters
therefore correspond to
\begin{align}
    \text{MBC:}\quad
    &\{\nu_i\}^{(1)}
    =
    \bigl(\nu_1,\nu_2,\nu_3,\nu_4\bigr)
    =
    \bigl(\nu_1,0,0,1\bigr),
    \label{eq:MBCsetnu}\\
    &\{\tilde{\nu}_i\}^{(1)}
    =
    \bigl(
        \tilde{\nu}_1,\tilde{\nu}_2,
        \tilde{\nu}_3,\tilde{\nu}_4
    \bigr)
    =
    \bigl(
        \tilde{\nu}_1,\tfrac12,\tfrac12,\tfrac12
    \bigr),
    \label{eq:MBCsettnu}
\end{align}
and the NMBC parameters correspond to
\begin{align}
    \text{NMBC:}\quad
    &\{\nu_i\}^{(0)}
    =
    \bigl(\nu_0,\nu_1,\nu_2,\nu_3,\nu_4\bigr)
    =
    \bigl(\nu_0,\nu_1,0,0,1\bigr),
    \label{eq:NMBCsetnu}\\
    &\{\tilde{\nu}_i\}^{(0)}
    =
    \bigl(
        \tilde{\nu}_0,\tilde{\nu}_1,
        \tilde{\nu}_2,\tilde{\nu}_3,\tilde{\nu}_4
    \bigr)
    =
    \bigl(
        \tilde{\nu}_0,\tilde{\nu}_1,
        \tfrac12,\tfrac12,\tfrac12
    \bigr).
    \label{eq:NMBCsettnu}
\end{align}
Here the zero entries denote the limiting
values $\nu_2,\nu_3\to0^{-}$ associated
with the swift-bounce approximation.
These equivalent parameter sets connect
the equations of state directly to the
indices of the analytic tensor mode solutions.

\subsection{Bounce energy scale and characteristic GW frequencies}

The two transition scales in NMBC have
distinct observational roles:
$\eta_{s\downarrow}$ determines the bounce
energy scale $\rho_{s\downarrow}^{1/4}$
and characteristic cutoff frequency
$f_{\rm cut}$, while $\eta_{0\downarrow}$
sets the break frequency $f_\star$ associated
with the additional contraction phase.

Following Refs.~\cite{Li:2025ilc,Lai:2025efh},
we characterize the bounce by the energy
scale at the onset of the radiation-dominated
Phase~IV:
\begin{equation}\label{eq:rhosdd}
    \rho_{s\downarrow}^{1/4}
    =
    \left[
        3H^2(\eta_{s\downarrow})m_{\rm pl}^2
    \right]^{1/4},
\end{equation}
where $m_{\rm pl}$ is the reduced Planck mass.
To relate this scale to $\eta_{s\downarrow}$,
we approximate the radiation background
by one normalized to the present photon
density,
$\rho_{\gamma0}=\Omega_{\gamma0}\rho_{c0}$,
with $a_0=1$.
Here $\rho_{c0}=3H_0^2m_{\rm pl}^2$ is
the present critical energy density and
$H_0$ is the present Hubble parameter.

Within this approximation, the Phase-IV
background gives
\begin{equation}\label{eq:etasrhos}
    \eta_{s\downarrow}
    =
    \frac{1}{H_0}
    \left(
        \frac{\rho_{c0}}
             {\Omega_{\gamma0}\rho_{s\downarrow}}
    \right)^{1/4},
\end{equation}
and the corresponding scale factor is
\begin{equation}\label{eq:asdrhosdeq}
    a_{s\downarrow}
    \equiv
    a(\eta_{s\downarrow})
    =
    H_0\sqrt{\Omega_{\gamma0}}\,
    \eta_{s\downarrow}
    =
    \left(
        \frac{\rho_{c0}\Omega_{\gamma0}}
             {\rho_{s\downarrow}}
    \right)^{1/4}.
\end{equation}
Specifying any one of
$\eta_{s\downarrow}$, $a_{s\downarrow}$,
and $\rho_{s\downarrow}^{1/4}$ therefore
fixes the other two.
The derivation is given in
Appendix~\ref{app:PhaseIVbackground}.

With the observed frequency
$f=k/(2\pi a_0)$, the Phase~0--I transition
defines the characteristic break frequency
\begin{equation}\label{eq:fstardef}
    f_\star
    =
    \frac{1}{2\pi a_0\eta_{0\downarrow}}.
\end{equation}
The bounce parameter $\eta_{s\downarrow}$
similarly sets the characteristic cutoff.
Using Eq.~\eqref{eq:etasrhos} and $a_0=1$,
we obtain
\begin{equation}
    f_{\rm cut}
    =
    \frac{1}{2\pi a_0\eta_{s\downarrow}}
    =
    \frac{H_0}{2\pi}
    \frac{\rho_{s\downarrow}^{1/4}}
         {(\rho_{c0}/\Omega_{\gamma0})^{1/4}}.
\end{equation}
Thus, the adopted cutoff rises linearly
with the bounce energy scale.

At $f=f_{\rm cut}$,
$k\eta_{s\downarrow}=1$.
The leading-order expressions derived below
apply in the small-argument regime at
the bounce boundaries,
\begin{equation}
    k\eta_{s\downarrow}\ll1,
\end{equation}
or equivalently $f\ll f_{\rm cut}$,
together with the other matching conditions
specified above.
We adopt $f_{\rm cut}$ as the upper cutoff
of the asymptotic spectrum and neglect
contributions above it.
The detailed behavior near or above this
cutoff lies beyond the present approximation.

Together with the analytic SGWB spectra,
these relations translate GW bounds into
constraints on the bounce energy and allow
explicit NMBC backgrounds to be reconstructed
from prescribed spectral features.

\section{SGWB spectrum from the PGW propagation kernel}
\label{sec:SGWBPGWkernel}

The PGW propagation kernel connects the contraction
history and bounce dynamics to both the spectral
shape and normalization of the present-day SGWB.
For MBC and NMBC, the present-day spectrum is
given by~\cite{Caprini:2018mtu,Li:2025ilc}
\begin{equation}\label{eq:sgwbphfessential}
    \Omega_{\rm GW}^{(1/0)}(f)\,h^2
    \simeq
    \frac{\Omega_{\gamma0}\,h^2}{24}\,
    \mathcal{P}_h^{(1/0)}(f)\,
    \mathcal{T}_{\rm eq}(f).
\end{equation}
The superscripts $(1)$ and $(0)$ denote MBC
and NMBC, respectively, and $\mathcal P_h$
is the primordial tensor power spectrum
including both polarizations.
The adopted approximate transfer function,
$\mathcal T_{\rm eq}(f)
=1+\tfrac{9}{32}(f_{\rm eq}/f)^2$,
accounts for matter--radiation equality,
with $f_{\rm eq}=2.01\times10^{-17}\,\mathrm{Hz}$.
The observed frequency $f=k/(2\pi a_0)$
relates the frequency and wavenumber
representations through
\begin{equation}\label{eq:pfk2pifessential}
    \mathcal{P}_h^{(1/0)}(f)
    =
    \left.
        \mathcal{P}_h^{(1/0)}(k)
    \right|_{k=2\pi a_0 f}.
\end{equation}
We set $a_0=1$ throughout.

Using the matrix-representation method developed
in Refs.~\cite{Li:2024dce,Li:2025ilc}, with
continuity of $h_k$ and $dh_k/d\tau$ at each
boundary, we obtain the primordial tensor
spectrum for an initial Bunch--Davies vacuum:
\begin{align}
    \mathcal{P}_h^{(1/0)}(k)
    &\simeq
    \frac{4}{\pi^2}\,
    \frac{k^4}
         {H_0^2\Omega_{\gamma0}m_{\rm pl}^2}\,
    \mathcal{N}_{22}^{(1/0)}
    \bigl(
        \{\tilde{\nu}_i\},
        \{\eta_{i\downarrow/\uparrow}\}
    \bigr).
    \label{eq:phkn22essential}
\end{align}
Here $\mathcal{N}_{22}^{(1/0)}$ is the $(2,2)$
entry of the propagation kernel matrix
\begin{equation}\label{eq:nxxdefineessential}
    \mathcal{N}^{(1/0)}
    \bigl(
        \{\tilde{\nu}_i\},
        \{\eta_{i\downarrow/\uparrow}\}
    \bigr)
    \equiv
    \mathcal{X}^{(1/0)\dagger}
    \mathcal{X}^{(1/0)}.
\end{equation}
Its wavenumber dependence enters through
the dimensionless boundary arguments
$k\eta_{i\downarrow/\uparrow}$.
The transfer matrix $\mathcal{X}^{(1/0)}$
evolves the mode coefficients from the initial
contraction phase to the radiation-dominated
Phase~IV.
It is an ordered product of the
basis-transformation matrices $T_i$,
their inverses, and the boundary-matching
matrices $\mathcal{M}_{i\uparrow/\downarrow}$.
The matching prescription is detailed in
Appendix~\ref{sec:pgwpropagationmatrix},
and the derivation of
Eq.~\eqref{eq:phkn22essential}, including
the vacuum and normalization conventions,
is given in Appendix~\ref{sec:SGWBandmatrix}.

For MBC, the ordered product is
\begin{align}
    \mathcal{X}^{(1)}
    &=
    T_4^{-1}\mathcal{M}_{3\downarrow}T_3
    \mathcal{M}_{2\uparrow}T_2^{-1}
    \mathcal{M}_{1\downarrow}T_1.
    \label{eq:x1defMBCessential}
\end{align}
The additional Phase~0 in NMBC gives
\begin{align}
    \mathcal{X}^{(0)}
    &=
    \begin{cases}
        \mathcal{X}^{(0)\ge}
        =
        \mathcal{X}^{(1)}\mathcal{M}_{0\uparrow}
        \simeq\mathcal{X}^{(1)},
        & f\ge f_\star,\\[4pt]
        \mathcal{X}^{(0)<}
        =
        \mathcal{X}^{(1)}T_1^{-1}
        \mathcal{M}_{0\downarrow}T_0,
        & f<f_\star.
    \end{cases}
    \label{eq:x0defNMBCessential}
\end{align}
At high frequencies, the diagonal phases
of $\mathcal{M}_{0\uparrow}$ can be absorbed
into the initial mode basis.
These phases leave $\mathcal{N}_{22}$
unchanged at leading order, explaining
the recovery of the MBC spectrum.
The two branches use large- and small-argument
matching at the Phase~0--I boundary, respectively.
Their controlled regimes are $f\gg f_\star$
and $f\ll f_\star$; resolving the transition
near $f=f_\star$ requires the full matching
calculation.

To obtain explicit spectra, we substitute
the MBC and NMBC background parameters
into the general matrices in
Eqs.~\eqref{eq:tranmatrix},
\eqref{eqsi:matchmatrup}, and
\eqref{eq:matchmatrdown}, and evaluate the
ordered products in
Eqs.~\eqref{eq:x1defMBCessential}
and \eqref{eq:x0defNMBCessential}.
The resulting $\mathcal{N}_{22}^{(1/0)}$
determines the primordial tensor spectrum
through Eq.~\eqref{eq:phkn22essential}
and the present-day SGWB through
Eq.~\eqref{eq:sgwbphfessential}.
Averaging over the bounce-induced phase
then gives the phase-averaged spectra
$\overline{\Omega}_{\rm GW}^{(1/0)}(f)h^2$.

The explicit results for MBC and the two
NMBC branches are presented in
Eqs.~\eqref{eq:MBCoverlineSGWBessential},
\eqref{eq:NMBCgeoverlineSGWBessential},
and \eqref{eq:NMBClowoverlineSGWBessential}.
By fixing both the spectral tilts and
normalization, this calculation provides
the analytic basis for the bounce-energy
constraints and background reconstruction
discussed below.
Detailed derivations are given in
Appendices~\ref{sec:pgwpropagationmatrix},
\ref{sec:mbcdetailapp}, and
\ref{sec:NMBCalldetail}.

\section{Minimal Bouncing Cosmology}
\label{sec:mbcessential}

We first establish the MBC spectrum and its
observational implications using continuity of
$h_k$ and $dh_k/d\tau$ at the phase boundaries.
Applying the same matching prescription to MBC
and NMBC allows us to isolate the effect of
the additional contraction phase.
The matching formalism is developed in
Appendix~\ref{sec:pgwpropagationmatrix}.

\subsection{SGWB of MBC}
\label{subsec:SGWBMBC}

In MBC, the contraction equation of state
determines the primordial tensor tilt,
while the analytic normalization connects
the SGWB amplitude to the bounce energy.
For the swift and symmetric background
specified in Eqs.~\eqref{eq:MBCsetw}
and \eqref{eq:MBCseteta}, the propagation-matrix
calculation gives the leading phase-averaged
spectrum
\begin{align}
    \overline{\Omega}_{\rm GW}^{(1)}(f)h^2
    &\simeq
    \frac{h^2}{24}\,
    \frac{8}{\pi}
    \left(
        \frac{f_{H_0}}{f_{m_{\rm pl}}}
    \right)^2
    \frac{\overline{\mathcal{C}}^{(1)}}
         {(2\pi)^{-n_T^{(1)}+1}}
    \left(
        \frac{f}{f_{H_0}}
    \right)^{n_T^{(1)}}
    \nonumber\\
    &\quad\times
    \left[
        \frac{\rho_{s\downarrow}^{1/4}}
             {(\rho_{c0}/\Omega_{\gamma0})^{1/4}}
    \right]^{4-n_T^{(1)}}
    \mathcal{T}_{\rm eq}(f),
    \label{eq:MBCoverlineSGWBessential}
\end{align}
valid for $k\eta_{s\downarrow}\ll1$ and
$k\eta_{2\uparrow}\gg1$.
The detailed derivation is given in
Appendix~\ref{sec:mbcdetailapp}.

The overbar denotes averaging over the
bounce phase $2k\eta_{2\uparrow}$, with
$\langle\sin^2(2k\eta_{2\uparrow})\rangle=1/2$.
The resulting amplitude coefficient is
\begin{equation}\label{eq:C1MBCLeadingessetial}
    \overline{\mathcal{C}}^{(1)}
    \equiv
    \frac12\mathcal{C}^{(1)}
    =
    \frac{
        2^{2\nu_1-1}
        \Gamma^2\!\left(\nu_1+\tfrac12\right)
    }{2\pi}
    =
    \frac{2^{\frac{4}{3w_1+1}-1}}{2\pi}
    \Gamma^2\!\left(
        \frac12+\frac{2}{3w_1+1}
    \right),
\end{equation}
and the primordial tensor spectral index is
\begin{equation}\label{eq:MBCntw1essential}
    n_T^{(1)}
    \equiv
    \frac{
        d\ln\overline{\mathcal{P}}_h^{(1)}(k)
    }{d\ln k}
    =
    2-2\nu_1
    =
    \frac{2(w_1-\tfrac13)}{w_1+\tfrac13},
    \qquad
    w_1>-\frac13.
\end{equation}
The separate asymptotic solutions at the
phase nodes,
$\sin^2(2k\eta_{2\uparrow})=0$,
are discussed in
Appendix~\ref{sec:phasenodeall}.

We use the reference inverse-time scales
$f_{H_0}\equiv H_0
\simeq2.2\times10^{-18}\,\mathrm{Hz}$
and
$f_{m_{\rm pl}}\equiv m_{\rm pl}c^2/\hbar
\simeq3.7\times10^{42}\,\mathrm{Hz}$.
We adopt $\Omega_{\gamma0}h^2=2.474\times10^{-5}$
and $h=0.677$, where $h$ is the reduced
Hubble parameter.
In natural units, the present critical
energy density is
$\rho_{c0}=3H_0^2m_{\rm pl}^2
\simeq3.7\times10^{-59}\,\mathrm{TeV}^4$.

Within the stated asymptotic regime,
a single primordial tilt links the amplitudes
across frequency bands, with
$\mathcal T_{\rm eq}(f)$ accounting for
matter--radiation equality.
At fixed $w_1$, the signal in one band
therefore cannot be adjusted independently
of that in another.
This restriction is central to the observational
tests of MBC and motivates the additional
contraction phase in NMBC.
We next examine the dependence on $w_1$
and the resulting bounds on the bounce energy.

\subsection{Tensor spectral index and amplitude coefficient in MBC}
\label{subsec:TsindexaCmbc}

The contraction equation of state determines
both the primordial tensor tilt and the
amplitude coefficient of the MBC spectrum.
Equation~\eqref{eq:MBCntw1essential} places
the transition between red and blue spectra
at $w_1=\tfrac13$:
the spectrum is red-tilted for
$-\tfrac13<w_1<\tfrac13$,
scale-invariant at $w_1=\tfrac13$,
and blue-tilted for $w_1>\tfrac13$.
Its limiting behavior is
\begin{equation}
    n_T^{(1)}\bigl(w_1=\tfrac13\bigr)=0,
    \qquad
    \lim_{w_1\to+\infty}n_T^{(1)}=2,
    \qquad
    \lim_{w_1\to(-1/3)^+}n_T^{(1)}=-\infty.
\end{equation}
Thus, increasing $w_1$ makes the primordial
spectrum progressively bluer, with its tilt
approaching $2$ at large $w_1$.
The present-day SGWB also includes the
frequency dependence of
$\mathcal T_{\rm eq}(f)$, so its slope
coincides with the primordial tilt only
where the equality correction is negligible.

\begin{figure}[htbp]
    \centering
    \begin{subfigure}[t]{0.48\textwidth}
        \centering
        \includegraphics[width=\linewidth]
            {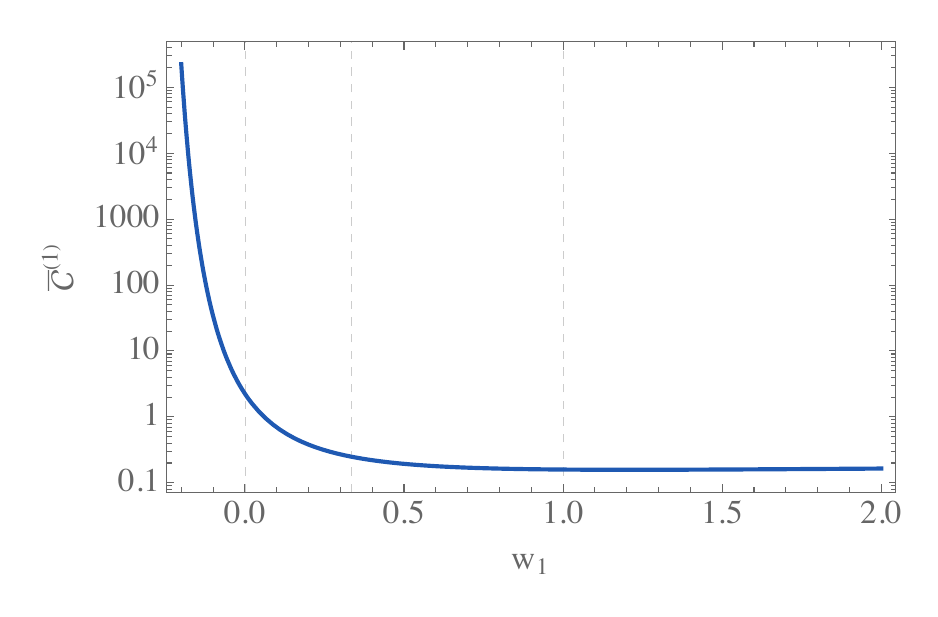}
        \caption{
            Phase-averaged amplitude coefficient
            $\overline{\mathcal C}^{(1)}$,
            shown on a logarithmic scale.
        }
        \label{fig:MBC_Cbarlogesstial}
    \end{subfigure}
    \hfill
    \begin{subfigure}[t]{0.48\textwidth}
        \centering
        \includegraphics[width=\linewidth]
            {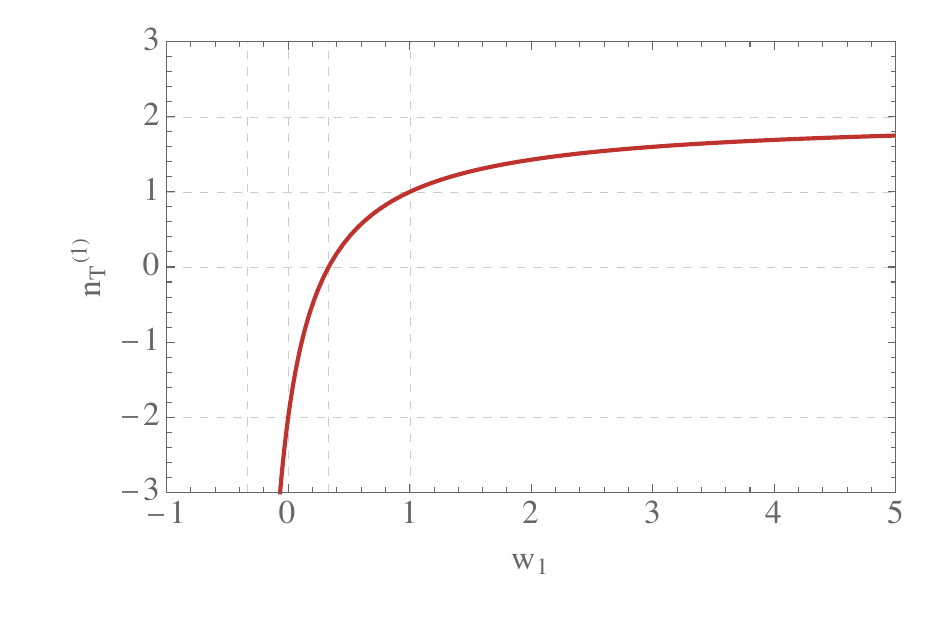}
        \caption{
            Primordial tensor spectral index
            $n_T^{(1)}$.
        }
        \label{fig:MBC_nTessential}
    \end{subfigure}
    \caption{
        Dependence of the MBC amplitude coefficient
        and primordial tensor spectral index on
        the contraction equation of state $w_1$,
        calculated from
        Eqs.~\eqref{eq:C1MBCLeadingessetial}
        and \eqref{eq:MBCntw1essential}.
    }
    \label{fig:MBC_Cbar_nTessential}
\end{figure}

Figure~\ref{fig:MBC_Cbar_nTessential}
summarizes the dependence of the amplitude
coefficient and spectral index on $w_1$.

The amplitude coefficient introduces a further
dependence on the contraction history.
As shown in Fig.~\ref{fig:MBC_Cbarlogesstial},
$\overline{\mathcal C}^{(1)}$ is of order
unity for $w_1>0$ and approaches
$\tfrac14$ as $w_1\to+\infty$.
Toward the opposite endpoint,
$w_1\to(-\tfrac13)^+$, the analytic coefficient
grows rapidly:
\begin{equation}
    \overline{\mathcal C}^{(1)}
    \sim
    \frac12
    \left[
        \frac{4}{e(3w_1+1)}
    \right]^{\frac{4}{3w_1+1}}
    \longrightarrow+\infty,
    \qquad
    w_1\to(-\tfrac13)^+.
\end{equation}
At fixed frequency and bounce energy,
this growth combines with the increasingly
red tilt to enhance the predicted SGWB
in the regime $k\eta_{s\downarrow}\ll1$.
An amplitude ceiling is therefore reached
at a lower bounce energy as $w_1$ approaches
$-\tfrac13$.
This behavior explains the tightening of
the energy bounds discussed below.

\subsection{GW constraints on the bounce energy scale in MBC}
\label{subsec:GWconstraintMBC}

An upper limit on the MBC signal translates
into an upper bound on the bounce energy
because the predicted amplitude increases
monotonically with $\rho_{s\downarrow}^{1/4}$.
At fixed $w_1>-\tfrac13$ and frequency $f$,
Eq.~\eqref{eq:MBCoverlineSGWBessential} gives
$\overline{\Omega}_{\rm GW}^{(1)}h^2
\propto(\rho_{s\downarrow}^{1/4})^{4-n_T^{(1)}}$,
with
$4-n_T^{(1)}=2+4/(3w_1+1)>0$.
Through
$a_{s\downarrow}
=(\rho_{c0}\Omega_{\gamma0}/\rho_{s\downarrow})^{1/4}$
[Eq.~\eqref{eq:asdrhosdeq}],
the same limit gives a lower bound on
$a_{s\downarrow}$.

We evaluate these bounds using the reference
amplitudes and projected sensitivities in
Table~\ref{tab:SGWB_benchmarksessential}.
Each entry is applied at its quoted frequency
as a pointwise benchmark.
The PTA amplitude is used as an approximate
ceiling on the primordial contribution,
whereas the projected sensitivities indicate
the reach of future searches.
For comparison, we consider the energy interval
$1\,\mathrm{TeV}\lesssim
\rho_{s\downarrow}^{1/4}\lesssim m_{\rm pl}$,
with $m_{\rm pl}=2.43\times10^{15}\,\mathrm{TeV}$.

\begin{table}[htbp]
    \centering
    \small
    \renewcommand{\arraystretch}{1.15}
    \begin{tabular}{l c c l}
        \toprule
        Entry
        & $f\,[\mathrm{Hz}]$
        & $\Omega_{\rm GW}h^2$
        & Interpretation \\
        \midrule
        CMB ($r=0.07$)
        & $7.73\times10^{-17}$
        & $1.5441\times10^{-16}$
        & Adopted upper limit \\
        BICEP ($r=0.036$)
        & $7.73\times10^{-17}$
        & $7.9413\times10^{-17}$
        & Adopted upper limit \\
        PTA benchmark
        & $1\times10^{-8}$
        & $1\times10^{-9}$
        & Amplitude benchmark \\
        LVK benchmark
        & $3\times10^{1}$
        & $1\times10^{-4}$
        & Adopted upper limit \\
        \midrule
        CMB-S4
        & $7.73\times10^{-17}$
        & $1.1030\times10^{-18}$
        & Projection \\
        aLVK
        & $2\times10^{1}$
        & $2\times10^{-5}$
        & Projection \\
        IPTA
        & $2\times10^{-9}$
        & $1\times10^{-14}$
        & Projection \\
        SKA
        & $2\times10^{-9}$
        & $2\times10^{-16}$
        & Projection \\
        DECIGO
        & $1.5\times10^{-1}$
        & $2\times10^{-13}$
        & Projection \\
        BBO
        & $2\times10^{-1}$
        & $3\times10^{-14}$
        & Projection \\
        LISA
        & $2.5\times10^{-3}$
        & $5\times10^{-12}$
        & Projection \\
        Cosmic Explorer
        & $1.5\times10^{1}$
        & $9\times10^{-10}$
        & Projection \\
        Einstein Telescope
        & $5$
        & $3\times10^{-9}$
        & Projection \\
        \bottomrule
    \end{tabular}
    \caption{
        Reference frequencies and SGWB amplitudes
        for the adopted upper limits, PTA amplitude
        benchmark, and projected sensitivities
        \cite{Schmitz:2020syl,Annis:2022xgg,
        Bi:2023tib,Li:2025ilc}.
        Each entry is treated as a pointwise
        benchmark.
        The CMB, BICEP, and CMB-S4 amplitudes
        are converted from the adopted
        tensor-to-scalar ratios using
        $A_s=2.1\times10^{-9}$ and
        $\Omega_{\gamma0}h^2=2.474\times10^{-5}$,
        including the matter--radiation equality
        transfer factor.
        For CMB-S4, we adopt $r=5\times10^{-4}$.
    }
    \label{tab:SGWB_benchmarksessential}
\end{table}

\begin{figure}[htbp]
    \centering
    \includegraphics[width=\textwidth]
        {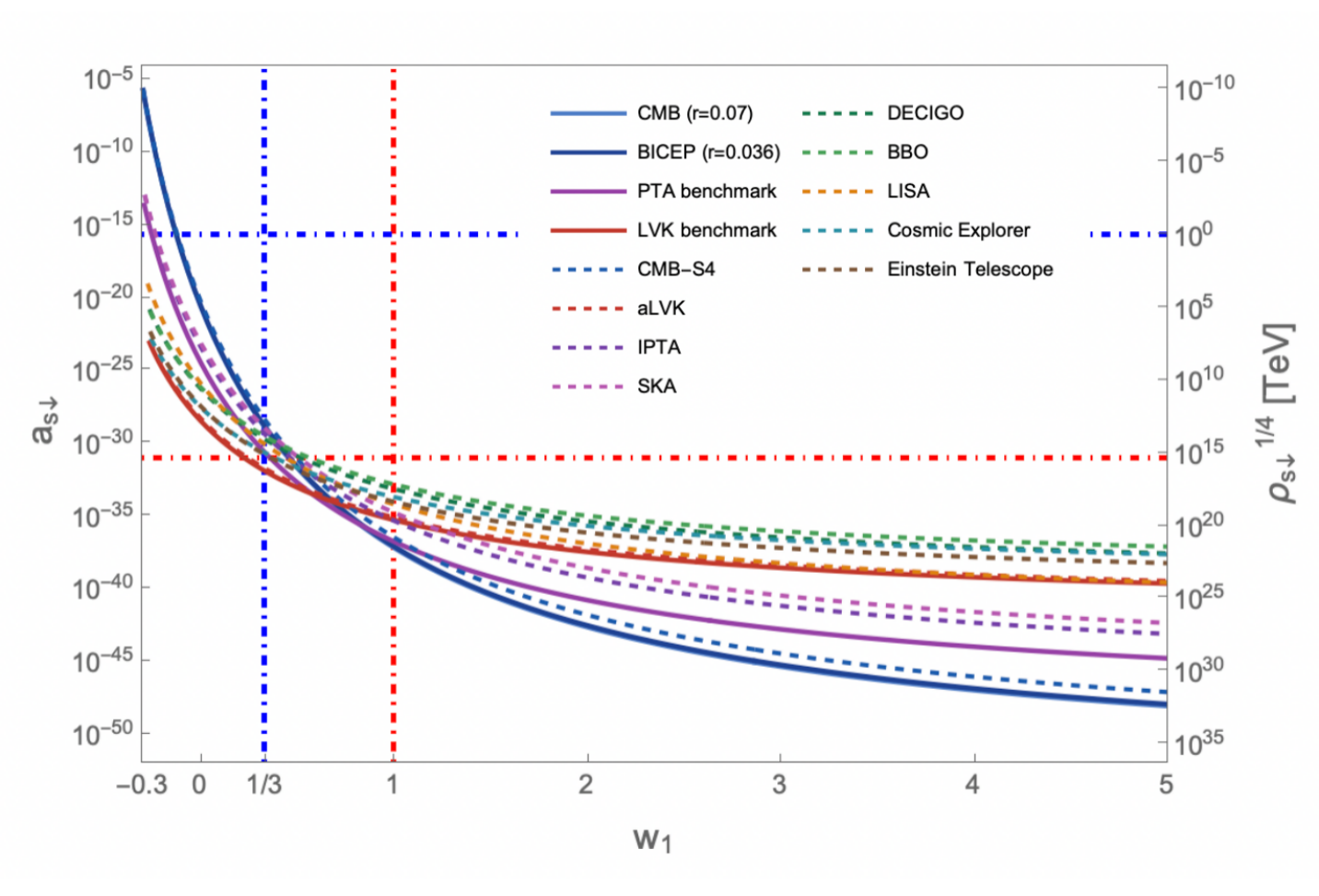}
    \caption{
        MBC bounds and projected sensitivities
        for $a_{s\downarrow}$ (left axis) and
        $\rho_{s\downarrow}^{1/4}$ (right axis),
        obtained from
        Table~\ref{tab:SGWB_benchmarksessential}
        using Eqs.~\eqref{eq:MBCoverlineSGWBessential}
        and \eqref{eq:asdrhosdeq}.
        Regions below solid upper-limit curves
        exceed the adopted bounds.
        The solid PTA curve is interpreted
        conditionally as an approximate ceiling
        on the primordial SGWB contribution.
        Regions below dashed curves would be
        excluded under the projected sensitivities
        if no signal were detected.
        Blue and red dot-dashed horizontal lines
        mark $\rho_{s\downarrow}^{1/4}
        =1\,\mathrm{TeV}$ and
        $m_{\rm pl}=2.43\times10^{15}\,\mathrm{TeV}$,
        respectively.
        Blue and red vertical lines mark
        $w_1=\tfrac13$ and $w_1=1$.
    }
    \label{fig:eMBC_Tau_Constraintsessential}
\end{figure}

Figure~\ref{fig:eMBC_Tau_Constraintsessential}
shows the resulting bounds and projected
sensitivities as functions of $w_1$.
The left and right axes express the same
constraints in terms of $a_{s\downarrow}$
and $\rho_{s\downarrow}^{1/4}$, respectively.
Since smaller $a_{s\downarrow}$ corresponds
to higher bounce energy, the region below
an amplitude-ceiling curve exceeds that
ceiling.

\begin{figure}[htbp]
    \centering
    \includegraphics[width=\textwidth]
        {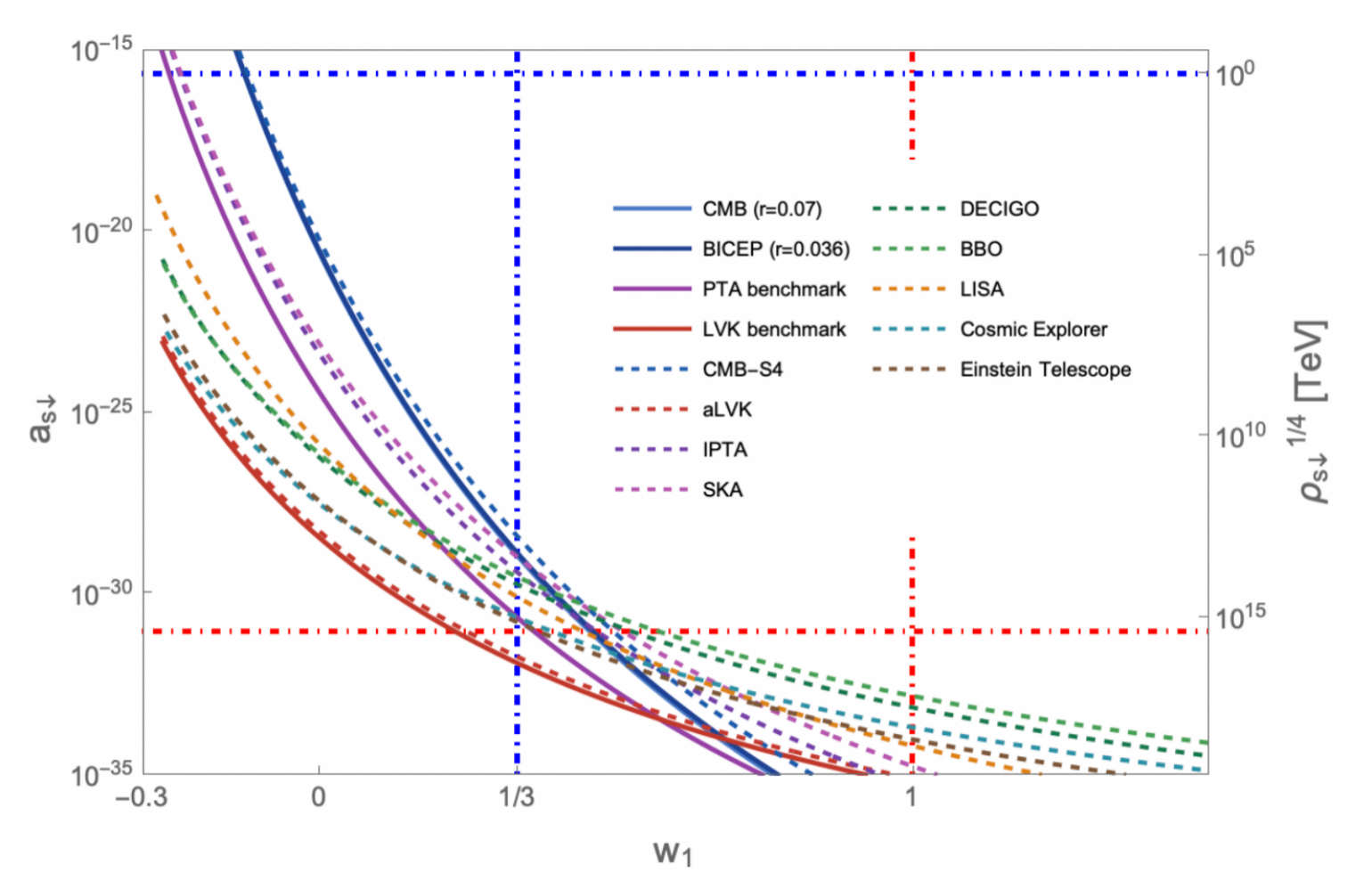}
    \caption{
        Zoomed view of
        Fig.~\ref{fig:eMBC_Tau_Constraintsessential}
        over $-0.3\le w_1\le1.5$,
        with the same axes and line conventions.
        The horizontal reference lines delimit
        the adopted energy interval
        $1\,\mathrm{TeV}\lesssim
        \rho_{s\downarrow}^{1/4}\lesssim m_{\rm pl}$.
        Intersections with these lines determine
        where each benchmark constrains this
        interval.
    }
    \label{fig:eMBC_Tau_Constraints_Zoominessential}
\end{figure}

The zoomed view in
Fig.~\ref{fig:eMBC_Tau_Constraints_Zoominessential}
shows how the benchmarks restrict the
selected energy interval.
The adopted CMB limit with $r=0.07$ reaches
$\rho_{s\downarrow}^{1/4}=1\,\mathrm{TeV}$
at $w_1\simeq-0.13$.
If the bounce energy is at least
$1\,\mathrm{TeV}$, this benchmark therefore
excludes $-\tfrac13<w_1\lesssim-0.13$.
The strong constraint on this red-tilted
region follows from the enhanced
low-frequency amplitude discussed above.

The adopted LVK benchmark instead reaches
$\rho_{s\downarrow}^{1/4}=m_{\rm pl}$
at $w_1\simeq0.229$.
For larger $w_1$, its upper energy bound lies
above the selected interval.
This crossing marks the loss of constraining
power of that benchmark for
$\rho_{s\downarrow}^{1/4}\le m_{\rm pl}$;
it does not impose an upper bound on $w_1$.

Within this benchmark comparison, candidate
signals must satisfy all adopted amplitude
ceilings, while those exceeding a projected
sensitivity are potential targets for the
corresponding search.
The bounds and sensitivity estimates apply
only where the asymptotic spectrum is valid.

~

\section{Next-to-Minimal Bouncing Cosmology}
\label{sec:NMBCessential}

NMBC allows the low- and high-frequency spectra
to probe different stages of contraction.
Low-frequency modes first exit the horizon
during the additional Phase~0, whereas
sufficiently high-frequency modes first exit
during Phase~I and follow the MBC evolution
at leading order.
The equations of state of these two phases
can therefore determine separate spectral tilts.

The transition between Phases~0 and~I sets
the break frequency
$f_\star=(2\pi a_0\eta_{0\downarrow})^{-1}$
[Eq.~\eqref{eq:fstardef}],
as illustrated in Fig.~\ref{fig:NMBC}.
The background parameters are specified in
Eqs.~\eqref{eq:NMBCsetw}
and \eqref{eq:NMBCseteta}.
We use the same bounce matching regimes as
in MBC,
$k\eta_{s\downarrow}\ll1$ and
$k\eta_{2\uparrow}\gg1$.
The analytic expressions below describe the
asymptotic branches $f\ll f_\star$ and
$f\gg f_\star$; resolving the transition
near $f_\star$ requires the full
Phase~0--I matching calculation.

\subsection{High-frequency modes ($f\ge f_\star$) in NMBC}

Modes with $k\eta_{0\downarrow}\gg1$ remain
well inside the horizon throughout Phase~0.
Their leading spectrum is therefore insensitive
to this earlier contraction stage.
They first exit during Phase~I and subsequently
undergo the same evolution as in MBC,
so the high-frequency NMBC spectrum follows
directly from
Eq.~\eqref{eq:MBCoverlineSGWBessential}:
\begin{align}
    \overline{\Omega}_{\rm GW}^{(0)\ge}(f)h^2
    &\simeq
    \overline{\Omega}_{\rm GW}^{(1)}(f)h^2
    \nonumber\\
    &\simeq
    \frac{h^2}{24}\,
    \frac{8}{\pi}
    \left(
        \frac{f_{H_0}}{f_{m_{\rm pl}}}
    \right)^2
    \frac{
        \overline{\mathcal C}^{(0)\ge}
    }{
        (2\pi)^{-n_T^{(0)\ge}+1}
    }
    \left(
        \frac{f}{f_{H_0}}
    \right)^{n_T^{(0)\ge}}
    \nonumber\\
    &\quad\times
    \left[
        \frac{\rho_{s\downarrow}^{1/4}}
             {(\rho_{c0}/\Omega_{\gamma0})^{1/4}}
    \right]^{4-n_T^{(0)\ge}}
    \mathcal T_{\rm eq}(f).
    \label{eq:NMBCgeoverlineSGWBessential}
\end{align}
Its primordial tensor tilt and amplitude
coefficient are consequently those of MBC:
\begin{equation}\label{eq:NMBCgentw1essetial}
    n_T^{(0)\ge}
    =
    n_T^{(1)}
    =
    2-2\nu_1
    =
    \frac{2(w_1-\tfrac13)}{w_1+\tfrac13},
    \qquad
    w_1>-\frac13,
\end{equation}
and
\begin{equation}\label{eq:NMBCgeavec1MBCLeadingessential}
    \overline{\mathcal C}^{(0)\ge}
    =
    \overline{\mathcal C}^{(1)}
    =
    \frac{
        2^{2\nu_1-1}
        \Gamma^2\!\left(\nu_1+\tfrac12\right)
    }{2\pi}
    =
    \frac{2^{\frac{4}{3w_1+1}-1}}{2\pi}
    \Gamma^2\!\left(
        \frac12+\frac{2}{3w_1+1}
    \right).
\end{equation}

We carry out the NMBC matrix calculation
and spectral analysis for $\nu_1>\tfrac12$,
corresponding to $-\tfrac13<w_1<1$ and
$n_T^{(0)\ge}<1$.
For the applications below, we focus further
on scale-invariant or red-tilted high-frequency
branches,
$n_T^{(0)\ge}\le0$, corresponding to
$\nu_1\ge1$ and $-\tfrac13<w_1\le\tfrac13$.
This choice avoids the high-frequency growth
of a blue primordial spectrum.
Compatibility with the integrated
$\Delta N_{\rm eff}$ bound still depends
on the normalization and the frequency
range over which the spectrum contributes.

\subsection{Low-frequency modes ($f<f_\star$) in NMBC}

Modes with $k\eta_{0\downarrow}\ll1$
first exit the horizon during Phase~0
and are already superhorizon at the
transition to Phase~I.
Their spectrum thus carries the imprint
of the earlier contraction stage, together
with the effect of subsequent propagation
through Phase~I and the bounce.
The matrix calculation detailed in
Appendix~\ref{subsec:lowNMBCalldetail}
gives the leading phase-averaged spectrum
\begin{align}
    \overline{\Omega}_{\rm GW}^{(0)<}(f)h^2
    &\simeq
    \frac{h^2}{24}\,
    \frac{8}{\pi}
    \left(
        \frac{f_{H_0}}{f_{m_{\rm pl}}}
    \right)^2
    \frac{
        \overline{\mathcal C}^{(0)<}
    }{
        (2\pi)^{-n_T^{(0)<}+1}
    }
    \left(
        \frac{f}{f_{H_0}}
    \right)^{n_T^{(0)<}}
    \nonumber\\
    &\quad\times
    \left[
        \frac{\rho_{s\downarrow}^{1/4}}
             {(\rho_{c0}/\Omega_{\gamma0})^{1/4}}
    \right]^{4-n_T^{(0)<}}
    \mathcal T_{\rm eq}(f).
    \label{eq:NMBClowoverlineSGWBessential}
\end{align}
The amplitude coefficient is
\begin{equation}\label{eq:avec1NMBClowLeadingessential}
    \overline{\mathcal C}^{(0)<}
    \equiv
    \frac{\mathcal C^{(0)<}}{2}
    =
    \frac{
        2^{2\nu_0-1}
        \Gamma^2\!\left(\nu_0+\tfrac12\right)
    }{2\pi}
    \left(
        \frac{\eta_{0\downarrow}}
             {\eta_{s\downarrow}}
    \right)^{2\nu_1-2\nu_0},
\end{equation}
whereas the primordial tensor tilt is
\begin{equation}\label{eq:NMBClowntw0essential}
    n_T^{(0)<}
    \equiv
    \frac{
        d\ln\overline{\mathcal P}_h^{(0)<}(k)
    }{d\ln k}
    =
    2-2\nu_0
    =
    \frac{2(w_0-\tfrac13)}{w_0+\tfrac13},
    \qquad
    w_0>-\frac13.
\end{equation}

These expressions separate the information
carried by the spectral tilt and normalization.
The low-frequency tilt depends only on $w_0$,
while the amplitude also depends on $\nu_1$
and the ratio
$\eta_{0\downarrow}/\eta_{s\downarrow}$.
The two slopes can therefore be set separately,
but the branch amplitudes are linked by
the contraction history.
In particular, NMBC can combine a blue
low-frequency primordial spectrum with
a scale-invariant or red-tilted high-frequency
branch.
A rise toward the break need not continue
to higher frequencies, overcoming the
single-tilt restriction of MBC.
The present-day SGWB additionally includes
the equality transfer function
$\mathcal T_{\rm eq}(f)$.

The analytic normalization makes this
spectral freedom quantitatively useful.
Within the adopted parameterization,
the two tensor tilts determine $w_0$ and $w_1$,
the break frequency fixes
$\eta_{0\downarrow}$, and a specified
high-frequency pivot amplitude determines
the bounce energy scale.
Here the pivot amplitude is defined by
extrapolating the high-frequency asymptotic
expression to $f=f_\star$.
The low-frequency normalization then follows
from the same reconstructed background.
Measurements confined to one asymptotic branch
generally leave part of the contraction
history undetermined, motivating observations
that constrain both branches and their break.
\subsection{Tensor spectral index and amplitude coefficient in NMBC}
\label{subsec:TsindexaCnmbc}

\subsubsection{Dependence on the contraction history}

The two NMBC branches obey the same relation
between the primordial tensor tilt and the
equation of state of the phase in which
the modes first exit the horizon:
\begin{equation}\label{eq:samedependence}
    n_T^{(i)}
    =
    2-2\nu_i
    =
    \frac{2(w_i-\tfrac13)}{w_i+\tfrac13},
    \qquad
    w_i>-\frac13.
\end{equation}
Here $n_T^{(0)}\equiv n_T^{(0)<}$ and
$n_T^{(1)}=n_T^{(0)\ge}$, with $i=0,1$
identifying the two contraction stages.
The NMBC calculation uses the additional
restriction $\nu_1>\tfrac12$ specified above.
Different equations of state therefore
produce different asymptotic tilts, with
the break marking the transition between
their horizon-exit histories.

The low-frequency normalization contains
additional information about Phase~I.
Relative to the MBC coefficient in
Eq.~\eqref{eq:C1MBCLeadingessetial},
with $\nu_1$ replaced by $\nu_0$,
Eq.~\eqref{eq:avec1NMBClowLeadingessential}
contains the factor
\begin{equation}
    \left(
        \frac{\eta_{0\downarrow}}
             {\eta_{s\downarrow}}
    \right)^{2\nu_1-2\nu_0}
    =
    \left(
        \frac{f_{\rm cut}}{f_\star}
    \right)^{n_T^{(0)<}-n_T^{(0)\ge}}.
\end{equation}
This form connects the normalization directly
to the difference between the spectral tilts
and the separation of the break and cutoff
frequencies.
Since $\eta_{0\downarrow}/\eta_{s\downarrow}>1$,
the factor enhances the coefficient when
$\nu_1>\nu_0$, or equivalently
$n_T^{(0)<}>n_T^{(0)\ge}$, and suppresses it
when the inequality is reversed.
Thus, fixing the low-frequency tilt does
not fix its normalization: the subsequent
evolution through Phase~I also contributes.

\begin{figure}[htbp]
    \centering
    \begin{subfigure}[t]{0.48\textwidth}
        \centering
        \includegraphics[width=\linewidth]
            {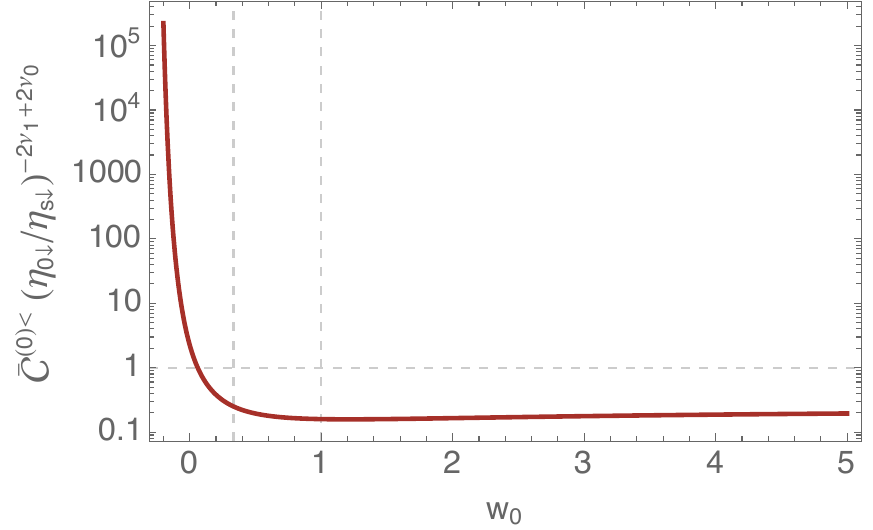}
        \caption{
            Normalized amplitude coefficient
            $\overline{\mathcal C}^{(0)<}
            (\eta_{0\downarrow}/\eta_{s\downarrow})
            ^{-2\nu_1+2\nu_0}$,
            shown on a logarithmic scale.
        }
        \label{fig:NMBC_CbarLow_Normalized_Plotsessential}
    \end{subfigure}
    \hfill
    \begin{subfigure}[t]{0.48\textwidth}
        \centering
        \includegraphics[width=\linewidth]
            {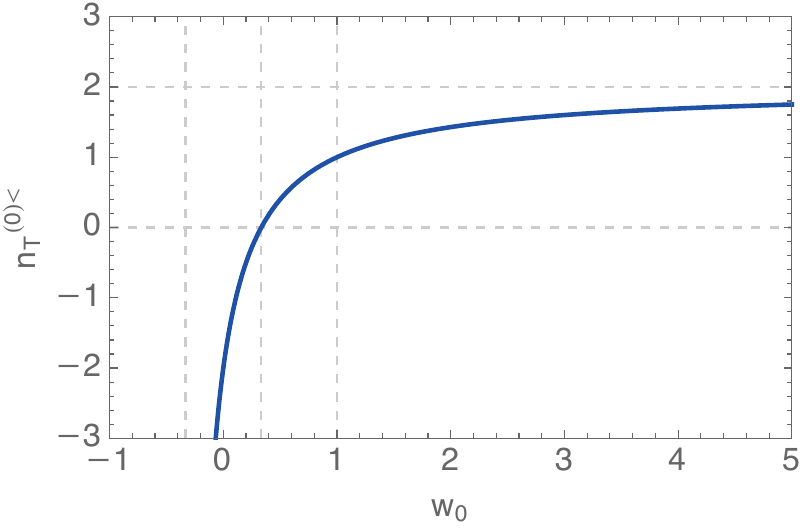}
        \caption{
            Low-frequency primordial tensor
            spectral index $n_T^{(0)<}$.
        }
        \label{fig:NMBC_nTLow_Plotsessential}
    \end{subfigure}
    \caption{
        Dependence of the normalized low-frequency
        amplitude coefficient and primordial tensor
        tilt on the Phase~0 equation of state $w_0$,
        calculated from
        Eqs.~\eqref{eq:avec1NMBClowLeadingessential}
        and \eqref{eq:NMBClowntw0essential}.
        Removing the transition factor from the
        coefficient reproduces the functional
        dependence found in MBC.
    }
    \label{fig:NMBC_Cbar_nTessential}
\end{figure}

Dividing out this transition factor isolates
the dependence on $w_0$.
The normalized coefficient
$\overline{\mathcal C}^{(0)<}
(\eta_{0\downarrow}/\eta_{s\downarrow})^{-2\nu_1+2\nu_0}$
and the tilt $n_T^{(0)<}$ then have the same
functional dependence on $w_0$ as their
MBC counterparts have on $w_1$.
Figure~\ref{fig:NMBC_Cbar_nTessential}
illustrates this correspondence with
Fig.~\ref{fig:MBC_Cbar_nTessential}.

\subsubsection{Two limits connecting MBC and NMBC}
\label{subsec:twolimitsmbcnmbc}

MBC is recovered when the distinction between
the two contraction stages disappears.
This occurs either when their equations of
state coincide or when Phase~I has zero
duration.
The two limits test both the spectral tilt
and normalization.
The kernel identities below refer to the
leading asymptotic expressions in
Eqs.~\eqref{eq:NMBClowN22LeadingUnified}
and \eqref{eq:N22MBCLeading}.

\begin{enumerate}
    \item \textbf{Identical contraction exponents:
    $\nu_0=\nu_1$.}

    When $\nu_0=\nu_1$, the Phase~0--I
    transition becomes trivial.
    The two tilts coincide, the additional
    normalization factor becomes unity,
    and the low-frequency kernel reduces to
    the MBC result with the common contraction
    exponent:
    \begin{equation}\label{eq:nu0nu1equallimit}
        \left.
        \mathcal N_{22}^{(0)<}(\nu_0,\nu_1)
        \right|_{\nu_0=\nu_1}
        =
        \mathcal N_{22}^{(1)}(\nu_1),
        \qquad
        f\ll f_\star.
    \end{equation}
    Since the high-frequency branch already
    follows MBC, the spectral break disappears
    in this limit.

    \item \textbf{Coincident transition boundaries:
    $\eta_{0\downarrow}\to\eta_{s\downarrow}^{+}$.}

    Bringing the two boundaries together
    eliminates Phase~I.
    Phase~0 then becomes the sole initial
    contraction stage, so the corresponding
    MBC background has contraction exponent
    $\nu_0$.
    The transition factor again approaches
    unity, and
    \begin{equation}\label{eq:eta0eta1equallimit}
        \lim_{\eta_{0\downarrow}\to\eta_{s\downarrow}^{+}}
        \mathcal N_{22}^{(0)<}(\nu_0,\nu_1)
        =
        \left.
        \mathcal N_{22}^{(1)}(\nu_1)
        \right|_{\nu_1=\nu_0},
        \qquad
        f\ll f_{\rm cut}.
    \end{equation}
    In this limit, $f_\star\to f_{\rm cut}$,
    so the distinct high-frequency NMBC
    asymptotic interval also disappears.
\end{enumerate}

Both limits therefore recover the corresponding
MBC spectrum, including its normalization:
$\overline{\Omega}_{\rm GW}^{(0)}(f)h^2
\simeq
\overline{\Omega}_{\rm GW}^{(1)}(f)h^2$.
The MBC contraction exponent is the common
value $\nu_0=\nu_1$ in the first limit and
the surviving Phase~0 exponent $\nu_0$
in the second.

~
\section{Cosmological Constraints and Applications of NMBC}

\subsection{Constraint on $\rho_{s\downarrow}^{1/4}$ from $\Delta N_{\rm eff}$}

The freedom to shape the two spectral branches
is limited by the total GW radiation they produce.
The analytic normalization therefore allows
the additional-radiation bound to constrain
the bounce energy directly.
For the adopted benchmark
$\Delta N_{\rm eff}^{\max}=0.30$
\cite{LISACosmologyWorkingGroup:2022jok},
the integrated SGWB must satisfy
\cite{Caprini:2018mtu}
\begin{equation}\label{eq:dneffcmbbbnintegralessential}
    \int_{f_{\min}}^{f_{\max}}
    d\ln f\,
    \overline{\Omega}_{\rm GW}^{(0)}(f)h^2
    \lesssim
    5.6\times10^{-6}\Delta N_{\rm eff}^{\max}
    \simeq
    1.7\times10^{-6}.
\end{equation}
Here, we adopt the phase-averaged spectrum in this
integral.
For a physical application, $f_{\min}$ is
the horizon frequency at the relevant CMB
or BBN epoch, redshifted to today, and
$f_{\max}$ is the upper cutoff of the
primordial spectrum.

Following the choice made above, we focus
on scale-invariant or red-tilted
high-frequency branches,
\begin{equation}\label{eq:ntbound13essential}
    n_T^{(0)\ge}\le0
    \quad\Longleftrightarrow\quad
    -\frac13<w_1\le\frac13.
\end{equation}
This restriction avoids the high-frequency
growth of a blue primordial spectrum.
It does not by itself ensure compatibility
with Eq.~\eqref{eq:dneffcmbbbnintegralessential}:
the normalization and frequency range
remain essential.

To illustrate the resulting energy bound,
we integrate the high-frequency expression
in Eq.~\eqref{eq:NMBCgeoverlineSGWBessential}
over the fixed interval
$f_{\min}=10^{-10}\,\mathrm{Hz}$ to
$f_{\max}=10^9\,\mathrm{Hz}$.
These limits are illustrative rather than
derived from the bounce dynamics.
Their NMBC interpretation assumes that
the high-frequency approximation applies
throughout the selected band and that
this band contributes to the relevant
radiation budget.
Under these conditions, assigning the entire
budget to this contribution gives a necessary
bound: adding the remaining positive GW
contribution can only strengthen it.
The low-frequency branch must also satisfy
the adopted CMB/BICEP limits.

At $w_1=\tfrac13$, the high-frequency
primordial spectrum is scale invariant.
The equality correction is negligible
over the selected band, so integration
multiplies its amplitude by
$\ln(f_{\max}/f_{\min})
=\ln(10^{19})\simeq43.75$.
The resulting bounce-energy bound is
\begin{equation}\label{eq:rhosdvalueintessential}
    \rho_{s\downarrow}^{1/4}
    \lesssim
    2.50\times10^{15}\,\mathrm{TeV}
    \simeq1.03\,m_{\rm pl}.
\end{equation}
Thus, even a flat high-frequency spectrum
can exhaust the radiation budget through
its broad frequency coverage.
For the adopted interval, this limits
the bounce energy to order the reduced
Planck scale.

For comparison, we apply the same numerical
ceiling at individual reference frequencies:
\begin{equation}\label{eq:dneffcmbbbnessential}
    \overline{\Omega}_{\rm GW}^{(0)}
    (f_{\rm typical})h^2
    \lesssim
    5.6\times10^{-6}\Delta N_{\rm eff}^{\max}
    \simeq1.7\times10^{-6}.
\end{equation}
We choose
$f_{\rm typical}
=7.75\times10^{-17},\,10^{-8},\,30,\,10^7$,
and $10^9\,\mathrm{Hz}$,
corresponding to the reference frequency
scales of CMB/BICEP, PTAs, LVK,
superconducting LC circuits, and
superconducting cavities, respectively.
These pointwise ceilings provide comparison
benchmarks; they do not replace the
integrated radiation constraint.

\begin{figure}[htbp]
    \centering
    \includegraphics[width=\textwidth]
        {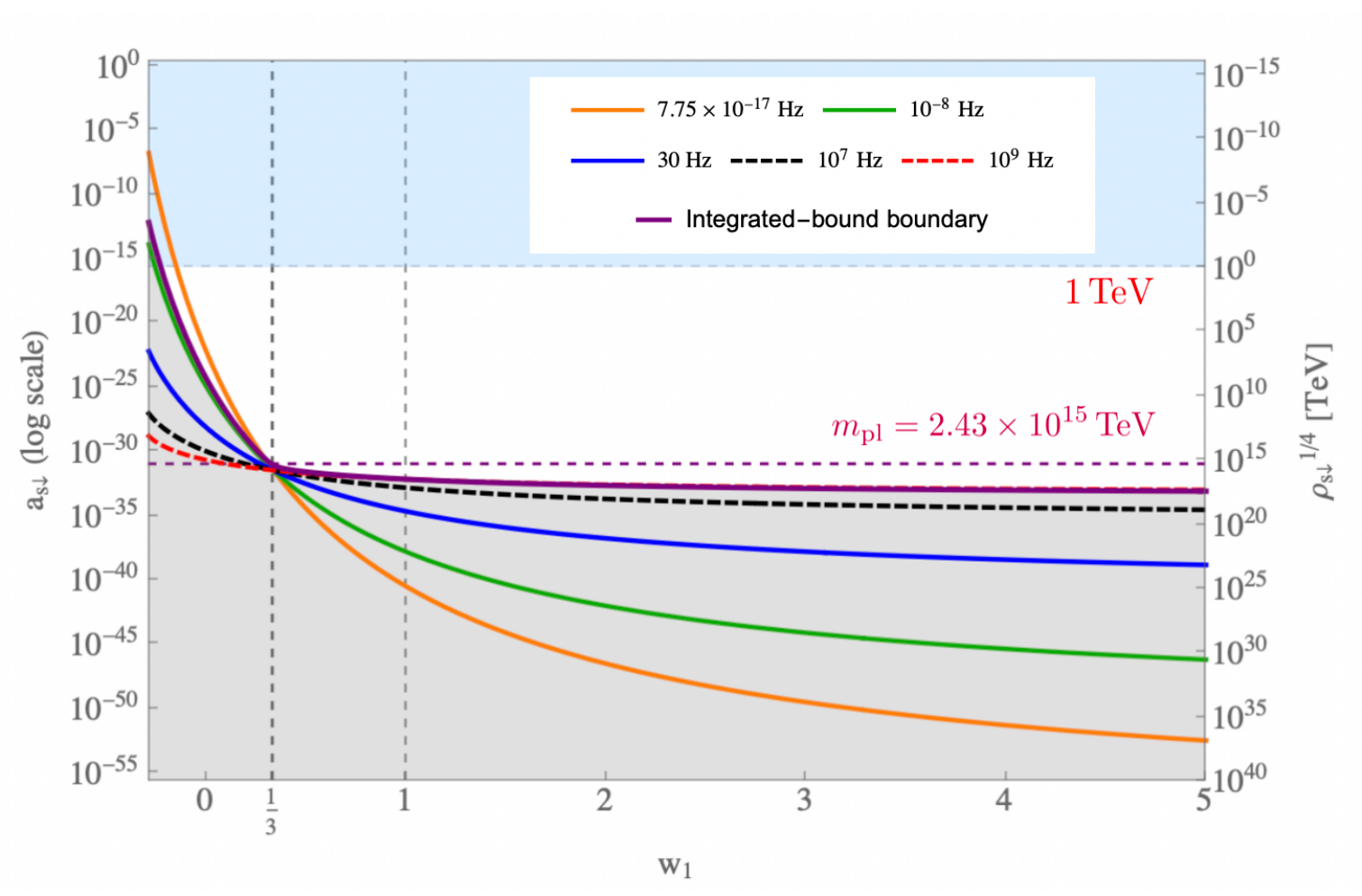}
    \caption{
        Integrated radiation bound and pointwise
        comparison benchmarks obtained from the
        high-frequency NMBC spectrum in
        Eq.~\eqref{eq:NMBCgeoverlineSGWBessential}.
        The left axis shows $a_{s\downarrow}$,
        and the right axis shows
        $\rho_{s\downarrow}^{1/4}$.
        The purple curve saturates the integrated
        budget $1.7\times10^{-6}$ over the
        illustrative interval
        $10^{-10}\le f\le10^9\,\mathrm{Hz}$;
        gray shading marks parameters that
        exceed this budget.
        The other curves apply the same ceiling
        pointwise at the reference frequencies
        specified in the text.
        The NMBC interpretation is conditional
        on the high-frequency approximation
        being valid over the integration band
        or at the selected reference frequency.
        The frequency limits are fixed for this
        comparison and are not model-derived
        cutoffs.
    }
    \label{fig:intMBC_Tau_DeltaNeff_Contours_Completeessential}
\end{figure}

Figure~\ref{fig:intMBC_Tau_DeltaNeff_Contours_Completeessential}
compares the integrated bound with these
pointwise benchmarks.
The left and right axes express the results
as bounds on $a_{s\downarrow}$ and
$\rho_{s\downarrow}^{1/4}$, respectively,
using Eq.~\eqref{eq:asdrhosdeq}.

At the scale-invariant point, the pointwise
curves nearly coincide.
Neglecting the small equality correction
at the lowest reference frequency gives
\begin{equation}
    \rho_{s\downarrow,\max}^{1/4}
    \left(w_1=\tfrac13\right)
    \simeq
    6.43\times10^{15}\,\mathrm{TeV}
    \simeq2.65\,m_{\rm pl}.
\end{equation}
For $w_1<\tfrac13$, the increasingly red
spectra give tighter benchmark bounds
within the applicable asymptotic regime.
The approximate upper envelope of the
pointwise comparison is therefore
\begin{equation}\label{eq:rhosdlerhosipointmpl}
    \rho_{s\downarrow}^{1/4}
    \lesssim
    \rho_{s\downarrow,\max}^{1/4}
    \left(w_1=\tfrac13\right)
    \simeq2.65\,m_{\rm pl}.
\end{equation}

The difference between the integrated and
pointwise results follows from the bandwidth
and the amplitude--energy scaling.
At fixed frequency and $w_1$,
Eq.~\eqref{eq:NMBCgeoverlineSGWBessential} gives
\begin{equation}
    \overline{\Omega}_{\rm GW}^{(0)\ge}(f)h^2
    \propto
    \left(\rho_{s\downarrow}^{1/4}\right)^{
        4-n_T^{(0)\ge}
    }.
\end{equation}
For a scale-invariant branch, fixed integration
limits, and negligible equality corrections,
using the same ceiling in both prescriptions
therefore gives
\begin{equation}\label{eq:dneffFlatEnergyRatioessential}
    \left(\rho_{s\downarrow}^{1/4}\right)_{\rm int}^{\max}
    \simeq
    \left(\rho_{s\downarrow}^{1/4}\right)_{\rm point}^{\max}
    \left[
        \ln\!\left(\frac{f_{\max}}{f_{\min}}\right)
    \right]^{-1/4}.
\end{equation}
For the selected band, this factor is
approximately $0.389$, accounting for
the reduction from $2.65\,m_{\rm pl}$
to $1.03\,m_{\rm pl}$.

The numerical energy limit is consequently
conditional on the spectral shape and
integration range.
For an individual NMBC background, the
integral must follow the appropriate
low- and high-frequency branches.
If its upper limit is identified with
$f_{\rm cut}$, the integration range also
depends on the bounce energy and the
bound must be solved self-consistently.
Appendix~\ref{sec:DeltaNeffintpoi}
develops these dependences in detail.

\subsection{Cosmological Applications of NMBC}

The analytic spectrum also allows the
background to be reconstructed from prescribed
spectral features.
Within the adopted parameterization, the inputs
$\{n_T^{(0)<},\,n_T^{(0)\ge},\,f_\star,\,
\overline{\Omega}_{\rm GW}^{(0)\ge}(f_\star)h^2\}$
determine
$\{w_0,\,w_1,\,\eta_{0\downarrow},\,
\eta_{s\downarrow}\}$,
together with the bounce energy
$\rho_{s\downarrow}^{1/4}$ and cutoff
$f_{\rm cut}$.
The high-frequency pivot amplitude is defined
by extrapolating the corresponding asymptotic
expression to $f=f_\star$.
The reconstruction procedure is given in
Appendix~\ref{sec:parameterrecon}.

\begin{figure}[htbp]
    \centering
    \includegraphics[width=\textwidth]
        {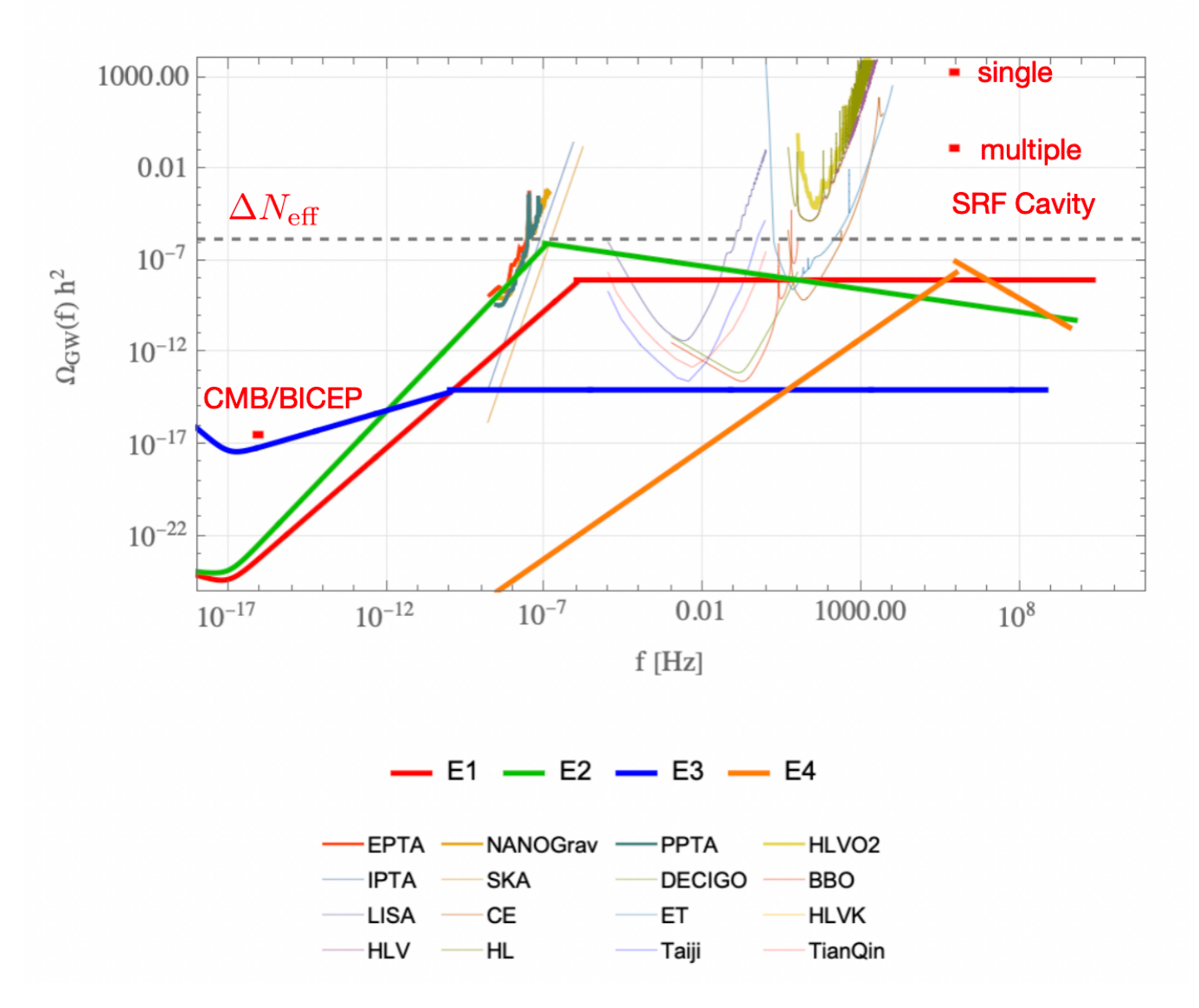}
    \caption{
        Phase-averaged SGWB spectra for the four
        illustrative NMBC backgrounds E1--E4,
        specified in
        Tables~\ref{tab:NMBC_example_inputs}
        and \ref{tab:NMBC_example_backgrounds}.
        The curves use the low- and high-frequency
        asymptotic expressions, extrapolated toward
        the break and cutoff.
        Apparent jumps at $f_\star$ reflect
        the mismatch between these approximations.
        E2 exceeds the adopted integrated
        radiation budget under this spectral
        prescription.
    }
    \label{fig:FourExamples_Tau}
\end{figure}

We illustrate this inverse relation with
four examples, E1--E4, whose inputs and
reconstructed backgrounds are listed in
Tables~\ref{tab:NMBC_example_inputs}
and \ref{tab:NMBC_example_backgrounds}.
Their spectra, obtained from
Eqs.~\eqref{eq:NMBCgeoverlineSGWBessential}
and \eqref{eq:NMBClowoverlineSGWBessential},
are shown in Fig.~\ref{fig:FourExamples_Tau}.
The examples span different break frequencies
and amplitudes chosen near the observational
benchmarks adopted in this work.

\begin{table}[htbp]
    \centering
    \small
    \renewcommand{\arraystretch}{1.2}
    \setlength{\tabcolsep}{5pt}
    \begin{tabular}{l c c c c}
        \toprule
        Input
        & E1 (red)
        & E2 (green)
        & E3 (blue)
        & E4 (orange) \\
        \midrule
        $n_T^{(0)<}$
        & $1.5$
        & $1.8$
        & $0.5$
        & $1.2$ \\
        $n_T^{(0)\ge}$
        & $0$
        & $-\tfrac14$
        & $0$
        & $-1$ \\
        $f_\star\,[\mathrm{Hz}]$
        & $10^{-6}$
        & $10^{-7}$
        & $10^{-10}$
        & $10^6$ \\
        $\overline{\Omega}_{\rm GW}^{(0)\ge}(f_\star)h^2$
        & $10^{-8}$
        & $10^{-6}$
        & $10^{-14}$
        & $10^{-7}$ \\
        \bottomrule
    \end{tabular}
    \caption{
        Phenomenological inputs for the four
        illustrative NMBC examples.
        The pivot amplitude is the high-frequency
        asymptotic amplitude extrapolated to
        $f=f_\star$.
        Colors identify the spectra in
        Fig.~\ref{fig:FourExamples_Tau}.
    }
    \label{tab:NMBC_example_inputs}
\end{table}

\begin{table}[htbp]
    \centering
    \small
    \renewcommand{\arraystretch}{1.2}
    \setlength{\tabcolsep}{5pt}
    \begin{tabular}{l c c c c}
        \toprule
        Parameter or scale
        & E1
        & E2
        & E3
        & E4 \\
        \midrule
        $w_0$
        & $\tfrac73$
        & $\tfrac{19}{3}$
        & $\tfrac59$
        & $\tfrac43$ \\
        $w_1$
        & $\tfrac13$
        & $\tfrac7{27}$
        & $\tfrac13$
        & $\tfrac19$ \\
        $\eta_{0\downarrow}\,[\mathrm{s}]$
        & $1.59\times10^5$
        & $1.59\times10^6$
        & $1.59\times10^9$
        & $1.59\times10^{-7}$ \\
        $\eta_{s\downarrow}\,[\mathrm{s}]$
        & $7.35\times10^{-12}$
        & $2.73\times10^{-11}$
        & $2.32\times10^{-10}$
        & $4.12\times10^{-11}$ \\
        $\rho_{s\downarrow}^{1/4}/m_{\rm pl}$
        & $0.732$
        & $0.197$
        & $2.32\times10^{-2}$
        & $0.131$ \\
        $f_{\rm cut}\,[\mathrm{Hz}]$
        & $2.17\times10^{10}$
        & $5.84\times10^9$
        & $6.85\times10^8$
        & $3.87\times10^9$ \\
        \bottomrule
    \end{tabular}
    \caption{
        Background parameters and physical scales
        reconstructed from
        Table~\ref{tab:NMBC_example_inputs}.
        Transition times, bounce energies, and
        cutoff frequencies are approximate.
    }
    \label{tab:NMBC_example_backgrounds}
\end{table}

The reconstructed bounce energies are
sub-Planckian in all four examples.
E1 and E3 have flat high-frequency branches,
but their plateau amplitudes differ by
six orders of magnitude.
Because the flat-branch amplitude scales
as $(\rho_{s\downarrow}^{1/4})^4$,
this corresponds to a factor of approximately
$32$ in the reconstructed bounce energy.
E4 instead places the break at
$f_\star=10^6\,\mathrm{Hz}$ and combines
a blue low-frequency primordial branch
with a high-frequency tilt of $-1$.
These examples show how the spectral shape
and normalization select different
contraction histories and physical scales.

The integrated radiation budget supplies
a separate test of these reconstructions.
In particular, E2 has a mildly red
high-frequency tilt, yet its normalization
is too large for the adopted budget.
Under the asymptotic extrapolation used
for the illustrative spectra, its
high-frequency branch alone contributes
approximately $4.0\times10^{-6}$ to
$\int_{f_\star}^{f_{\rm cut}}
d\ln f\,
\overline{\Omega}_{\rm GW}^{(0)\ge}(f)h^2$,
exceeding the allowed $1.7\times10^{-6}$.
E2 therefore demonstrates that a red
high-frequency tilt and a sub-Planckian
bounce energy do not by themselves
guarantee compatibility with the
integrated radiation constraint.

The apparent discontinuities in
Fig.~\ref{fig:FourExamples_Tau}
also require a distinction between the
asymptotic representation and the physical
transition near $f_\star$.
We quantify the mismatch between the
extrapolated branch amplitudes by
\begin{equation}\label{eq:flefgeNMBCessential}
    \overline{\Omega}_{\rm GW}^{(0)\ge}(f_\star)h^2
    =
    \epsilon_g(\nu_0,\nu_1)\,
    \overline{\Omega}_{\rm GW}^{(0)<}(f_\star)h^2.
\end{equation}
For E1--E4, respectively,
\begin{equation}\label{eq:epsilongvalueessential}
    \epsilon_g
    \simeq
    \left\{
        1.48,\,
        1.50,\,
        1.35,\,
        4.02
    \right\}.
\end{equation}
These factors describe the mismatch between
two asymptotic approximations, rather than
a physical discontinuity in the SGWB.
Resolving the transition requires the full
Phase~0--I matching calculation.
The mismatch factors are derived in
Appendix~\ref{subsec:nmbcfstarmismatch}.

The examples thus illustrate both the
constructive and restrictive uses of the
analytic spectrum.
Prescribed spectral features determine
candidate NMBC backgrounds and their
bounce scales; the integrated radiation
budget and frequency-specific amplitude
limits then test those candidates.
The additional contraction phase expands
the available spectral shapes, while the
analytic normalization makes their
observational consequences quantitative.

~
~
\section{Conclusions and outlook}
\label{sec:conclusions}

In this work, we have introduced next-to-minimal
bouncing cosmology (NMBC) by adding an early
contraction phase to the four-phase MBC background.
This extension allows low- and high-frequency
primordial gravitational waves to first exit
the horizon during different stages of contraction.
Different equations of state in these stages
therefore generate a broken power-law SGWB,
providing greater freedom to shape an observable
signal across widely separated frequency bands.

By extending our PGW propagation matrix method
with the improved boundary-matching prescription,
we have derived explicit analytic expressions
for both the spectral tilts and amplitudes.
For the swift and symmetric bounce considered
here, we derive closed-form analytic expressions
for the leading-order, phase-averaged NMBC
SGWB spectrum in the high- and low-frequency
regimes, given in
Eqs.~\eqref{eq:NMBCgeoverlineSGWBessential}
and \eqref{eq:NMBClowoverlineSGWBessential},
respectively.
Combining these two asymptotic expressions
gives the following compact form:
\begin{align}
    \overline{\Omega}_{\rm GW}^{(0)}(f)h^2
    &\simeq
    \frac{h^2}{6\pi^2}
    \left(
        \frac{f_{H_0}}{f_{m_{\rm pl}}}
    \right)^2
    \left(
        \frac{2\pi f_\star}{f_{H_0}}
    \right)^{n_T^{(0)\ge}}
    \nonumber\\
    &\quad\times
    \left[
        \frac{\rho_{s\downarrow}^{1/4}}
             {(\rho_{c0}/\Omega_{\gamma0})^{1/4}}
    \right]^{4-n_T^{(0)\ge}}
    \mathcal{T}_{\rm eq}(f)
    \nonumber\\
    &\quad\times
    \begin{cases}
        \mathcal{A}\!\left(n_T^{(0)<}\right)
        \left(\dfrac{f}{f_\star}\right)^{n_T^{(0)<}},
        & f\ll f_\star,\\[6pt]
        \mathcal{A}\!\left(n_T^{(0)\ge}\right)
        \left(\dfrac{f}{f_\star}\right)^{n_T^{(0)\ge}},
        & f\gg f_\star,
    \end{cases}
    \label{eq:NMBCsummary}
\end{align}
where
\begin{equation}
    \mathcal{A}(n)
    \equiv
    \frac{2^{-n}}{\pi}
    \Gamma^2\!\left(\frac{3-n}{2}\right),
    \label{eq:NMBCsummaryA}
\end{equation}
and
\begin{equation}
    n_T^{(0)<}
    =
    \frac{2(w_0-\tfrac13)}{w_0+\tfrac13},
    \qquad
    n_T^{(0)\ge}
    =
    \frac{2(w_1-\tfrac13)}{w_1+\tfrac13}.
    \label{eq:NMBCsummaryTilts}
\end{equation}
These expressions apply in the parameter domain
$n_T^{(0)<}<2$ ($w_0>-\tfrac13$) and $n_T^{(0)\ge}<1$ ($-\tfrac13<w_1<1$),
within the specified asymptotic regimes
well below the bounce cutoff.

The analytic spectrum reveals a key freedom
of NMBC: the low- and high-frequency tilts
are determined separately by the equations
of state during Phases~0 and~I, while the
amplitude also records the transition-time scales.
A blue low-frequency branch can therefore
coexist with a scale-invariant or red-tilted
high-frequency branch.
This addresses a central restriction of MBC:
enhancing a signal in one frequency band
no longer requires extending the same
spectral tilt to all higher frequencies.
The high-frequency branch recovers the
leading MBC prediction.
The limits of equal contraction exponents
and vanishing Phase~I duration also recover
the corresponding MBC backgrounds, providing
two independent consistency checks on
our analytic calculation.

The analytic normalization turns the
$\Delta N_{\rm eff}$ bound into a direct
constraint on the bounce energy.
For the scale-invariant or red-tilted
high-frequency branch considered here,
the adopted integrated benchmark limits
this energy to order the reduced Planck scale.
At the scale-invariant point, the radiation
budget $1.7\times10^{-6}$ over the illustrative
high-frequency interval
$10^{-10}\le f\le10^9\,\mathrm{Hz}$ gives
$\rho_{s\downarrow}^{1/4}
\lesssim2.50\times10^{15}\,\mathrm{TeV}
\simeq1.03\,m_{\rm pl}$, provided that the adopted interval lies within the regime of validity of the high-frequency spectrum.
For comparison, the illustrative pointwise
benchmarks give
\begin{equation}
    \rho_{s\downarrow}^{1/4}
    \lesssim2.65\,m_{\rm pl}.
\end{equation}
The integrated bound depends on the spectral
shape and frequency range.
When the upper limit is set by $f_{\rm cut}$,
it also depends on the bounce energy scale
$\rho_{s\downarrow}^{1/4}$.
We leave a systematic evaluation with these
model-dependent integration limits to future work.
Including the low-frequency contribution
can only strengthen the necessary bound
obtained from the high-frequency branch alone.
Within the adopted setup, this result provides
an observational basis for restricting bounce
energies far above the Planck scale and guides
the construction of observationally compatible
bouncing models.

Our analytical result
also makes NMBC backgrounds reconstructible
from prescribed spectral features.
Using the two tilts, break frequency, and
pivot amplitude, we reconstruct the background
parameters and physical scales of four
illustrative examples.
Their analytic spectra reproduce the prescribed
features, providing an internal consistency
check of the reconstruction.
The examples in Fig.~\ref{fig:FourExamples_Tau}
show how different contraction histories
produce signals near observational benchmarks
across several frequency bands.
They also highlight the need to assess
these benchmarks together with the integrated
radiation budget, making the framework useful
for both constructing and testing candidate
bouncing backgrounds.

This reconstruction can be taken further
through a Bayesian analysis of GW observations
to infer the joint posterior distribution
of the four NMBC parameters,
$(w_0,\eta_{0\downarrow},w_1,
\rho_{s\downarrow}^{1/4})$.
Marginalizing over the remaining parameters
then gives two-dimensional credible regions.
This approach quantifies parameter degeneracies
and identifies regions compatible with the
adopted observations and priors;
see our recent work~\cite{OurFutureWork1}.

Two further calculations would sharpen these
observational tests.
Full Phase~0--I matching would resolve the
spectrum near $f_\star$, where the extrapolated
asymptotic amplitudes generally differ.
A specified smooth bounce would test the
instantaneous-transition approximation and
determine the spectrum near the characteristic
cutoff.
These developments would improve the integrated
radiation constraint and support more precise
inference from observations across multiple
frequency bands.

Beyond NMBC, the matrix-representation method
can be extended to other multiphase inflationary
and bouncing histories, including epochs
of kination and reheating.
For analytically solvable phases, it can yield
closed-form spectral expressions that complement
numerical calculations and expose how successive
stages of cosmic evolution shape the SGWB.
Dark-sector phase transitions offer another
direction: their effects on background evolution
can be represented by an appropriate sequence
of phases, while the GWs they generate require
explicit source terms.
Incorporating these sources would bring GW
production and subsequent propagation into
a common framework.

Looking ahead, complementary GW searches
across multiple frequency bands offer
opportunities to test a wide range of NMBC
backgrounds, with proposed high-frequency
detectors extending the observational reach
\cite{Aggarwal:2020olq,Bian:2025ifp}.
The analytic relations developed here provide
a route from measured spectral features to
the contraction history and bounce energy,
including scenarios with energies approaching
the Planck scale.
Connecting these effective backgrounds to
concrete quantum-gravity realizations would
further allow observational constraints
to inform proposed mechanisms for resolving
the initial singularity.

~
\acknowledgments  We thank Chang Rao and Junrong Lai
for reading the manuscript and helping us
correct a typographical error. C.L. is supported by the NSFC under Grants No.11963005 and No. 11603018, by Yunnan Provincial Foundation under Grants No.202401AT070459, No.2019FY003005, and No.2016FD006, by Young and Middle-aged Academic and Technical Leaders in Yunnan Province Program, by Yunnan Provincial High level Talent Training Support Plan Youth Top Program, by Yunnan University Donglu Talent Young Scholar, and by the NSFC under Grant No.11847301 and by the Fundamental Research Funds for the Central Universities under Grant No. 2019CDJDWL0005.

\newpage
~
\appendix
~
\section{Derivation and normalization of the primordial tensor spectrum}
\label{sec:SGWBandmatrix}

For the backgrounds specified in
Sec.~\ref{sec:effparamceBC}, we derive the
kernel-to-spectrum relation in
Eq.~\eqref{eq:phkn22essential}.
The polarization normalization, radiation-era
oscillation average, and initial-state convention
together fix the predicted SGWB amplitude.
The superscripts $(1)$ and $(0)$ denote MBC
and NMBC, respectively.

\paragraph{Primordial normalization in Phase IV.}

Assuming equal power in the two tensor
polarizations and a negligible decaying mode,
the radiation-era solution is
\begin{equation}
    h_{k,\lambda}^{[4]}(\eta)
    \simeq
    h_{k,\lambda}^{[4],\mathrm{prim}}
    \frac{\sin(k\eta)}{k\eta},
\end{equation}
where $\lambda=+,\times$ and
$h_{k,\lambda}^{[4],\mathrm{prim}}$ is the
constant superhorizon amplitude.
With the polarization normalization
$\sum_{i,j}e_{ij}^{\lambda}
e_{ij}^{\lambda'\ast}
=2\delta_{\lambda\lambda'}$,
the primordial tensor spectrum becomes
\begin{align}
    \mathcal{P}_h^{(1/0)}(k)
    &\equiv
    \frac{k^3}{2\pi^2}
    \sum_{i,j}
    \left|
        h_{ij}^{[4],\mathrm{prim}}(\boldsymbol{k})
    \right|^2
    \nonumber\\
    &=
    \frac{k^3}{2\pi^2}\,
    2\sum_{\lambda=+,\times}
    \left|h_{k,\lambda}^{[4],\mathrm{prim}}\right|^2
    \nonumber\\
    &\simeq
    \frac{k^3}{\pi^2}\,
    \frac{32\pi G\,\eta_k}{a^2(\eta_k)}
    \frac{2}{\pi}
    \begin{pmatrix}
        A_4^\ast & B_4^\ast
    \end{pmatrix}
    \begin{pmatrix}
        A_4\\[2pt]
        B_4
    \end{pmatrix}
    \kappa.
    \label{eq:defpha4b4new}
\end{align}
Here $A_4$ and $B_4$ are the Hankel-mode
coefficients of the normalized mode $u_k^{[4]}$,
and $\eta_k=k^{-1}$ is the radiation-era
horizon re-entry time.
The factor $\kappa=2$ restores the primordial
normalization from the oscillation-averaged
power, using
$\langle\sin^2(k\eta)\rangle=1/2$.
This average is distinct from the bounce-phase
average used in the main text.
Although $\eta_k$ serves as a reference time,
the spectrum is defined by the constant
primordial amplitudes.

\paragraph{Initial state and propagation kernel.}

Following Ref.~\cite{Li:2024dce},
the transfer matrix $\mathcal{X}^{(1/0)}$
relates the initial mode coefficients to
those in Phase IV:
\begin{equation}\label{eq:a4b4toa10b10}
    \begin{pmatrix}
        A_4\\[2pt]
        B_4
    \end{pmatrix}
    =
    \mathcal{X}^{(1/0)}
    \begin{pmatrix}
        A_{1/0}\\[2pt]
        B_{1/0}
    \end{pmatrix}.
\end{equation}
The initial phase is Phase~I for MBC and
Phase~0 for NMBC.
The explicit transfer matrices are given in
Eqs.~\eqref{eq:x1defMBCessential}
and \eqref{eq:x0defNMBCessential},
with their constituent matrices derived in
Appendix~\ref{sec:pgwpropagationmatrix}.

The Bunch--Davies mode satisfies
$u_k\sim e^{-ik\tau}/\sqrt{2k}$
in the initial subhorizon regime.
Since $\sigma_0=\sigma_1=-1$, this
positive-frequency mode selects the
$A_{1/0}$ branch.
To retain the $\mathcal{N}_{22}$ notation,
we use its complex conjugate for the
power-spectrum calculation:
\begin{equation}\label{eq:a10b10exp}
    \begin{pmatrix}
        A_{1/0}\\[2pt]
        B_{1/0}
    \end{pmatrix}
    =
    \begin{pmatrix}
        0\\[2pt]
        \displaystyle
        \frac{\sqrt{\pi}}{2}
        e^{-i\left(
            \frac{\tilde{\nu}_{1/0}\pi}{2}
            +\frac{\pi}{4}
        \right)}
    \end{pmatrix},
\end{equation}
up to an irrelevant overall phase.
Because the evolution equations and matching
conditions are real, conjugation leaves
the squared mode amplitude unchanged.

Defining the propagation kernel matrix by
\begin{equation}\label{eq:nxxdefineapp}
    \mathcal{N}^{(1/0)}
    \bigl(
        \{\tilde{\nu}_i\},
        \{\eta_{i\downarrow/\uparrow}\}
    \bigr)
    \equiv
    \mathcal{X}^{(1/0)\dagger}
    \mathcal{X}^{(1/0)},
\end{equation}
Eqs.~\eqref{eq:a4b4toa10b10}
and \eqref{eq:a10b10exp} give
$|A_4|^2+|B_4|^2
=(\pi/4)\mathcal{N}_{22}^{(1/0)}$.
The same conjugation symmetry implies
$\mathcal{N}_{11}^{(1/0)}
=\mathcal{N}_{22}^{(1/0)}$,
so either initial representation gives
the same spectrum.

Substituting into
Eq.~\eqref{eq:defpha4b4new}, with
$\nu_4=1$ and $k\eta_k=1$, yields
\begin{align}
    \mathcal{P}_h^{(1/0)}(k)
    &\simeq
    \frac{2\kappa}{\pi^2}\,
    \frac{H^2(\eta_k)}{m_{\rm pl}^2}\,
    \mathcal{N}_{22}^{(1/0)}
    \bigl(
        \{\tilde{\nu}_i\},
        \{\eta_{i\downarrow/\uparrow}\}
    \bigr)
    \label{eq:phfpketa}\\
    &=
    \frac{4}{\pi^2}\,
    \frac{k^4}
         {H_0^2\Omega_{\gamma0}m_{\rm pl}^2}\,
    \mathcal{N}_{22}^{(1/0)}
    \bigl(
        \{\tilde{\nu}_i\},
        \{\eta_{i\downarrow/\uparrow}\}
    \bigr).
    \label{eq:phfpketarad}
\end{align}
The second line uses $\kappa=2$ and
Eq.~\eqref{eq:Hubblek4}, with the radiation
background normalized by the present photon
density.
This recovers Eq.~\eqref{eq:phkn22essential}
with its normalization fixed explicitly.

Combining Eq.~\eqref{eq:phfpketarad} with
the frequency conversion in
Eq.~\eqref{eq:pfk2pifessential} and the
SGWB relation in Eq.~\eqref{eq:sgwbphfessential}
then connects the propagation kernel
to the present-day SGWB amplitude.

~
\section{Improved propagation matrix method for primordial gravitational waves}
\label{sec:pgwpropagationmatrix}

We derive the PGW transformation and matching
matrices within the framework of
Ref.~\cite{Li:2024dce}, with an improved
boundary-matching prescription that accounts
for the time orientation in each phase.
The essential change concerns the matching
of time derivatives across the bounce.

In Ref.~\cite{Li:2024dce}, continuity is
imposed on the derivative with respect to
the positive local conformal-time argument
$\eta$ [see Eq.~(24) therein],
\begin{equation}\label{eq:matchveleta}
    \left.
    \frac{d h_k^{[i+1]}}{d\eta^{[i+1]}}
    \right|_{m_i}
    =
    \left.
    \frac{d h_k^{[i]}}{d\eta^{[i]}}
    \right|_{m_i}.
\end{equation}
Here $\left.\cdots\right|_{m_i}$ denotes
evaluation at the boundary between phases
$i$ and $i+1$, approached from the
corresponding phase.

In this work, assuming a continuous, nonzero
scale factor and no impulsive tensor source,
we impose continuity of $h_k$ and its
derivative with respect to cosmic time $t$.
Since $d\tau=dt/a$, derivative continuity
in $t$ is equivalent to continuity in
the future-directed conformal time $\tau$:
\begin{equation}
    \left.
    \frac{d h_k^{[i+1]}}{dt}
    \right|_{m_i}
    =
    \left.
    \frac{d h_k^{[i]}}{dt}
    \right|_{m_i}
    \quad\Longleftrightarrow\quad
    \left.
    \frac{d h_k^{[i+1]}}{d\tau}
    \right|_{m_i}
    =
    \left.
    \frac{d h_k^{[i]}}{d\tau}
    \right|_{m_i}.
\end{equation}

The positive local argument
$\eta^{[i]}=|\tau-\tau_i(0)|>0$
can increase or decrease as $\tau$ increases.
Accordingly,
\begin{equation}
    \frac{d}{d\tau}
    =
    \sigma_i\frac{d}{d\eta^{[i]}},
    \qquad
    \sigma_i
    \equiv
    \frac{d\eta^{[i]}}{d\tau}
    =
    \operatorname{sgn}\!\left[\tau-\tau_i(0)\right],
\end{equation}
where $\sigma_i=\pm1$ is constant within
each phase.
The matching condition therefore becomes
\begin{equation}\label{eq:matchveletasigma}
    \sigma_{i+1}
    \left.
    \frac{d h_k^{[i+1]}}{d\eta^{[i+1]}}
    \right|_{m_i}
    =
    \sigma_i
    \left.
    \frac{d h_k^{[i]}}{d\eta^{[i]}}
    \right|_{m_i}.
\end{equation}

    For a power-law phase
    $a^{[i]}=a_i(\eta^{[i]})^{\nu_i}$
    with finite, nonzero $\nu_i$, we have
    \begin{equation}
        \dot a^{[i]}
        =
        \frac{\sigma_i\nu_i}{\eta^{[i]}},
        \qquad
        \left|a^{[i]}H^{[i]}\right|^{-1}
        =
        \frac{\eta^{[i]}}{|\nu_i|}.
    \end{equation}
 With $\nu_0,\nu_1,\nu_4>0$ and
$\nu_2,\nu_3<0$, the contraction and
expansion directions specified in
Sec.~\ref{sec:effparamceBC} give
    \begin{equation}\label{eq:sigmas}
        (\sigma_0,\sigma_1,\sigma_2,\sigma_3,\sigma_4)
        =
        (-1,-1,+1,-1,+1).
    \end{equation}
    These signs reproduce the successive
    shrinking and growing comoving Hubble
    radii and are retained when taking
    the swift-bounce limit
    $\nu_2,\nu_3\to0^{-}$.

The prescriptions in
Eqs.~\eqref{eq:matchveleta}
and \eqref{eq:matchveletasigma}
coincide when $\sigma_{i+1}=\sigma_i$.
When $\sigma_{i+1}=-\sigma_i$, continuity
in $\tau$ instead requires the local
$\eta$ derivatives to have opposite signs.
These orientation factors must therefore
be retained in both the subhorizon and
superhorizon matching matrices derived below.

\subsection{Equation of motion and general solution}

The perturbed spatially flat FLRW metric is
\begin{equation}
    ds^2 = a^2(\tau)
    \left[-d\tau^2 + (\delta_{ij}+h_{ij})\,dx^i dx^j\right],
\end{equation}
where $\tau$ is the future-directed conformal
time, $d\tau=dt/a$, and $h_{ij}$ is the
transverse and traceless tensor perturbation.
Following Ref.~\cite{Caprini:2018mtu}, we
denote the two polarization amplitudes by
$h_r$, with $r=+,\times$.

Neglecting tensor anisotropic stress, the
linearized Einstein equations give
\begin{equation}\label{eq:eomvk}
    \frac{d^2u_k}{d\tau^2}
    + \left(k^2-\frac{1}{a}\frac{d^2a}{d\tau^2}\right)u_k
    = 0,
\end{equation}
where $u_k$ is the Fourier mode with comoving
wavenumber $k$ of the normalized tensor
perturbation
$u_r\equiv a(\tau)h_r/\sqrt{16\pi G}$.
We suppress the polarization index because
both polarizations obey the same equation.

For a phase with constant equation-of-state
parameter $w_i\ne-\tfrac13$, we write the
scale factor in terms of the positive local
argument $\eta^{[i]}=|\tau-\tau_i(0)|>0$ as
\begin{equation}\label{eq:ataueta}
    a^{[i]}(\tau)
    = a_i\left(\eta^{[i]}\right)^{\nu_i},
\end{equation}
where $a_i$ is a normalization constant
and $[i]$ labels the cosmic phase.
The exponent is related to $w_i$ by
\begin{equation}\label{eq:nuw}
    \nu_i = \frac{2}{3w_i+1}.
\end{equation}

Within each phase, we write
$\eta=\eta^{[i]}$ for simplicity.
Since $\sigma_i$ is constant and
$\sigma_i^2=1$, the second derivatives satisfy
$d^2/d\tau^2=d^2/d\eta^2$ within that phase.
Thus, the orientation signs leave the
mode equation unchanged within each phase,
although they remain essential to
the boundary matching.
Substituting Eq.~\eqref{eq:ataueta}
into Eq.~\eqref{eq:eomvk} gives
\begin{equation}\label{eq:eomvksim}
    \frac{d^2u_k^{[i]}}{d\eta^2}
    + \left[
        k^2-\frac{1}{\eta^2}
        \left(\tilde{\nu}_i^2-\frac{1}{4}\right)
    \right]u_k^{[i]}
    = 0,
\end{equation}
where
\begin{equation}
    \tilde{\nu}_i
    \equiv \left|\nu_i-\frac{1}{2}\right|
    = \left|\frac{3(1-w_i)}{2(3w_i+1)}\right|.
\end{equation}

The general solution is
\begin{equation}\label{eqsi:Hankelsols}
    u_k^{[i]}(\eta)
    = \sqrt{\eta}\left[
        A_i H_{\tilde{\nu}_i}^{(1)}(k\eta)
        + B_i H_{\tilde{\nu}_i}^{(2)}(k\eta)
    \right],
\end{equation}
where $H_{\tilde{\nu}_i}^{(1)}$ and
$H_{\tilde{\nu}_i}^{(2)}$ are Hankel functions
of the first and second kind, respectively.
For each mode $k$, the coefficients $A_i$
and $B_i$ are constant within each phase
and are determined by the initial state
and boundary-matching conditions.

\subsection{Solutions in the subhorizon and superhorizon limits}

We match the solutions across phase boundaries using their asymptotic
forms in the appropriate subhorizon or superhorizon limit.

\textbf{Subhorizon limit ($k\eta\gg1$).}
For fixed $\tilde{\nu}_i$, the leading asymptotic form of
Eq.~\eqref{eqsi:Hankelsols} is
\begin{equation}\label{eqsi:gensolsub}
    u_{k\uparrow}^{[i]}
    \simeq \sqrt{\frac{2}{\pi k}}
    \left[
        A_i e^{i\left(k\eta-\frac{\tilde{\nu}_i\pi}{2}
        -\frac{\pi}{4}\right)}
        + B_i e^{-i\left(k\eta-\frac{\tilde{\nu}_i\pi}{2}
        -\frac{\pi}{4}\right)}
    \right],
\end{equation}
where the subscript $\uparrow$ denotes the solution in the
subhorizon limit.

Using $h_k^{[i]}=\sqrt{16\pi G}\,a^{-1}(\eta)u_k^{[i]}$, we obtain
the corresponding tensor mode,
\begin{equation}
    h_{k\uparrow}^{[i]}
    \simeq \frac{\sqrt{16\pi G}}{a(\eta)}
    \sqrt{\frac{2}{\pi k}}
    \left[
        A_i e^{i\left(k\eta-\frac{\tilde{\nu}_i\pi}{2}
        -\frac{\pi}{4}\right)}
        + B_i e^{-i\left(k\eta-\frac{\tilde{\nu}_i\pi}{2}
        -\frac{\pi}{4}\right)}
    \right].
\end{equation}
Differentiating this approximation with respect to the future-directed
conformal time $\tau$ gives
\begin{equation}
\begin{aligned}
    \frac{d h_{k\uparrow}^{[i]}}{d\tau}
    = \sigma_i\frac{d h_{k\uparrow}^{[i]}}{d\eta}
    \simeq
    &\sigma_i\frac{\sqrt{16\pi G}}{a(\eta)}
    \sqrt{\frac{2}{\pi k}}
    \Biggl[
        A_i\left(-\frac{\nu_i}{\eta}+ik\right)
        e^{i\left(k\eta-\frac{\tilde{\nu}_i\pi}{2}
        -\frac{\pi}{4}\right)}
    \\
    &   + B_i\left(-\frac{\nu_i}{\eta}-ik\right)
        e^{-i\left(k\eta-\frac{\tilde{\nu}_i\pi}{2}
        -\frac{\pi}{4}\right)}
    \Biggr].
\end{aligned}
\end{equation}
The terms proportional to $\nu_i/\eta$ arise from differentiating
$a^{-1}(\eta)$; higher-order accuracy also requires retaining the
subleading terms in the Hankel expansion.

\textbf{Superhorizon limit ($k\eta\ll1$).}
For noninteger $\tilde{\nu}_i>0$, retaining the leading term of each
independent small-argument branch gives
\begin{equation}\label{eqsi:gensolsup}
    u_{k\downarrow}^{[i]}
    \simeq \sqrt{\eta}
    \left[
        E_i(k\eta)^{\tilde{\nu}_i}
        + F_i(k\eta)^{-\tilde{\nu}_i}
    \right],
\end{equation}
where the subscript $\downarrow$ denotes the solution in the
superhorizon limit. The coefficients $E_i$ and $F_i$ are related
to $A_i$ and $B_i$ by
\begin{equation}\label{eqsi:tranmatr}
    \begin{pmatrix}
        E_i\\
        F_i
    \end{pmatrix}
    =
    T_i
    \begin{pmatrix}
        A_i\\
        B_i
    \end{pmatrix},
\end{equation}
where the transformation matrix is
\begin{equation}\label{eq:tranmatrix}
    T_i \equiv
    \begin{pmatrix}
        \alpha_i & \alpha_i^\ast\\
        \beta_i & \beta_i^\ast
    \end{pmatrix}.
\end{equation}
Its elements follow from the small-argument expansions of the
Hankel functions,
\begin{equation}\label{eqsi:alphaexp}
\begin{aligned}
    \alpha_i
    \equiv
    2^{-\tilde{\nu}_i}
    \left[
        \frac{1}{\Gamma(\tilde{\nu}_i+1)}
        -i\frac{\Gamma(-\tilde{\nu}_i)
        \cos(\pi\tilde{\nu}_i)}{\pi}
    \right]
    =
    \frac{i\,2^{-\tilde{\nu}_i}e^{-i\pi\tilde{\nu}_i}}
    {\sin(\pi\tilde{\nu}_i)\Gamma(\tilde{\nu}_i+1)},
\end{aligned}
\end{equation}
and
\begin{equation}
    \beta_i
    \equiv -\frac{i\,2^{\tilde{\nu}_i}
    \Gamma(\tilde{\nu}_i)}{\pi}.
\end{equation}
Here, $\Gamma(z)$ is the gamma function, defined by
$\Gamma(z)=\int_0^\infty t^{z-1}e^{-t}\,dt$ for
$\operatorname{Re}z>0$ and extended elsewhere by analytic continuation.
The second equality in Eq.~\eqref{eqsi:alphaexp} follows from the
reflection formula
$\Gamma(z)\Gamma(1-z)=\pi/\sin(\pi z)$.

The corresponding tensor mode and its derivative with respect to
$\tau$ are
\begin{align}
    h_{k\downarrow}^{[i]}
    &\simeq
    \frac{\sqrt{16\pi G}}{a(\eta)}\sqrt{\eta}
    \left[
        E_i(k\eta)^{\tilde{\nu}_i}
        + F_i(k\eta)^{-\tilde{\nu}_i}
    \right],
    \\
    \frac{d h_{k\downarrow}^{[i]}}{d\tau}
    &= \sigma_i\frac{d h_{k\downarrow}^{[i]}}{d\eta}
    \simeq
    \sigma_i\frac{\sqrt{16\pi G}}{a(\eta)\sqrt{\eta}}
    \left[
        E_i\chi_i(k\eta)^{\tilde{\nu}_i}
        + F_i\bar{\chi}_i(k\eta)^{-\tilde{\nu}_i}
    \right],
\end{align}
where
\begin{align} \label{eq:chiibdef}
    \chi_i
    \equiv
    \tilde{\nu}_i-\left(\nu_i-\frac{1}{2}\right),
    \quad 
    \bar{\chi}_i
    \equiv
    -\tilde{\nu}_i-\left(\nu_i-\frac{1}{2}\right).
\end{align}

For integer $\tilde{\nu}_i$, the small-argument Hankel expansions
contain logarithmic terms and must be treated separately.
In particular, when $\tilde{\nu}_i=0$, the two independent leading
behaviors of $u_k^{[i]}$ are $\sqrt{\eta}$ and
$\sqrt{\eta}\ln(k\eta)$.

\subsection{Matching conditions and their matrix representations}

At the boundary between phases $i$ and $i+1$, we impose continuity
of the tensor mode and its future-directed conformal-time derivative.
These conditions assume a continuous scale factor and no impulsive
tensor source:
\begin{equation}
\label{eqsi:matconone}
    h_k^{[i+1]}(\eta_{m_i})
    =
    h_k^{[i]}(\eta_{m_i}),
\end{equation}
and
\begin{equation}
\label{eqsi:matcontwotau}
    \left.
    \frac{d h_k^{[i+1]}}{d\tau}
    \right|_{\eta=\eta_{m_i}}
    =
    \left.
    \frac{d h_k^{[i]}}{d\tau}
    \right|_{\eta=\eta_{m_i}}.
\end{equation}
Here $\eta$ denotes the positive local conformal-time magnitude,
and $\eta_{m_i}>0$ is its common value on the two sides of the
interface. The coordinate $\tau$ increases throughout the physical
evolution. Since $d/d\tau=\sigma_j d/d\eta$ within phase $j$,
Eq.~\eqref{eqsi:matcontwotau} is equivalent to
\begin{equation}
    \left.
    \sigma_{i+1}\frac{d h_k^{[i+1]}}{d\eta}
    \right|_{\eta=\eta_{m_i}}
    =
    \left.
    \sigma_i\frac{d h_k^{[i]}}{d\eta}
    \right|_{\eta=\eta_{m_i}}.
\end{equation}
We denote the matching magnitude by $\eta_{i\uparrow}$ in the
subhorizon regime and by $\eta_{i\downarrow}$ in the superhorizon
regime.

These conditions determine the subhorizon and superhorizon matching
matrices, denoted by $\mathcal{M}_{i\uparrow}$ and
$\mathcal{M}_{i\downarrow}$, respectively.

\begin{enumerate}

\item \textbf{Subhorizon matching matrix
$\mathcal{M}_{i\uparrow}$.}

For a mode with comoving wavenumber $k$, the subhorizon matching
matrix relates the Hankel coefficients across an interface satisfying
$k\eta_{i\uparrow}\gg1$:
\begin{equation}
    \begin{pmatrix}
        A_{i+1}\\
        B_{i+1}
    \end{pmatrix}
    =
    \mathcal{M}_{i\uparrow}
    \begin{pmatrix}
        A_i\\
        B_i
    \end{pmatrix}.
\end{equation}
The transition between phases II and III in
Fig.~\ref{fig:NMBC} provides an example.

At this interface, define
\begin{equation}
    L\equiv\eta_{i\uparrow}>0,
    \qquad
    \phi_j\equiv
    kL-\frac{\pi\tilde{\nu}_j}{2}-\frac{\pi}{4},
    \qquad j=i,i+1.
\end{equation}
Using Eqs.~\eqref{eqsi:matconone} and
\eqref{eqsi:matcontwotau}, together with the asymptotic solution
in Eq.~\eqref{eqsi:gensolsub}, the matching conditions take the form
\begin{equation}
    \mathcal{K}_{i+1,\uparrow}(L)
    \begin{pmatrix}
        A_{i+1}\\
        B_{i+1}
    \end{pmatrix}
    =
    \mathcal{K}_{i,\uparrow}(L)
    \begin{pmatrix}
        A_i\\
        B_i
    \end{pmatrix}.
\end{equation}
Both coefficient matrices are evaluated at the same boundary
magnitude $L$. Common prefactors cancel because the scale factor
is continuous at the interface.

At leading subhorizon order, the coefficient matrices are
\begin{equation}
    \mathcal{K}_{j,\uparrow}(L)
    \simeq
    \begin{pmatrix}
        e^{i\phi_j}
        &
        e^{-i\phi_j}
        \\[4pt]
        \sigma_j ik\,e^{i\phi_j}
        &
        -\sigma_j ik\,e^{-i\phi_j}
    \end{pmatrix},
    \qquad j=i,i+1.
\end{equation}
Differentiating the scale-factor prefactor in the leading
asymptotic solution produces $-\nu_j/L$. We retain only the
leading derivative terms, using
\begin{equation}
    \sigma_j\left(-\frac{\nu_j}{L}\pm ik\right)
    \simeq
    \pm\sigma_j ik,
\end{equation}
in the limit $kL\gg1$ at fixed background exponents and Hankel
orders.

Defining the relative orientation
\begin{equation}\label{eq:ridefinesigmai}
    r_i\equiv\frac{\sigma_i}{\sigma_{i+1}},
\end{equation}
we obtain
\begin{equation}
\label{eqsi:matchmatrup}
    \mathcal{M}_{i\uparrow}
    =
    \mathcal{K}_{i+1,\uparrow}^{-1}(L)
    \mathcal{K}_{i,\uparrow}(L)
    \simeq
    \frac{1}{2}
    \begin{pmatrix}
        (1+r_i)e^{i(\phi_i-\phi_{i+1})}
        &
        (1-r_i)e^{-i(\phi_i+\phi_{i+1})}
        \\[4pt]
        (1-r_i)e^{i(\phi_i+\phi_{i+1})}
        &
        (1+r_i)e^{-i(\phi_i-\phi_{i+1})}
    \end{pmatrix}.
\end{equation}
Thus equal orientations, $r_i=+1$, give a diagonal matrix,
whereas opposite orientations, $r_i=-1$, give an off-diagonal
matrix at this order.

For the phase-II-to-phase-III transition, the time orientations
satisfy
\begin{equation}
    \sigma_2=+1,
    \qquad
    \sigma_3=-1,
    \qquad
    r_2=-1.
\end{equation}
Consequently, for $k\eta_{2\uparrow}\gg1$,
\begin{align}
    \mathcal{M}_{2\uparrow}
    &\simeq
    \begin{pmatrix}
        0 & e^{-i(\phi_2+\phi_3)}\\[4pt]
        e^{i(\phi_2+\phi_3)} & 0
    \end{pmatrix}
    \nonumber\\[6pt]
    &=
    \begin{pmatrix}
        0
        &
        \exp\!\left[
            -2ik\eta_{2\uparrow}
            +\frac{i\pi}{2}
            (\tilde{\nu}_2+\tilde{\nu}_3+1)
        \right]
        \\[6pt]
        \exp\!\left[
            2ik\eta_{2\uparrow}
            -\frac{i\pi}{2}
            (\tilde{\nu}_2+\tilde{\nu}_3+1)
        \right]
        &
        0
    \end{pmatrix}.
    \label{eq:m2exp}
\end{align}

  \item \textbf{Superhorizon matching matrix
$\mathcal{M}_{i\downarrow}$.}

At the transition between phases $i$ and $i+1$, we assume that the
scale factor is continuous and that both local conformal-time
arguments equal $\eta_{i\downarrow}$. Within the leading superhorizon
approximation, the matching matrix $\mathcal{M}_{i\downarrow}$
relates the coefficients in the adjacent phases through
\begin{equation}
    \begin{pmatrix}
        E_{i+1}\\
        F_{i+1}
    \end{pmatrix}
    =
    \mathcal{M}_{i\downarrow}
    \begin{pmatrix}
        E_i\\
        F_i
    \end{pmatrix},
    \qquad k\eta_{i\downarrow}\ll1.
\end{equation}
Examples include the transitions from Phase I to Phase II and
from Phase III to Phase IV in Fig.~\ref{fig:NMBC}.

Substituting Eq.~\eqref{eqsi:gensolsup} into the matching conditions,
Eqs.~\eqref{eqsi:matconone} and~\eqref{eqsi:matcontwotau}, and
canceling the common background factors gives
\begin{equation}
    \mathcal{K}_{i+1\downarrow}(\eta_{i\downarrow})
    \begin{pmatrix}
        E_{i+1}\\
        F_{i+1}
    \end{pmatrix}
    =
    \mathcal{K}_{i\downarrow}(\eta_{i\downarrow})
    \begin{pmatrix}
        E_i\\
        F_i
    \end{pmatrix},
\end{equation}
where both coefficient matrices are evaluated at the same boundary,
\begin{equation}
    \mathcal{K}_{j\downarrow}(\eta_{i\downarrow})
    \equiv
    \begin{pmatrix}
        \left(k\eta_{i\downarrow}\right)^{\tilde{\nu}_j}
        &
        \left(k\eta_{i\downarrow}\right)^{-\tilde{\nu}_j}
        \\[1ex]
        \sigma_j\chi_j
        \left(k\eta_{i\downarrow}\right)^{\tilde{\nu}_j}
        &
        \sigma_j\bar{\chi}_j
        \left(k\eta_{i\downarrow}\right)^{-\tilde{\nu}_j}
    \end{pmatrix},
    \qquad j=i,i+1.
\end{equation}
Defining the relative sign $r_i\equiv\sigma_i/\sigma_{i+1}$,
we obtain the general superhorizon matching matrix,
\begin{equation}\label{eq:matchmatrdown}
\begin{aligned}
    \mathcal{M}_{i\downarrow}
    &=
    \mathcal{K}_{i+1\downarrow}^{-1}(\eta_{i\downarrow})
    \mathcal{K}_{i\downarrow}(\eta_{i\downarrow})
    \\
    &=
    \frac{1}{\bar{\chi}_{i+1}-\chi_{i+1}}
    \begin{pmatrix}
        (\bar{\chi}_{i+1}-r_i\chi_i)
        \left(k\eta_{i\downarrow}\right)^{
            \tilde{\nu}_i-\tilde{\nu}_{i+1}}
        &
        (\bar{\chi}_{i+1}-r_i\bar{\chi}_i)
        \left(k\eta_{i\downarrow}\right)^{
            -(\tilde{\nu}_i+\tilde{\nu}_{i+1})}
        \\[1ex]
        (-\chi_{i+1}+r_i\chi_i)
        \left(k\eta_{i\downarrow}\right)^{
            \tilde{\nu}_i+\tilde{\nu}_{i+1}}
        &
        (-\chi_{i+1}+r_i\bar{\chi}_i)
        \left(k\eta_{i\downarrow}\right)^{
            -(\tilde{\nu}_i-\tilde{\nu}_{i+1})}
    \end{pmatrix}.
\end{aligned}
\end{equation}
These expressions assume $\tilde{\nu}_i,\tilde{\nu}_{i+1}>0$;
a logarithmic basis is required whenever either order vanishes.

Using the identities in
Eqs.~\eqref{eq:chibcge}--\eqref{eq:chisquare}, presented in the next
subsection, we obtain explicit expressions for the relevant transitions.
For the five-phase sign convention adopted here,
$r_0=1$ and $r_1=r_3=-1$. In phases II and III,
$w_2,w_3<-\tfrac{1}{3}$ imply $\nu_2,\nu_3<0$,
with $\nu_2,\nu_3\to0^-$ as $w_2,w_3\to-\infty$.

\begin{enumerate}

\item \textbf{Transition from Phase I to Phase II.}

For $\nu_2<\tfrac{1}{2}$ and $r_1=-1$, the matching matrix becomes
\begin{equation}\label{eq:m1down}
    \mathcal{M}_{1\downarrow}(\nu_2<\tfrac{1}{2})
    =
    \begin{cases}
        \begin{pmatrix}
            0
            &
            \dfrac{\tilde{\nu}_1}{\tilde{\nu}_2}
            \left(k\eta_{1\downarrow}\right)^{
                -(\tilde{\nu}_1+\tilde{\nu}_2)}
            \\[1ex]
            \left(k\eta_{1\downarrow}\right)^{
                \tilde{\nu}_1+\tilde{\nu}_2}
            &
            \left(1-\dfrac{\tilde{\nu}_1}{\tilde{\nu}_2}\right)
            \left(k\eta_{1\downarrow}\right)^{
                -(\tilde{\nu}_1-\tilde{\nu}_2)}
        \end{pmatrix},
        & \nu_1>\tfrac{1}{2},
        \\[3ex]
        \begin{pmatrix}
            -\dfrac{\tilde{\nu}_1}{\tilde{\nu}_2}
            \left(k\eta_{1\downarrow}\right)^{
                \tilde{\nu}_1-\tilde{\nu}_2}
            &
            0
            \\[1ex]
            \left(1+\dfrac{\tilde{\nu}_1}{\tilde{\nu}_2}\right)
            \left(k\eta_{1\downarrow}\right)^{
                \tilde{\nu}_1+\tilde{\nu}_2}
            &
            \left(k\eta_{1\downarrow}\right)^{
                -(\tilde{\nu}_1-\tilde{\nu}_2)}
        \end{pmatrix},
        & \nu_1<\tfrac{1}{2}.
    \end{cases}
\end{equation}

\item \textbf{Transition from Phase III to Phase IV.}

For $\nu_3<\tfrac{1}{2}$ and $r_3=-1$, the matching matrix becomes
\begin{equation}\label{eq:m3down}
    \mathcal{M}_{3\downarrow}(\nu_3<\tfrac{1}{2})
    =
    \begin{cases}
        \begin{pmatrix}
            \left(1-\dfrac{\tilde{\nu}_3}{\tilde{\nu}_4}\right)
            \left(k\eta_{3\downarrow}\right)^{
                \tilde{\nu}_3-\tilde{\nu}_4}
            &
            \left(k\eta_{3\downarrow}\right)^{
                -(\tilde{\nu}_3+\tilde{\nu}_4)}
            \\[1ex]
            \dfrac{\tilde{\nu}_3}{\tilde{\nu}_4}
            \left(k\eta_{3\downarrow}\right)^{
                \tilde{\nu}_3+\tilde{\nu}_4}
            &
            0
        \end{pmatrix},
        & \nu_4>\tfrac{1}{2},
        \\[3ex]
        \begin{pmatrix}
            -\dfrac{\tilde{\nu}_3}{\tilde{\nu}_4}
            \left(k\eta_{3\downarrow}\right)^{
                \tilde{\nu}_3-\tilde{\nu}_4}
            &
            0
            \\[1ex]
            \left(1+\dfrac{\tilde{\nu}_3}{\tilde{\nu}_4}\right)
            \left(k\eta_{3\downarrow}\right)^{
                \tilde{\nu}_3+\tilde{\nu}_4}
            &
            \left(k\eta_{3\downarrow}\right)^{
                -(\tilde{\nu}_3-\tilde{\nu}_4)}
        \end{pmatrix},
        & \nu_4<\tfrac{1}{2}.
    \end{cases}
\end{equation}

\item \textbf{Transition from Phase 0 to Phase I.}

For this transition, we restrict Phase I to
$-\tfrac{1}{3}<w_1<1$, corresponding to $\nu_1>\tfrac{1}{2}$.
With $r_0=1$, the matching matrix takes the form
\begin{equation}\label{eq:NMBClowm0down}
    \mathcal{M}_{0\downarrow}(\nu_1>\tfrac{1}{2})
    =
    \begin{cases}
        \begin{pmatrix}
            \left(k\eta_{0\downarrow}\right)^{
                \tilde{\nu}_0-\tilde{\nu}_1}
            &
            \left(1-\dfrac{\tilde{\nu}_0}{\tilde{\nu}_1}\right)
            \left(k\eta_{0\downarrow}\right)^{
                -(\tilde{\nu}_0+\tilde{\nu}_1)}
            \\[1ex]
            0
            &
            \dfrac{\tilde{\nu}_0}{\tilde{\nu}_1}
            \left(k\eta_{0\downarrow}\right)^{
                -(\tilde{\nu}_0-\tilde{\nu}_1)}
        \end{pmatrix},
        & \nu_0>\tfrac{1}{2},
        \\[3ex]
        \begin{pmatrix}
            \left(1+\dfrac{\tilde{\nu}_0}{\tilde{\nu}_1}\right)
            \left(k\eta_{0\downarrow}\right)^{
                \tilde{\nu}_0-\tilde{\nu}_1}
            &
            \left(k\eta_{0\downarrow}\right)^{
                -(\tilde{\nu}_0+\tilde{\nu}_1)}
            \\[1ex]
            -\dfrac{\tilde{\nu}_0}{\tilde{\nu}_1}
            \left(k\eta_{0\downarrow}\right)^{
                \tilde{\nu}_0+\tilde{\nu}_1}
            &
            0
        \end{pmatrix},
        & \nu_0<\tfrac{1}{2}.
    \end{cases}
\end{equation}
We leave the branch $\nu_1<\tfrac{1}{2}$ for future work.

\end{enumerate}
\end{enumerate}

\subsection{Algebraic identities for $\chi_i$ and $\bar{\chi}_i$}
\label{sec:inequality}

The definitions in Eq.~\eqref{eq:chiibdef} imply the following
algebraic identities, also given in Ref.~\cite{Li:2024dce}.

\begin{enumerate}

\item \textbf{Values in the two branches.}
\begin{align}
    \chi_i &= 0,
    \qquad \bar{\chi}_i = -2\tilde{\nu}_i,
    \qquad \nu_i>\tfrac{1}{2},
    \label{eq:chibcge}
    \\
    \chi_i &= 2\tilde{\nu}_i,
    \qquad \bar{\chi}_i = 0,
    \qquad \nu_i\le\tfrac{1}{2}.
    \label{eq:chibcle}
\end{align}

\item \textbf{Vanishing product.}
\begin{equation}
    \chi_i\bar{\chi}_i=0,
    \qquad \text{for all }\nu_i.
\end{equation}

\item \textbf{Squares.}
\begin{align}
    \chi_i^2 &= 0,
    \qquad \bar{\chi}_i^2 = 4\tilde{\nu}_i^2,
    \qquad \nu_i>\tfrac{1}{2},
    \\
    \chi_i^2 &= 4\tilde{\nu}_i^2,
    \qquad \bar{\chi}_i^2 = 0,
    \qquad \nu_i\le\tfrac{1}{2}.
    \label{eq:chisquare}
\end{align}

\end{enumerate}

These identities simplify the superhorizon matching matrix in
Eq.~\eqref{eq:matchmatrdown} and make explicit the two branches
that enter the SGWB spectrum. At $\nu_i=\tfrac{1}{2}$, one has
$\tilde{\nu}_i=\chi_i=\bar{\chi}_i=0$. The algebraic identities
remain valid at this point, but the matching problem requires
the logarithmic superhorizon basis.

~
\section{Radiation-era background relations}
\label{app:PhaseIVbackground}

We derive the Phase-IV relations connecting
the transition parameter $\eta_{s\downarrow}$
to the bounce energy scale and scale factor.
We approximate the radiation background
using the present photon density, with
$a_0=1$:
\begin{equation}
    \rho_\gamma(a)
    =
    \rho_{\gamma0}a^{-4},
    \qquad
    \rho_{\gamma0}
    =
    \Omega_{\gamma0}\rho_{c0}.
\end{equation}
This approximation neglects contributions
from other relativistic species and changes
in the thermal degrees of freedom.

Using $\rho_{c0}=3H_0^2m_{\rm pl}^2$,
the Friedmann equation becomes
\begin{equation}
    H^2(a)
    =
    \frac{\rho_{\gamma0}a^{-4}}
         {3m_{\rm pl}^2}
    =
    H_0^2\Omega_{\gamma0}a^{-4}.
\end{equation}
With the local conformal-time origin adopted
in Phase~IV, the radiation-era solution is
\begin{equation}
    a(\eta)
    =
    H_0\sqrt{\Omega_{\gamma0}}\,\eta,
    \qquad
    H^2(\eta)
    =
    \frac{1}
         {H_0^2\Omega_{\gamma0}\eta^4}.
\end{equation}

For modes re-entering the horizon during
Phase~IV, the condition
$k=a(\eta_k)H(\eta_k)$ gives
\begin{equation}
    k
    =
    \frac{H_0\sqrt{\Omega_{\gamma0}}}
         {a(\eta_k)}
    =
    \frac{1}{\eta_k},
    \qquad
    a(\eta_k)
    =
    \frac{H_0\sqrt{\Omega_{\gamma0}}}{k}.
\end{equation}
Substituting into the Friedmann equation yields
\begin{equation}\label{eq:Hubblek4}
    H^2(\eta_k)
    =
    \frac{k^4}{H_0^2\Omega_{\gamma0}}
    =
    \frac{1}
         {H_0^2\Omega_{\gamma0}\eta_k^4}.
\end{equation}

At the onset of Phase~IV, the bounce energy
density defined in Eq.~\eqref{eq:rhosdd}
therefore satisfies
\begin{equation}
    \frac{\rho_{s\downarrow}}{3m_{\rm pl}^2}
    =
    H^2(\eta_{s\downarrow})
    =
    \frac{1}
         {H_0^2\Omega_{\gamma0}
          \eta_{s\downarrow}^4}.
\end{equation}
Solving for $\eta_{s\downarrow}$ gives
Eq.~\eqref{eq:etasrhos}, while evaluating
$a(\eta)$ at the same boundary yields
the scale-factor relation in
Eq.~\eqref{eq:asdrhosdeq}.

\section{Derivation of the SGWB spectrum in MBC}
\label{sec:mbcdetailapp}

We derive the MBC spectrum presented in
Eq.~\eqref{eq:MBCoverlineSGWBessential}.
For the background specified in
Eqs.~\eqref{eq:MBCsetw} and \eqref{eq:MBCseteta}
and illustrated in Fig.~\ref{fig:NMBC},
the modes first exit the horizon during Phase~I.
Their subsequent propagation through
Phases~II and~III gives
\begin{equation}\label{eq:hfa4a1}
\begin{aligned}
    \begin{pmatrix}
        A_4\\
        B_4
    \end{pmatrix}
    &=
    T_4^{-1}
    \begin{pmatrix}
        E_4\\
        F_4
    \end{pmatrix}
    \\
    &=
    T_4^{-1}\mathcal{M}_{3\downarrow}
    \begin{pmatrix}
        E_3\\
        F_3
    \end{pmatrix}
    \\
    &=
    T_4^{-1}\mathcal{M}_{3\downarrow}T_3
    \begin{pmatrix}
        A_3\\
        B_3
    \end{pmatrix}
    \\
    &=
    T_4^{-1}\mathcal{M}_{3\downarrow}T_3
    \mathcal{M}_{2\uparrow}
    \begin{pmatrix}
        A_2\\
        B_2
    \end{pmatrix}
    \\
    &=
    T_4^{-1}\mathcal{M}_{3\downarrow}T_3
    \mathcal{M}_{2\uparrow}T_2^{-1}
    \begin{pmatrix}
        E_2\\
        F_2
    \end{pmatrix}
    \\
    &=
    T_4^{-1}\mathcal{M}_{3\downarrow}T_3
    \mathcal{M}_{2\uparrow}T_2^{-1}
    \mathcal{M}_{1\downarrow}
    \begin{pmatrix}
        E_1\\
        F_1
    \end{pmatrix}
    \\
    &=
    T_4^{-1}\mathcal{M}_{3\downarrow}T_3
    \mathcal{M}_{2\uparrow}T_2^{-1}
    \mathcal{M}_{1\downarrow}T_1
    \begin{pmatrix}
        A_1\\
        B_1
    \end{pmatrix}.
\end{aligned}
\end{equation}
The total transfer matrix is therefore
\begin{equation}\label{eq:mbcx1def}
    \mathcal{X}^{(1)}
    =
    T_4^{-1}\mathcal{M}_{3\downarrow}T_3
    \mathcal{M}_{2\uparrow}T_2^{-1}
    \mathcal{M}_{1\downarrow}T_1.
\end{equation}
Using the complex-conjugate Bunch--Davies
representation explained in
Appendix~\ref{sec:SGWBandmatrix},
the Phase-I coefficients are
\begin{equation}\label{eq:MBCa1b1exp}
    \begin{pmatrix}
        A_1\\[2pt]
        B_1
    \end{pmatrix}
    =
    \begin{pmatrix}
        0\\[2pt]
        \displaystyle
        \frac{\sqrt{\pi}}{2}
        e^{-i\left(
            \frac{\tilde{\nu}_1\pi}{2}
            +\frac{\pi}{4}
        \right)}
    \end{pmatrix}.
\end{equation}

\subsection{Propagation kernel for $\sin(2k\eta_{2\uparrow})\neq0$}

To evaluate $\mathcal{X}^{(1)}$, we substitute
the MBC parameters in
Eqs.~\eqref{eq:MBCsetw}, \eqref{eq:MBCseteta},
\eqref{eq:MBCsetnu}, and \eqref{eq:MBCsettnu}
into the transformation matrix in
Eq.~\eqref{eq:tranmatrix} and the matching
matrices in Eqs.~\eqref{eq:m1down},
\eqref{eq:m2exp}, and \eqref{eq:m3down}.
For $k\eta_{s\downarrow}\ll1$ and
$k\eta_{2\uparrow}\gg1$, the leading
propagation kernel is
\begin{equation}\label{eq:N22MBCLeading}
    \mathcal{N}_{22}^{(1)}
    =
    \left[
        \mathcal{X}^{(1)\dagger}\mathcal{X}^{(1)}
    \right]_{22}
    \simeq
    \mathcal{C}^{(1)}
    \frac{
        \sin^2\!\left(2k\eta_{2\uparrow}\right)
    }{
        \left(k\eta_{s\downarrow}\right)^{n^{(1)}}
    }.
\end{equation}
For $w_1>-\tfrac13$, the coefficient and
exponent are
\begin{equation}\label{eq:C1MBCLeading}
    \mathcal{C}^{(1)}
    =
    \frac{
        2^{2\nu_1-1}
        \Gamma^2\!\left(\nu_1+\tfrac12\right)
    }{\pi}
    =
    \frac{2^{\frac{4}{3w_1+1}-1}}{\pi}
    \Gamma^2\!\left(
        \frac12+\frac{2}{3w_1+1}
    \right)
\end{equation}
and
\begin{equation}\label{eq:n1MBCLeading}
    n^{(1)}
    =
    2+2\nu_1
    =
    \frac{6(w_1+1)}{3w_1+1}.
\end{equation}
At integer values of $\tilde{\nu}_1$,
including $\tilde{\nu}_1=0$ ($w_1=1$),
the intermediate Hankel expansions require
the corresponding logarithmic terms;
the expressions for $\mathcal{C}^{(1)}$
and $n^{(1)}$ remain valid by continuity.
Near the zeros of
$\sin(2k\eta_{2\uparrow})$, subleading
contributions must be retained.
The phase-node solutions are treated in
Appendix~\ref{sec:phasenodeall}.

Uniform averaging over the bounce phase gives
\begin{equation}\label{eq:MBCsin2ave}
    \left\langle
        \sin^2\!\left(2k\eta_{2\uparrow}\right)
    \right\rangle
    =
    \frac12.
\end{equation}
The phase-averaged leading kernel is therefore
\begin{equation}\label{eq:aveN22MBCLeading}
    \left\langle\mathcal{N}_{22}^{(1)}\right\rangle
    \simeq
    \frac{\overline{\mathcal{C}}^{(1)}}
         {\left(k\eta_{s\downarrow}\right)^{n^{(1)}}},
\end{equation}
with
\begin{equation}\label{eq:avec1MBCLeading}
    \overline{\mathcal{C}}^{(1)}
    \equiv
    \frac{\mathcal{C}^{(1)}}{2}.
\end{equation}

\subsection{SGWB spectrum in MBC}

Substituting Eq.~\eqref{eq:N22MBCLeading}
into Eq.~\eqref{eq:phfpketarad} gives
the leading primordial tensor spectrum,
\begin{equation}\label{eq:unavephf1mbc}
    \mathcal{P}_h^{(1)}(k)
    \simeq
    \frac{4}{\pi^2}\,
    \frac{k^4}
         {H_0^2\Omega_{\gamma0}m_{\rm pl}^2}\,
    \frac{\mathcal{C}^{(1)}}
         {(k\eta_{s\downarrow})^{n^{(1)}}}\,
    \sin^2\!\left(2k\eta_{2\uparrow}\right).
\end{equation}
Averaging over the bounce phase yields
\begin{equation}\label{eq:phf1mbc}
    \overline{\mathcal{P}}_h^{(1)}(k)
    \simeq
    \frac{4}{\pi^2}\,
    \frac{k^4}
         {H_0^2\Omega_{\gamma0}m_{\rm pl}^2}\,
    \frac{\overline{\mathcal{C}}^{(1)}}
         {(k\eta_{s\downarrow})^{n^{(1)}}}.
\end{equation}
The overbar denotes averaging over
$2k\eta_{2\uparrow}$, distinct from the
radiation-era oscillation average used
to normalize the primordial spectrum.

We then substitute Eq.~\eqref{eq:phf1mbc}
into Eq.~\eqref{eq:sgwbphfessential},
use $k=2\pi f$ for $a_0=1$, and express
$\eta_{s\downarrow}$ in terms of the bounce
energy scale $\rho_{s\downarrow}^{1/4}$
through Eq.~\eqref{eq:etasrhos}.
This gives the phase-averaged MBC
SGWB spectrum:%
\footnote{
    The frequency prefactor in the SGWB
    expression of Ref.~\cite{Li:2025ilc}
    contains a typographical error:
    $1/(2\pi)^{-n_T-1}$ should be
    $1/(2\pi)^{-n_T+1}$ for the
    reference-scale convention used here.
    We thank Chang Rao and Junrong Lai
    for pointing out this error and helping
    us correct the same error in this manuscript.
    This correction is separate from the
    normalization factor $8/\pi$ included
    in the present calculation.
}
\begin{align}
    \overline{\Omega}_{\rm GW}^{(1)}(f)h^2
    &\simeq
    \frac{h^2}{24}\,
    \frac{8}{\pi}
    \left(
        \frac{f_{H_0}}{f_{m_{\rm pl}}}
    \right)^2
    \frac{\overline{\mathcal{C}}^{(1)}}
         {(2\pi)^{-n_T^{(1)}+1}}
    \left(\frac{f}{f_{H_0}}\right)^{n_T^{(1)}}
    \nonumber\\
    &\quad\times
    \left[
        \frac{\rho_{s\downarrow}^{1/4}}
             {(\rho_{c0}/\Omega_{\gamma0})^{1/4}}
    \right]^{4-n_T^{(1)}}
    \mathcal{T}_{\rm eq}(f).
    \label{eq:MBCoverlineSGWB}
\end{align}
The primordial tensor spectral index follows
directly from Eq.~\eqref{eq:phf1mbc}:
\begin{equation}\label{eq:MBCntw1}
    n_T^{(1)}
    \equiv
    \frac{d\ln\overline{\mathcal{P}}_h^{(1)}(k)}
         {d\ln k}
    =
    4-n^{(1)}
    =
    \frac{2(w_1-\tfrac13)}{w_1+\tfrac13},
    \qquad w_1>-\frac13.
\end{equation}
We use the reference inverse-time scales
\begin{equation}
    f_{H_0}\equiv H_0
    \simeq2.2\times10^{-18}\,\mathrm{Hz},
    \qquad
    f_{m_{\rm pl}}\equiv
    \frac{m_{\rm pl}c^2}{\hbar}
    \simeq3.7\times10^{42}\,\mathrm{Hz}.
\end{equation}
Thus, $f_{m_{\rm pl}}=m_{\rm pl}$ in
natural units, while the observed frequency
is $f=k/(2\pi)$.

Equation~\eqref{eq:MBCoverlineSGWB} reproduces
the main-text result in
Eq.~\eqref{eq:MBCoverlineSGWBessential}.
Within the stated asymptotic regime,
the spectrum follows a power law apart
from the equality transfer factor.
Its dependence on $w_1$ and the resulting
observational constraints are discussed in
Secs.~\ref{subsec:TsindexaCmbc}
and \ref{subsec:GWconstraintMBC}.
Earlier bounds and forecasts
\cite{Li:2025ilc,Lai:2025efh,Li:2025mxj}
were obtained using the previous matching
prescription and spectral normalization.
Within their common regime of validity,
the corresponding leading MBC spectrum
agrees with the phase-node result in
Appendix~\ref{sec:phasenodeall},
up to an overall normalization factor.
In this work, we use the spectrum derived
with the improved matching prescription
to obtain the updated MBC bounds presented
in Sec.~\ref{subsec:GWconstraintMBC}.

~
\section{Derivation of the SGWB spectrum in NMBC}
\label{sec:NMBCalldetail}

We derive the high- and low-frequency NMBC
spectra presented in
Eqs.~\eqref{eq:NMBCgeoverlineSGWBessential}
and \eqref{eq:NMBClowoverlineSGWBessential}.
The background is specified in
Eqs.~\eqref{eq:NMBCsetw} and \eqref{eq:NMBCseteta}
and illustrated in Fig.~\ref{fig:NMBC}.
The additional Phase~0--I transition sets
the break frequency
$f_\star=(2\pi a_0\eta_{0\downarrow})^{-1}$
defined in Eq.~\eqref{eq:fstardef}.
We retain the branch labels $f\ge f_\star$
and $f<f_\star$, while evaluating their
spectra in the controlled limits
$f\gg f_\star$ and $f\ll f_\star$,
respectively.

\subsection{High-frequency modes ($f\ge f_\star$) in NMBC}
\label{subsec:highNMBCalldetail}

Modes sufficiently far above the break satisfy
$k\eta_{0\downarrow}\gg1$ and remain well
inside the horizon at the Phase~0--I transition.
They first exit the horizon during Phase~I,
as in MBC.
Their subsequent propagation to Phase~IV
follows Eq.~\eqref{eq:hfa4a1}.
Including the preceding Phase~0--I transition
gives
\begin{equation}\label{eq:nmbchfa4a1}
\begin{aligned}
    \begin{pmatrix}
        A_4\\
        B_4
    \end{pmatrix}
    &=
    T_4^{-1}
    \begin{pmatrix}
        E_4\\
        F_4
    \end{pmatrix}
    \\
    &=
    T_4^{-1}\mathcal{M}_{3\downarrow}
    \begin{pmatrix}
        E_3\\
        F_3
    \end{pmatrix}
    \\
    &=
    T_4^{-1}\mathcal{M}_{3\downarrow}T_3
    \begin{pmatrix}
        A_3\\
        B_3
    \end{pmatrix}
    \\
    &=
    T_4^{-1}\mathcal{M}_{3\downarrow}T_3
    \mathcal{M}_{2\uparrow}
    \begin{pmatrix}
        A_2\\
        B_2
    \end{pmatrix}
    \\
    &=
    T_4^{-1}\mathcal{M}_{3\downarrow}T_3
    \mathcal{M}_{2\uparrow}T_2^{-1}
    \begin{pmatrix}
        E_2\\
        F_2
    \end{pmatrix}
    \\
    &=
    T_4^{-1}\mathcal{M}_{3\downarrow}T_3
    \mathcal{M}_{2\uparrow}T_2^{-1}
    \mathcal{M}_{1\downarrow}
    \begin{pmatrix}
        E_1\\
        F_1
    \end{pmatrix}
    \\
    &=
    T_4^{-1}\mathcal{M}_{3\downarrow}T_3
    \mathcal{M}_{2\uparrow}T_2^{-1}
    \mathcal{M}_{1\downarrow}T_1
    \begin{pmatrix}
        A_1\\
        B_1
    \end{pmatrix}
    \\
    &=
    T_4^{-1}\mathcal{M}_{3\downarrow}T_3
    \mathcal{M}_{2\uparrow}T_2^{-1}
    \mathcal{M}_{1\downarrow}T_1
    \mathcal{M}_{0\uparrow}
    \begin{pmatrix}
        A_0\\
        B_0
    \end{pmatrix},
    \qquad f>f_\star.
\end{aligned}
\end{equation}
The corresponding transfer matrix is
\begin{equation}\label{eq:nmbcx0gedef}
    \mathcal{X}^{(0)\ge}
    =
    T_4^{-1}\mathcal{M}_{3\downarrow}T_3
    \mathcal{M}_{2\uparrow}T_2^{-1}
    \mathcal{M}_{1\downarrow}T_1
    \mathcal{M}_{0\uparrow}.
\end{equation}
Comparing with Eq.~\eqref{eq:mbcx1def},
we obtain
\begin{equation}
    \mathcal{X}^{(0)\ge}
    =
    \mathcal{X}^{(1)}\mathcal{M}_{0\uparrow}.
\end{equation}
At leading subhorizon order,
$\mathcal{M}_{0\uparrow}$ is diagonal
and unitary, as follows from
Eq.~\eqref{eqsi:matchmatrup} together with
Eqs.~\eqref{eq:ridefinesigmai}
and \eqref{eq:sigmas}.
The additional transition therefore changes
only the phases of the mode coefficients,
leaving $\mathcal{N}_{22}$ unchanged
at this order.
Physically, these high-frequency modes
remain well inside the horizon during
Phase~0 and enter Phase~I with the same
vacuum normalization as in MBC.

Absorbing these phases into the initial
coefficients, we use the Phase-I basis
and choose the complex-conjugate
Bunch--Davies representation
\begin{equation}\label{eq:NMBCgea10b10exp}
    \begin{pmatrix}
        A_1\\[2pt]
        B_1
    \end{pmatrix}
    =
    \begin{pmatrix}
        0\\[2pt]
        \displaystyle
        \frac{\sqrt{\pi}}{2}
        e^{-i\left(
            \frac{\tilde{\nu}_1\pi}{2}
            +\frac{\pi}{4}
        \right)}
    \end{pmatrix},
\end{equation}
up to an irrelevant overall phase.
In this rephased basis,
\begin{equation}\label{eq:MBCgeX0geX1}
    \mathcal{X}^{(0)\ge}
    \simeq
    \mathcal{X}^{(1)}.
\end{equation}
Thus, these modes enter Phase~I with the
same vacuum normalization as in MBC and
undergo the same subsequent evolution
through the bounce.

The leading kernel therefore follows
from Eqs.~\eqref{eq:N22MBCLeading},
\eqref{eq:C1MBCLeading}, and
\eqref{eq:n1MBCLeading}.
For generic bounce phases,
\begin{equation}\label{eq:NMBCgeN22MBCLeading}
    \mathcal{N}_{22}^{(0)\ge}
    \simeq
    \mathcal{N}_{22}^{(1)}
    \simeq
    \frac{\mathcal{C}^{(0)\ge}}
         {(k\eta_{s\downarrow})^{n^{(0)\ge}}}
    \sin^2\!\left(2k\eta_{2\uparrow}\right),
\end{equation}
where
\begin{equation}\label{eq:NMBCgeC1MBCLeading}
    \mathcal{C}^{(0)\ge}
    =
    \mathcal{C}^{(1)}
    =
    \frac{2^{\frac{4}{3w_1+1}-1}}{\pi}
    \Gamma^2\!\left(
        \frac12+\frac{2}{3w_1+1}
    \right)
\end{equation}
and
\begin{equation}\label{eq:NMBCgen1MBCLeading}
    n^{(0)\ge}
    =
    n^{(1)}
    =
    \frac{6(w_1+1)}{3w_1+1},
    \qquad w_1>-\frac13.
\end{equation}

Using Eqs.~\eqref{eq:MBCoverlineSGWB},
\eqref{eq:MBCntw1}, and
\eqref{eq:avec1MBCLeading}, we obtain
the phase-averaged high-frequency spectrum:
\begin{align}
    \overline{\Omega}_{\rm GW}^{(0)\ge}(f)h^2
    \simeq
    \overline{\Omega}_{\rm GW}^{(1)}(f)h^2
    &\simeq
    \frac{h^2}{24}\,
    \frac{8}{\pi}
    \left(
        \frac{f_{H_0}}{f_{m_{\rm pl}}}
    \right)^2
    \frac{\overline{\mathcal{C}}^{(0)\ge}}
         {(2\pi)^{-n_T^{(0)\ge}+1}}
    \left(\frac{f}{f_{H_0}}\right)^{n_T^{(0)\ge}}
    \nonumber\\
    &\quad\times
    \left[
        \frac{\rho_{s\downarrow}^{1/4}}
             {(\rho_{c0}/\Omega_{\gamma0})^{1/4}}
    \right]^{4-n_T^{(0)\ge}}
    \mathcal{T}_{\rm eq}(f).
    \label{eq:NMBCgeoverlineSGWB}
\end{align}
The tensor spectral index is
\begin{equation}\label{eq:NMBCgentw1}
    n_T^{(0)\ge}
    =
    n_T^{(1)}
    =
    4-n^{(0)\ge}
    =
    2-2\nu_1
    =
    \frac{2(w_1-\tfrac13)}{w_1+\tfrac13},
    \qquad w_1>-\frac13,
\end{equation}
and the phase-averaged amplitude coefficient is
\begin{equation}\label{eq:NMBCgeavec1MBCLeading}
    \overline{\mathcal{C}}^{(0)\ge}
    =
    \overline{\mathcal{C}}^{(1)}
    =
    \frac{\mathcal{C}^{(1)}}{2}.
\end{equation}

\subsection{Low-frequency modes ($f<f_\star$) in NMBC}
\label{subsec:lowNMBCalldetail}

We derive the low-frequency branch in the
regime $k\eta_{0\downarrow}\ll1$, or
$f\ll f_\star$.
These modes first exit the horizon during
Phase~0 and are already outside it at
the Phase~0--I transition.
Including this superhorizon transition,
described by $\mathcal{M}_{0\downarrow}$,
in the MBC propagation sequence gives
\begin{equation}\label{eq:NMBClelfa4a0}
\begin{aligned}
    \begin{pmatrix}
        A_4\\
        B_4
    \end{pmatrix}
    &=
    T_4^{-1}
    \begin{pmatrix}
        E_4\\
        F_4
    \end{pmatrix}
    \\
    &=
    T_4^{-1}\mathcal{M}_{3\downarrow}
    \begin{pmatrix}
        E_3\\
        F_3
    \end{pmatrix}
    \\
    &=
    T_4^{-1}\mathcal{M}_{3\downarrow}T_3
    \begin{pmatrix}
        A_3\\
        B_3
    \end{pmatrix}
    \\
    &=
    T_4^{-1}\mathcal{M}_{3\downarrow}T_3
    \mathcal{M}_{2\uparrow}
    \begin{pmatrix}
        A_2\\
        B_2
    \end{pmatrix}
    \\
    &=
    T_4^{-1}\mathcal{M}_{3\downarrow}T_3
    \mathcal{M}_{2\uparrow}T_2^{-1}
    \begin{pmatrix}
        E_2\\
        F_2
    \end{pmatrix}
    \\
    &=
    T_4^{-1}\mathcal{M}_{3\downarrow}T_3
    \mathcal{M}_{2\uparrow}T_2^{-1}
    \mathcal{M}_{1\downarrow}
    \begin{pmatrix}
        E_1\\
        F_1
    \end{pmatrix}
    \\
    &=
    T_4^{-1}\mathcal{M}_{3\downarrow}T_3
    \mathcal{M}_{2\uparrow}T_2^{-1}
    \mathcal{M}_{1\downarrow}\mathcal{M}_{0\downarrow}
    \begin{pmatrix}
        E_0\\
        F_0
    \end{pmatrix}
    \\
    &=
    T_4^{-1}\mathcal{M}_{3\downarrow}T_3
    \mathcal{M}_{2\uparrow}T_2^{-1}
    \mathcal{M}_{1\downarrow}\mathcal{M}_{0\downarrow}T_0
    \begin{pmatrix}
        A_0\\
        B_0
    \end{pmatrix},
    \qquad f<f_\star.
\end{aligned}
\end{equation}
In the complex-conjugate Bunch--Davies
representation used for the power-spectrum
calculation, the Phase-0 coefficients are
\begin{equation}\label{eqsi:a0b0exp}
    \begin{pmatrix}
        A_0\\[2pt]
        B_0
    \end{pmatrix}
    =
    \begin{pmatrix}
        0\\[2pt]
        \displaystyle
        \frac{\sqrt{\pi}}{2}
        e^{-i\left(
            \frac{\tilde{\nu}_0\pi}{2}
            +\frac{\pi}{4}
        \right)}
    \end{pmatrix},
\end{equation}
up to an irrelevant overall phase.

The transfer matrix is therefore
\begin{equation}\label{eq:nmbcx0ltdef}
    \mathcal{X}^{(0)<}
    =
    T_4^{-1}\mathcal{M}_{3\downarrow}T_3
    \mathcal{M}_{2\uparrow}T_2^{-1}
    \mathcal{M}_{1\downarrow}\mathcal{M}_{0\downarrow}T_0.
\end{equation}
Comparing with Eq.~\eqref{eq:mbcx1def}
gives
\begin{equation}
    \mathcal{X}^{(0)<}
    =
    \mathcal{X}^{(1)}T_1^{-1}
    \mathcal{M}_{0\downarrow}T_0,
    \qquad f<f_\star.
\end{equation}
The additional factor
$T_1^{-1}\mathcal{M}_{0\downarrow}T_0$,
with $\mathcal{M}_{0\downarrow}$ given
in Eq.~\eqref{eq:NMBClowm0down},
encodes Phase~0 and its transition to Phase~I.
It modifies the low-frequency tilt and
normalization, producing the broken
power-law spectrum discussed in
Sec.~\ref{subsec:TsindexaCnmbc}.

\subsection{Low-frequency propagation kernel for generic phases}

For $\sin(2k\eta_{2\uparrow})\neq0$,
we evaluate $\mathcal{X}^{(0)<}$ using
the NMBC parameters in
Eqs.~\eqref{eq:NMBCsetw} and \eqref{eq:NMBCseteta},
with the exponents in
Eqs.~\eqref{eq:NMBCsetnu} and \eqref{eq:NMBCsettnu}.
Substituting these into the transformation
matrix in Eq.~\eqref{eq:tranmatrix} and
the matching matrices in
Eqs.~\eqref{eq:m1down}, \eqref{eq:m2exp},
\eqref{eq:m3down}, and \eqref{eq:NMBClowm0down},
we calculate the ordered product in
Eq.~\eqref{eq:nmbcx0ltdef}.

For $\nu_1>\tfrac12$, in the regime
$0<k\eta_{s\downarrow}\le k\eta_{0\downarrow}\ll1$
and $k\eta_{2\uparrow}\gg1$, and sufficiently
far from the phase nodes, this gives
\begin{align}\label{eq:NMBClowN22Leading}
    \mathcal{N}_{22}^{(0)<}
    &=
    \left[
        \left(\mathcal{X}^{(0)<}\right)^\dagger
        \mathcal{X}^{(0)<}
    \right]_{22}
    \nonumber\\
    &\simeq
    \frac{\sin^2(2k\eta_{2\uparrow})}
         {(k\eta_{s\downarrow})^{2\tilde\nu_1+3}}
    \begin{cases}
        \displaystyle
        \frac{
            2^{2\tilde\nu_0}
            \Gamma^2(1+\tilde\nu_0)
        }{\pi}
        (k\eta_{0\downarrow})^{2\tilde\nu_1-2\tilde\nu_0},
        & \nu_0>\tfrac12,\\[10pt]
        \displaystyle
        \frac{
            \pi\,2^{-2\tilde\nu_0}
            \csc^2(\pi\tilde\nu_0)
        }{\Gamma^2(\tilde\nu_0)}
        (k\eta_{0\downarrow})^{2\tilde\nu_1+2\tilde\nu_0},
        & 0<\nu_0<\tfrac12.
    \end{cases}
\end{align}
Here $\tilde\nu_0=|\nu_0-\tfrac12|$
and $\tilde\nu_1=\nu_1-\tfrac12$.
For the lower branch, the reflection identity
$\Gamma(\tilde\nu_0)\Gamma(1-\tilde\nu_0)
=\pi/\sin(\pi\tilde\nu_0)$ gives
\begin{equation}
    \frac{
        \pi\,2^{-2\tilde\nu_0}
        \csc^2(\pi\tilde\nu_0)
    }{\Gamma^2(\tilde\nu_0)}
    =
    \frac{
        2^{-2\tilde\nu_0}
        \Gamma^2(1-\tilde\nu_0)
    }{\pi}.
\end{equation}
Both branches therefore reduce to
\begin{equation}\label{eq:NMBClowN22LeadingUnified}
    \mathcal{N}_{22}^{(0)<}
    \simeq
    \frac{
        2^{2\nu_0-1}
        \Gamma^2(\nu_0+\tfrac12)
    }{\pi}
    \frac{
        (k\eta_{0\downarrow})^{2\nu_1-2\nu_0}
    }{
        (k\eta_{s\downarrow})^{2\nu_1+2}
    }
    \sin^2(2k\eta_{2\uparrow}).
\end{equation}

This derivation uses the two-power Hankel
basis for positive, noninteger orders.
The case $\nu_0=\tfrac12$ requires a
logarithmic basis and is not established
by this calculation.
At positive integer $\tilde\nu_0$, the
full matrix expression likewise requires
the appropriate logarithmic expansion,
although the upper-branch leading
coefficient has a smooth limit.

We restrict the low-frequency calculation
to $\nu_1>\tfrac12$, corresponding to
$-\tfrac13<w_1<1$.%
\footnote{
    The omitted range $w_1>1$ gives
    $1<n_T^{(0)\ge}<2$ through
    Eq.~\eqref{eq:NMBCgentw1};
    see also Fig.~\ref{fig:MBC_nTessential}.
    Such strongly blue-tilted spectra
    enhance the high-frequency contribution
    to the integrated bound in
    Eq.~\eqref{eq:dneffcmbbbnintegralessential},
    particularly for a sizeable PTA-band
    signal extending over a broad frequency
    range.
    Their exclusion nevertheless depends
    on the normalization and upper cutoff.
    The condition $\nu_1>\tfrac12$ still
    allows a blue high-frequency branch;
    the constraint analysis in the main
    text adopts the stronger condition
    $n_T^{(0)\ge}\le0$, or $\nu_1\ge1$,
    as stated in Eq.~\eqref{eq:ntbound13essential}.
}

We rewrite Eq.~\eqref{eq:NMBClowN22LeadingUnified}
as
\begin{equation}\label{eq:NMBClowN22Cnexp}
    \mathcal{N}_{22}^{(0)<}
    \simeq
    \frac{\mathcal{C}^{(0)<}}
         {(k\eta_{s\downarrow})^{n^{(0)<}}}
    \sin^2\!\left(2k\eta_{2\uparrow}\right),
\end{equation}
where
\begin{equation}\label{eq:c0lowNMBC}
    \mathcal{C}^{(0)<}
    =
    \frac{
        2^{2\nu_0-1}
        \Gamma^2\!\left(\nu_0+\tfrac12\right)
    }{\pi}
    \left(
        \frac{\eta_{0\downarrow}}{\eta_{s\downarrow}}
    \right)^{2\nu_1-2\nu_0}
\end{equation}
and
\begin{equation}
    n^{(0)<}
    =
    2\nu_0+2
    =
    \frac{6(w_0+1)}{3w_0+1}.
\end{equation}

As in MBC, uniform averaging over the
bounce phase gives
\begin{equation}\label{eq:NMBClowsin2ave}
    \left\langle
        \sin^2\!\left(2k\eta_{2\uparrow}\right)
    \right\rangle
    =
    \frac12.
\end{equation}
The phase-averaged leading kernel is
\begin{equation}\label{eq:aveN22NMBClowLeading}
    \left\langle\mathcal{N}_{22}^{(0)<}\right\rangle
    \simeq
    \frac{\overline{\mathcal{C}}^{(0)<}}
         {(k\eta_{s\downarrow})^{n^{(0)<}}},
\end{equation}
with
\begin{equation}\label{eq:avec1NMBClowLeading}
    \overline{\mathcal{C}}^{(0)<}
    \equiv
    \frac{\mathcal{C}^{(0)<}}{2}.
\end{equation}

\subsection{Low-frequency SGWB spectrum in NMBC}

Substituting
Eq.~\eqref{eq:aveN22NMBClowLeading}
into Eq.~\eqref{eq:phfpketarad} gives
the leading phase-averaged primordial
tensor spectrum:
\begin{equation}\label{eq:phf1NMBClow}
    \overline{\mathcal{P}}_h^{(0)<}(k)
    \simeq
    \frac{4}{\pi^2}\,
    \frac{k^4}
         {H_0^2\Omega_{\gamma0}m_{\rm pl}^2}\,
    \frac{\overline{\mathcal{C}}^{(0)<}}
         {(k\eta_{s\downarrow})^{n^{(0)<}}}.
\end{equation}
The overbar again denotes averaging over
the bounce phase $2k\eta_{2\uparrow}$.

Using Eq.~\eqref{eq:sgwbphfessential},
$k=2\pi f$ for $a_0=1$, and the relation
between $\eta_{s\downarrow}$ and the
bounce energy scale $\rho_{s\downarrow}^{1/4}$
in Eq.~\eqref{eq:etasrhos}, we obtain
\begin{align}
    \overline{\Omega}_{\rm GW}^{(0)<}(f)h^2
    &\simeq
    \frac{h^2}{24}\,
    \frac{8}{\pi}
    \left(
        \frac{f_{H_0}}{f_{m_{\rm pl}}}
    \right)^2
    \frac{\overline{\mathcal{C}}^{(0)<}}
         {(2\pi)^{-n_T^{(0)<}+1}}
    \left(\frac{f}{f_{H_0}}\right)^{n_T^{(0)<}}
    \nonumber\\
    &\quad\times
    \left[
        \frac{\rho_{s\downarrow}^{1/4}}
             {(\rho_{c0}/\Omega_{\gamma0})^{1/4}}
    \right]^{4-n_T^{(0)<}}
    \mathcal{T}_{\rm eq}(f).
    \label{eq:NMBClowoverlineSGWB}
\end{align}
For fixed background parameters,
$\overline{\mathcal{C}}^{(0)<}$ is
independent of $k$, so the primordial
tensor spectral index is
\begin{equation}\label{eq:NMBClowntw0}
    n_T^{(0)<}
    \equiv
    \frac{d\ln\overline{\mathcal{P}}_h^{(0)<}(k)}
         {d\ln k}
    =
    4-n^{(0)<}
    =
    2-2\nu_0
    =
    \frac{2(w_0-\tfrac13)}{w_0+\tfrac13},
    \qquad w_0>-\frac13.
\end{equation}
Together with the coefficient in
Eq.~\eqref{eq:c0lowNMBC}, these results
give the low-frequency spectrum presented
in Eq.~\eqref{eq:NMBClowoverlineSGWBessential},
within the domain specified above.

~
\section{Constraints on the bounce energy scale from
$\Delta N_{\rm eff}$}
\label{sec:DeltaNeffintpoi}

We derive the bounce-energy bounds associated with
the integrated constraint in
Eq.~\eqref{eq:dneffcmbbbnintegralessential}
and compare them with the pointwise benchmark in
Eq.~\eqref{eq:dneffcmbbbnessential}.
We use the phase-averaged spectrum when the
integration spans many bounce-induced oscillations.
The pointwise benchmark does not replace the
integrated constraint and is not generally
a conservative approximation to it.

The integrated bound and pointwise benchmarks
are compared in
Fig.~\ref{fig:intMBC_Tau_DeltaNeff_Contours_Completeessential}.
The integration limits used there are illustrative.
Below, we derive how the bounce-energy bound
depends on the spectral shape and frequency range,
including the effect of a model-dependent upper cutoff.
\subsection{Dependence of the bounce-energy bound
on the spectral shape}

We first consider an integration interval contained
entirely within the high-frequency NMBC branch,
with $\mathcal{T}_{\rm eq}(f)\simeq1$.
In the power-law expressions below, amplitudes
at $f_\star$ denote the asymptotic branch
expressions extrapolated to the break.

For a scale-invariant branch,
$n_T^{(0)\ge}=0$, the spectrum is constant
over the interval, giving
\begin{equation}\label{eq:dneffFlatBranch}
    \int_{f_{\min}}^{f_{\max}}
    d\ln f\,
    \overline{\Omega}_{\rm GW}^{(0)\ge}(f)h^2
    \simeq
    \overline{\Omega}_{\rm GW}^{(0)\ge}(f_\star)h^2
    \ln\!\left(\frac{f_{\max}}{f_{\min}}\right).
\end{equation}
The same logarithmic dependence applies to
a scale-invariant MBC spectrum under the
same transfer-function approximation.

For fixed spectral tilt and frequency,
Eq.~\eqref{eq:NMBCgeoverlineSGWB} implies
\begin{equation}
    \overline{\Omega}_{\rm GW}^{(0)\ge}(f)h^2
    \propto
    \left(\rho_{s\downarrow}^{1/4}\right)^{
        4-n_T^{(0)\ge}
    }.
\end{equation}
For $n_T^{(0)\ge}=0$, using the same numerical
ceiling for the integrated and pointwise
prescriptions therefore gives
\begin{equation}\label{eq:dneffFlatEnergyRatio}
    \left(\rho_{s\downarrow}^{1/4}\right)_{\rm int}^{\max}
    \simeq
    \left(\rho_{s\downarrow}^{1/4}\right)_{\rm point}^{\max}
    \left[
        \ln\!\left(\frac{f_{\max}}{f_{\min}}\right)
    \right]^{-1/4},
\end{equation}
provided that the integration limits are fixed.
A broad frequency interval thus strengthens
the upper bound through its logarithmic bandwidth.

For a nonzero tilt, integration over a single
power-law branch gives
\begin{align}\label{eq:dneffTiltedBranch}
    \int_{f_{\min}}^{f_{\max}}
    d\ln f\,
    \overline{\Omega}_{\rm GW}^{(0)\ge}(f)h^2
    \simeq
    \overline{\Omega}_{\rm GW}^{(0)\ge}(f_\star)h^2
    \frac{
        \left(f_{\max}/f_\star\right)^{n_T^{(0)\ge}}
        -
        \left(f_{\min}/f_\star\right)^{n_T^{(0)\ge}}
    }{n_T^{(0)\ge}}.
\end{align}
This expression reduces continuously to
Eq.~\eqref{eq:dneffFlatBranch} as
$n_T^{(0)\ge}\to0$.

When $f_{\min}<f_\star<f_{\max}$, both NMBC
branches contribute.
Within the piecewise power-law approximation,
and taking $\mathcal{T}_{\rm eq}(f)\simeq1$
throughout the interval, we obtain
\begin{align}\label{eq:dneffBrokenPowerLaw}
    &\int_{f_{\min}}^{f_{\max}}
    d\ln f\,
    \overline{\Omega}_{\rm GW}^{(0)}(f)h^2
    \nonumber\\
    &\qquad\quad\simeq
    \overline{\Omega}_{\rm GW}^{(0)<}(f_\star)h^2
    \frac{
        1-\left(f_{\min}/f_\star\right)^{n_T^{(0)<}}
    }{n_T^{(0)<}}
    +
    \overline{\Omega}_{\rm GW}^{(0)\ge}(f_\star)h^2
    \frac{
        \left(f_{\max}/f_\star\right)^{n_T^{(0)\ge}}-1
    }{n_T^{(0)\ge}}.
\end{align}
For a vanishing low- or high-frequency tilt,
the corresponding fraction is replaced by
$\ln(f_\star/f_{\min})$ or
$\ln(f_{\max}/f_\star)$, respectively.
The two pivot amplitudes are retained separately
because the asymptotic branches need not match
at $f_\star$.

For $n_T^{(0)\ge}=0$ and $n_T^{(0)<}>0$,
using
$\overline{\Omega}_{\rm GW}^{(0)<}(f_\star)
=\overline{\Omega}_{\rm GW}^{(0)\ge}(f_\star)/\epsilon_g$
gives
\begin{align}\label{eq:dneffFlatHighBranch}
    \int_{f_{\min}}^{f_{\max}}
    d\ln f\,
    \overline{\Omega}_{\rm GW}^{(0)}(f)h^2
    \simeq
    \overline{\Omega}_{\rm GW}^{(0)\ge}(f_\star)h^2
    \left[
        \frac{
            1-\left(f_{\min}/f_\star\right)^{n_T^{(0)<}}
        }{\epsilon_g n_T^{(0)<}}
        +
        \ln\!\left(\frac{f_{\max}}{f_\star}\right)
    \right].
\end{align}
Thus, scale invariance of the high-frequency
branch alone does not reduce the full NMBC
integral to Eq.~\eqref{eq:dneffFlatBranch}.

If $f_{\max}$ is identified with the adopted
cutoff $f_{\rm cut}\propto\rho_{s\downarrow}^{1/4}$,
the integration range itself depends on the
bounce energy.
The integrated bound must then be solved
self-consistently, and
Eq.~\eqref{eq:dneffFlatEnergyRatio}
is an implicit scaling relation rather than
a fixed multiplicative correction.
Whenever equality effects are non-negligible,
$\mathcal{T}_{\rm eq}(f)$ must also be retained
inside the integral.

\subsection{$\Delta N_{\rm eff}$ bound on the
scale-invariant high-frequency branch of NMBC}

At leading order, the high-frequency NMBC
spectrum coincides with the MBC spectrum
under the $\tau$-derivative continuity prescription.
For the scale-invariant case,
$n_T^{(0)\ge}=n_T^{(1)}=0$, we have
$w_1=1/3$ and
$\overline{\mathcal C}^{(1)}=1/4$.

Using the same branch approximation and defining
$f_{\rm L}\equiv\max(f_\star,f_{\min})<f_{\max}$,
the high-frequency contribution must satisfy
\begin{equation}\label{eq:NMBC_HF_DeltaNeff}
    \int_{f_{\rm L}}^{f_{\max}}
    \overline{\Omega}_{\rm GW}^{(0)\ge}(f)h^2\,
    \mathrm{d}\ln f
    \le B_\nu\,\Delta N_{\rm eff}^{\max},
    \qquad
    B_\nu\equiv
    \frac{7}{8}\left(\frac{4}{11}\right)^{4/3}
    \Omega_{\gamma0}h^2
    \simeq5.6186\times10^{-6}.
\end{equation}
Retaining the adopted equality transfer function
gives the bandwidth factor
\begin{equation}\label{eq:NMBC_HF_bandwidth}
    \begin{aligned}
        J(f_{\rm L},f_{\max})
        \equiv
        \int_{f_{\rm L}}^{f_{\max}}
        \mathcal T_{\rm eq}(f)\,\mathrm{d}\ln f=
        \ln\!\left(\frac{f_{\max}}{f_{\rm L}}\right)
        +\frac{9f_{\rm eq}^{\,2}}{64}
        \left(f_{\rm L}^{-2}-f_{\max}^{-2}\right).
    \end{aligned}
\end{equation}
Substitution into
Eq.~\eqref{eq:NMBC_HF_DeltaNeff}
yields
\begin{equation}\label{eq:NMBC_HF_rhos_bound}
    \rho_{s\downarrow}^{1/4}
    \le
    \left(\frac{\rho_{c0}}{\Omega_{\gamma0}}\right)^{1/4}
    \left[
        \frac{24\pi^2}{h^2}
        \left(\frac{f_{m_{\rm pl}}}{f_{H_0}}\right)^2
        \frac{B_\nu\Delta N_{\rm eff}^{\max}}
             {J(f_{\rm L},f_{\max})}
    \right]^{1/4}.
\end{equation}

For comparison with
Fig.~\ref{fig:intMBC_Tau_DeltaNeff_Contours_Completeessential},
we use the illustrative limits
$f_{\rm L}=10^{-10}\,\mathrm{Hz}$ and
$f_{\max}=10^9\,\mathrm{Hz}$,
with the rounded radiation budget
$1.7\times10^{-6}$.
The equality correction is negligible over this
interval, so $J\simeq\ln(10^{19})\simeq43.75$.
The resulting bound is
\begin{equation}\label{eq:rhosdboundint}
    \rho_{s\downarrow}^{1/4}
    \lesssim
    2.50\times10^{15}\,\mathrm{TeV}
    \simeq1.03\,m_{\rm pl}.
\end{equation}
This estimate assumes that the high-frequency
approximation applies over the selected interval.
Assigning the entire radiation budget to this
branch gives a necessary bound; including
the low-frequency contribution can only
strengthen it.

\paragraph{Pointwise comparison.}
Using the same radiation budget in the pointwise
benchmark of Eq.~\eqref{eq:dneffcmbbbnessential}
instead gives
\begin{equation}
    \rho_{s\downarrow}^{1/4}
    \le
    \left(\frac{\rho_{c0}}{\Omega_{\gamma0}}\right)^{1/4}
    \left[
        \frac{24\pi^2}{h^2}
        \left(\frac{f_{m_{\rm pl}}}{f_{H_0}}\right)^2
        \frac{B_\nu\Delta N_{\rm eff}^{\max}}
             {\mathcal T_{\rm eq}(f_{\rm p})}
    \right]^{1/4}.
\end{equation}
For example, we take $f_{\rm p}=10^{-6}\,\mathrm{Hz}$,
assuming $f_{\rm p}\gg f_\star$ so that the
high-frequency approximation applies.
At this frequency, equality effects are negligible
and $\mathcal T_{\rm eq}(f_{\rm p})\simeq1$.
With the same rounded budget, this yields
\begin{equation}
    \rho_{s\downarrow}^{1/4}
    \lesssim
    6.43\times10^{15}\,\mathrm{TeV}.
\end{equation}
The integrated bound in
Eq.~\eqref{eq:rhosdboundint}
is stronger by the factor $J^{1/4}\simeq2.57$,
consistent with
Eq.~\eqref{eq:dneffFlatEnergyRatio}.

~
\section{Parameter reconstruction for four illustrative NMBC spectra}
\label{sec:parameterrecon}
Here we reconstruct the background parameters
underlying the four illustrative NMBC spectra
E1--E4 presented in the main text.

Each example is specified by four phenomenological
parameters:
\begin{equation}
    \left\{
        n_T^{(0)<},\,
        n_T^{(0)\ge},\,
        f_\star,\,
        \overline{\Omega}_{\rm GW}^{(0)\ge}(f_\star)h^2
    \right\}.
\end{equation}
The pivot amplitude denotes the phase-averaged
high-frequency expression extrapolated to
$f=f_\star$.
We use the normalization $a_0=1$ throughout
and retain the validity conditions of the
asymptotic spectra derived above.

Within the adopted NMBC parameterization,
these inputs determine the background parameters
\begin{equation}\label{eq:wietaiExample}
    \left\{
        w_0,\,
        w_1,\,
        \eta_{0\downarrow},\,
        \eta_{s\downarrow}
    \right\}
\end{equation}
and the physical scales
\begin{equation}\label{eq:rhosfcutExample}
    \left\{
        \rho_{s\downarrow}^{1/4},\,
        f_{\rm cut}
    \right\}
\end{equation}
through the following steps:
\begin{enumerate}
    \item Invert the spectral-index relations
    in Eqs.~\eqref{eq:NMBClowntw0}
    and \eqref{eq:NMBCgentw1}:
    \begin{equation}
        w_0
        =
        \frac{n_T^{(0)<}+2}
             {3\left(2-n_T^{(0)<}\right)},
        \qquad w_0>-\frac13,
    \end{equation}
    and
    \begin{equation}
        w_1
        =
        \frac{n_T^{(0)\ge}+2}
             {3\left(2-n_T^{(0)\ge}\right)},
        \qquad -\frac13<w_1<1.
    \end{equation}
    These ranges require
    $n_T^{(0)<}<2$ and $n_T^{(0)\ge}<1$,
    respectively.
    A scale-invariant or red-tilted high-frequency
    branch additionally requires
    $n_T^{(0)\ge}\le0$, corresponding to
    $w_1\le\tfrac13$.

    \item Determine the Phase~0--I transition
    scale from the pivot frequency:
    \begin{equation}
        \eta_{0\downarrow}
        =
        \frac{1}{2\pi a_0f_\star}.
    \end{equation}

    \item For the inferred $w_1$, obtain
    $\rho_{s\downarrow}^{1/4}$ by inverting
    Eq.~\eqref{eq:NMBCgeoverlineSGWB}
    at $f=f_\star$, using the specified
    high-frequency pivot amplitude.

    \item Compute the bounce transition scale
    from Eq.~\eqref{eq:etasrhos}:
    \begin{equation}
        \eta_{s\downarrow}
        =
        \frac{1}{H_0}
        \left(
            \frac{\rho_{c0}}
                 {\Omega_{\gamma0}\rho_{s\downarrow}}
        \right)^{1/4}.
    \end{equation}
    For the symmetric bounce,
    $\eta_{1\downarrow}=\eta_{3\downarrow}
    =\eta_{s\downarrow}$.

    \item Evaluate the characteristic cutoff
    frequency:
    \begin{equation}
        f_{\rm cut}
        =
        \frac{1}{2\pi a_0\eta_{s\downarrow}},
    \end{equation}
    corresponding to $k\eta_{s\downarrow}=1$.
    We adopt this scale as the upper cutoff
    of the asymptotic spectrum; the leading-order
    expressions apply sufficiently below it.
\end{enumerate}

The reconstructed parameters must satisfy
$\eta_{0\downarrow}>\eta_{s\downarrow}$,
or equivalently $f_\star<f_{\rm cut}$,
so that Phase~I has a nonzero duration.
The remaining scale $\eta_{2\uparrow}$,
which controls the bounce-induced oscillation
phase, is not determined by the phase-averaged
spectrum.
It must be specified separately so that
$k\eta_{2\uparrow}\gg1$ throughout the
frequency range considered.

~

~
\section{Phase-node solution ($2k\eta_{2\uparrow}=m\pi$)
for MBC and NMBC}
\label{sec:phasenodeall}

The generic-phase kernels derived above contain
a leading contribution proportional to
$\sin^2(2k\eta_{2\uparrow})$.
At the phase nodes  ($2k\eta_{2\uparrow}=m\pi$ or equivalently, $\sin^2(2k\eta_{2\uparrow})=0$), this contribution vanishes,
and terms that are subleading away from the nodes
determine the leading nonvanishing kernel.
The phase-node solutions describe these exceptional
frequencies of the same oscillatory spectrum,
complementing the generic-phase results and
clarifying their range of validity.
\subsection{Phase-node solution for MBC}

At $2k\eta_{2\uparrow}=m\pi$, or equivalently
$\sin^2(2k\eta_{2\uparrow})=0$, the leading
contribution in Eq.~\eqref{eq:N22MBCLeading}
vanishes.
To extract the surviving contribution,
we evaluate the ordered transfer matrix
in Eq.~\eqref{eq:mbcx1def},
\[
    \mathcal X^{(1)}
    =
    T_4^{-1}\mathcal M_{3\downarrow}T_3
    \mathcal M_{2\uparrow}T_2^{-1}
    \mathcal M_{1\downarrow}T_1,
\]
imposing the node condition before extracting
the leading small-argument term.

For $k\eta_{s\downarrow}\ll1$,
$k\eta_{2\uparrow}\gg1$, and $\nu_1>0$,
excluding $\nu_1=\tfrac12$ and $\nu_1=1$,
the leading nonvanishing kernel is
\begin{equation}\label{eq:N221node}
    \mathcal N_{22\,\mathrm{node}}^{(1)}
    \simeq
    \frac{
        \mathcal C_{\mathrm{node}}^{(1)}
    }{
        \left(k\eta_{s\downarrow}\right)^{
            n_{\mathrm{node}}^{(1)}
        }
    },
\end{equation}
where
\begin{equation}\label{eq:C1node}
    \mathcal C_{\mathrm{node}}^{(1)}
    =
    \frac{
        2^{2\tilde\nu_1-2}
        \Gamma^2(\tilde\nu_1)
    }{\pi}
    \begin{cases}
        (1-2\tilde\nu_1)^2,
        & \nu_1>\tfrac12,\\[2pt]
        1,
        & 0<\nu_1<\tfrac12,
    \end{cases}
\end{equation}
and
\begin{equation}\label{eq:n1node}
    n_{\mathrm{node}}^{(1)}
    =
    1+2\tilde\nu_1.
\end{equation}
Here $\tilde\nu_1=|\nu_1-\tfrac12|$.
The case $\nu_1=\tfrac12$ requires a logarithmic
superhorizon expansion.
At $\nu_1=1$, the coefficient in
Eq.~\eqref{eq:C1node} also vanishes,
requiring further subleading terms.

Substituting Eq.~\eqref{eq:N221node}
into Eq.~\eqref{eq:phfpketarad}, and then using
Eqs.~\eqref{eq:sgwbphfessential} and \eqref{eq:etasrhos},
gives the MBC phase-node SGWB envelope:
\begin{align}
    \Omega_{\rm GW,node}^{(1)}(f)h^2
    &\simeq
    \frac{h^2}{24}\,
    \frac{8}{\pi}
    \left(
        \frac{f_{H_0}}{f_{m_{\rm pl}}}
    \right)^2
    \frac{
        \mathcal C_{\mathrm{node}}^{(1)}
    }{
        (2\pi)^{-n_{T,\mathrm{node}}^{(1)}+1}
    }
    \left(
        \frac{f}{f_{H_0}}
    \right)^{n_{T,\mathrm{node}}^{(1)}}
    \nonumber\\
    &\quad\times
    \left[
        \frac{\rho_{s\downarrow}^{1/4}}
             {(\rho_{c0}/\Omega_{\gamma0})^{1/4}}
    \right]^{4-n_{T,\mathrm{node}}^{(1)}}
    \mathcal T_{\rm eq}(f),
    \label{eq:MBCNodeSGWB}
\end{align}
where $k=2\pi f$ for $a_0=1$.
Within the domain specified above, the
primordial node-envelope tilt is
\begin{equation}\label{eq:MBCNodeNT}
    n_{T,\mathrm{node}}^{(1)}
    =
    4-n_{\mathrm{node}}^{(1)}
    =
    3-2\tilde\nu_1
    =
    \begin{cases}
        4-2\nu_1,
        & \nu_1>\tfrac12,\\[2pt]
        2+2\nu_1,
        & 0<\nu_1<\tfrac12.
    \end{cases}
\end{equation}
No phase averaging is performed, so
$\mathcal C_{\mathrm{node}}^{(1)}$
enters without an additional factor of $1/2$.
For a fixed background,
Eq.~\eqref{eq:MBCNodeSGWB} describes a
continuous envelope through the discrete
phase-node values within the stated
asymptotic regime.

Within this regime, the coefficient and exponent
in Eqs.~\eqref{eq:C1node} and \eqref{eq:n1node}
reproduce Eqs.~(18) and (19) of
Ref.~\cite{Li:2025ilc}.
Those results used continuity of the derivative
with respect to the positive local conformal-time
argument $\eta$, as specified in
Eq.~\eqref{eq:matchveleta};
see also Eq.~(24) of Ref.~\cite{Li:2024dce}.

At an interface with $\sigma_{i+1}=-\sigma_i$,
continuity of $dh_k/d\eta$ reverses the sign
of $dh_k/d\tau$ across the interface.
In this limited sense, the earlier prescription
acts as a reflection of the derivative component
of the local mode data.
The agreement at the phase nodes provides
an algebraic consistency check of the leading
asymptotic kernels, but does not establish
equivalence between the two matching prescriptions.

\paragraph{Interpretation of the node envelope.}
For a fixed bounce, $\eta_{2\uparrow}$ is fixed,
so the phase-node condition holds only at
$f_m=m/(4\eta_{2\uparrow})$, where
$k=2\pi f$ for $a_0=1$.
Here $m$ is a positive integer, restricted to
frequencies satisfying the asymptotic conditions
stated above.
A continuous curve constructed from
Eq.~\eqref{eq:N221node} therefore represents
a node envelope, extending the nodal expression
to frequencies that need not satisfy the
phase-node condition.

Integrating this envelope provides a formal
comparison rather than the physical
$\Delta N_{\rm eff}$ constraint.
The latter requires the full frequency-dependent
SGWB spectrum, or its phase-averaged approximation
when the integration spans sufficiently many
oscillations.

\subsection{Phase-node solution for NMBC}

\subsubsection{Low-frequency phase-node kernel}

At the phase nodes,
\begin{equation}\label{eq:NMBClowN22NodeCondition}
    \sin^2(2k\eta_{2\uparrow})=0,
    \qquad
    2k\eta_{2\uparrow}=m\pi,
    \qquad
    m\in\mathbb{Z},
\end{equation}
the generic-phase leading term in
Eq.~\eqref{eq:NMBClowN22LeadingUnified} vanishes.
We therefore impose the node condition on the
ordered matrix product in
Eq.~\eqref{eq:nmbcx0ltdef} before extracting
the leading nonvanishing contribution.

We use $\tilde\nu_0=|\nu_0-\tfrac12|$ and
$\tilde\nu_1=\nu_1-\tfrac12$, taking
$\nu_1>\tfrac12$ and $\nu_0>0$ with positive,
noninteger $\tilde\nu_0$.
The boundary arguments satisfy
$0<k\eta_{s\downarrow}<k\eta_{0\downarrow}\ll1$
and $k\eta_{2\uparrow}\gg1$.

Define the second-column coefficients
\begin{equation}\label{eq:NMBClowN22NodeColumn}
    \begin{pmatrix}
        \mathcal E_1\\
        \mathcal F_1
    \end{pmatrix}
    \equiv
    \mathcal M_{0\downarrow}T_0
    \begin{pmatrix}
        0\\1
    \end{pmatrix}
    =
    \mathcal M_{0\downarrow}
    \begin{pmatrix}
        \alpha_0^*\\
        \beta_0^*
    \end{pmatrix},
\end{equation}
where
\begin{equation}\label{eq:NMBClowN22NodeHankelCoefficients}
    \alpha_0
    =
    \frac{2^{-\tilde\nu_0}}
         {\Gamma(1+\tilde\nu_0)}
    \bigl[1+i\cot(\pi\tilde\nu_0)\bigr],
    \qquad
    \beta_0
    =
    -\frac{i\,2^{\tilde\nu_0}
               \Gamma(\tilde\nu_0)}{\pi}.
\end{equation}
The node kernel is obtained from the second column
of the transfer matrix evaluated at a node:
\begin{equation}\label{eq:NMBClowNodeMatrixNorm}
    \mathcal N_{22,\mathrm{node}}^{(0)<}
    =
    \left[
        \left(\mathcal X_{\mathrm{node}}^{(0)<}\right)^\dagger
        \mathcal X_{\mathrm{node}}^{(0)<}
    \right]_{22}
    =
    \left|\mathcal X_{12,\mathrm{node}}^{(0)<}\right|^2
    +
    \left|\mathcal X_{22,\mathrm{node}}^{(0)<}\right|^2.
\end{equation}
Within the adopted subhorizon approximation,
the central phase matrix contributes an overall
factor $(-1)^m$, which cancels in this
quadratic combination.
Retaining all terms of the adopted asymptotic
matching matrices gives
\begin{align}
    \mathcal N_{22,\mathrm{node}}^{(0)<}
    &\simeq
    \frac{\pi}{4}\Bigl[
    \left|
        (k\eta_{s\downarrow})^{\nu_1-1}\mathcal E_1
        +2(1-\nu_1)
        (k\eta_{s\downarrow})^{-\nu_1}\mathcal F_1
    \right|^2
    \nonumber\\
    &\qquad+
    (2\nu_1-1)^2
    (k\eta_{s\downarrow})^{2-2\nu_1}
    |\mathcal F_1|^2
    \Bigr].
    \label{eq:NMBClowN22NodeComplete}
\end{align}

For $\nu_0>\tfrac12$, so that
$\tilde\nu_0=\nu_0-\tfrac12$,
the matching-column components are
\begin{align}
    \mathcal E_1
    &=
    \alpha_0^*
    (k\eta_{0\downarrow})^{\tilde\nu_0-\tilde\nu_1}
    +
    \beta_0^*
    \frac{\tilde\nu_1-\tilde\nu_0}{\tilde\nu_1}
    (k\eta_{0\downarrow})^{-\tilde\nu_0-\tilde\nu_1},
    \nonumber\\
    \mathcal F_1
    &=
    \beta_0^*
    \frac{\tilde\nu_0}{\tilde\nu_1}
    (k\eta_{0\downarrow})^{\tilde\nu_1-\tilde\nu_0}.
    \label{eq:NMBClowN22NodeColumnUpper}
\end{align}
Substituting these components and the coefficients
in Eq.~\eqref{eq:NMBClowN22NodeHankelCoefficients}
into Eq.~\eqref{eq:NMBClowN22NodeComplete}
reproduces the explicit upper-branch expansion
in Eq.~\eqref{eq:NMBClowN22NodeExpanded}.\footnote{
For $\nu_0>\tfrac12$, the expanded expression is
\begin{align}
    &\mathcal N_{22,\mathrm{node}}^{(0)<}
    \nonumber\\
    &\simeq
    \frac{\pi}{4}\Biggl\{
    \frac{
        2^{2+2\tilde\nu_0}
        \tilde\nu_0^2\Gamma^2(\tilde\nu_0)
    }{\pi^2}
    (k\eta_{0\downarrow})^{2\tilde\nu_1-2\tilde\nu_0}
    (k\eta_{s\downarrow})^{1-2\tilde\nu_1}
    \nonumber\\
    &\quad+
    \frac{
        (k\eta_{0\downarrow})^{-2\tilde\nu_0-2\tilde\nu_1}
    }{
        \tilde\nu_1^2
        (k\eta_{s\downarrow})^{1+2\tilde\nu_1}
    }
    \nonumber\\
    &\qquad\times
    \Biggl[
        \frac{
            2^{\tilde\nu_0}\Gamma(\tilde\nu_0)
        }{\pi}
        (\tilde\nu_1-\tilde\nu_0)
        (k\eta_{s\downarrow})^{2\tilde\nu_1}
        +
        \frac{
            2^{\tilde\nu_0}\Gamma(\tilde\nu_0)
        }{\pi}
        \tilde\nu_0(1-2\tilde\nu_1)
        (k\eta_{0\downarrow})^{2\tilde\nu_1}
        -
        \frac{
            2^{-\tilde\nu_0}\tilde\nu_1
            \cot(\pi\tilde\nu_0)
        }{
            \Gamma(1+\tilde\nu_0)
        }
        (k\eta_{0\downarrow})^{2\tilde\nu_0}
        (k\eta_{s\downarrow})^{2\tilde\nu_1}
    \Biggr]^2
    \nonumber\\
    &\quad+
    \frac{
        2^{-2\tilde\nu_0}
    }{
        \Gamma^2(1+\tilde\nu_0)
    }
    (k\eta_{0\downarrow})^{2\tilde\nu_0-2\tilde\nu_1}
    (k\eta_{s\downarrow})^{2\tilde\nu_1-1}
    \Biggr\}.
    \label{eq:NMBClowN22NodeExpanded}
\end{align}
Its equivalence to
Eq.~\eqref{eq:NMBClowN22NodeComplete}
is algebraic within the adopted asymptotic
matching approximation.
}
For $0<\nu_0<\tfrac12$, the compact kernel
in Eq.~\eqref{eq:NMBClowN22NodeComplete}
remains applicable, but the matching-column
components must be evaluated on that branch.

\paragraph{Leading node envelope at fixed boundary ratio.}
For a fixed background,
$\eta_{0\downarrow}/\eta_{s\downarrow}>1$
is independent of $k$.
Expanding Eq.~\eqref{eq:NMBClowN22NodeComplete}
jointly in the two small arguments at this
fixed ratio gives
\begin{equation}\label{eq:NMBClowN22NodeLeading}
    \mathcal N_{22,\mathrm{node}}^{(0)<}
    \simeq
    \begin{cases}
        \displaystyle
        \frac{
            2^{2\nu_0-1}
            \Gamma^2(\nu_0+\tfrac12)
        }{\pi}
        \frac{
            (k\eta_{0\downarrow})^{2\nu_1-2\nu_0}
        }{
            (k\eta_{s\downarrow})^{2\nu_1}
        }
        \mathcal B_{\mathrm{node}}^2,
        & \nu_0>\tfrac12,\\[12pt]
        \displaystyle
        \frac{
            2^{-2\nu_0-1}
            \Gamma^2(\tfrac12-\nu_0)
        }{\pi}
        \frac{
            (k\eta_{s\downarrow})^{2\nu_1-2}
        }{
            (k\eta_{0\downarrow})^{2\nu_1-2\nu_0}
        },
        & 0<\nu_0<\tfrac12,
    \end{cases}
\end{equation}
where
\begin{equation}\label{eq:NMBClowN22NodeInterference}
    \mathcal B_{\mathrm{node}}
    \equiv
    \frac{\nu_1-1}{\nu_1-\tfrac12}
    -
    \frac{\nu_1-\nu_0}{
        2(\nu_1-\tfrac12)(\nu_0-\tfrac12)
    }
    \left(
        \frac{\eta_{s\downarrow}}
             {\eta_{0\downarrow}}
    \right)^{2\nu_1-1}.
\end{equation}
The upper branch assumes
$\mathcal B_{\mathrm{node}}\neq0$;
an additional cancellation requires retaining
further terms.
The lower-branch coefficient is nonzero
throughout its stated domain.
Both branches then obey
\begin{equation}\label{eq:NMBClowN22NodeScaling}
    \mathcal N_{22,\mathrm{node}}^{(0)<}
    \propto k^{-2\tilde\nu_0-1}.
\end{equation}
For fixed $\eta_{2\uparrow}$, the nodes occur at
$k_m=m\pi/(2\eta_{2\uparrow})$, restricted to
the stated matching regime.
Equation~\eqref{eq:NMBClowN22NodeScaling}
describes an envelope through these discrete
nodes, rather than the spectrum between them.

\paragraph{Distinct limit at fixed $k\eta_{0\downarrow}$.}
If $k\eta_{s\downarrow}\to0$ while
$k\eta_{0\downarrow}\ll1$ is held fixed,
then for $\nu_1\neq1$ both branches give
\begin{equation}\label{eq:NMBClowN22NodeFixedArgument}
    \mathcal N_{22,\mathrm{node}}^{(0)<}
    \simeq
    \frac{
        2^{2\nu_0-1}
        \Gamma^2(\nu_0+\tfrac12)
    }{\pi}
    \left(
        \frac{\nu_1-1}{\nu_1-\tfrac12}
    \right)^2
    \frac{
        (k\eta_{0\downarrow})^{2\nu_1-2\nu_0}
    }{
        (k\eta_{s\downarrow})^{2\nu_1}
    }.
\end{equation}
The retained term dominates when
\begin{equation}\label{eq:NMBClowN22NodeHierarchy}
    \begin{gathered}
        (k\eta_{s\downarrow})^{2\nu_1-1}
        |\mathcal E_1|
        \ll
        2|1-\nu_1|\,|\mathcal F_1|,
        \\
        (2\nu_1-1)k\eta_{s\downarrow}
        \ll
        2|1-\nu_1|.
    \end{gathered}
\end{equation}
This limit is not interchangeable with the
fixed-ratio expansion in
Eq.~\eqref{eq:NMBClowN22NodeLeading},
particularly for $0<\nu_0<\tfrac12$,
where the competing terms have different
frequency scalings.

\subsubsection{Low-frequency phase-node SGWB
from the fixed-argument limit}

We use Eq.~\eqref{eq:NMBClowN22NodeFixedArgument}
for $\nu_1\neq1$, subject to
Eq.~\eqref{eq:NMBClowN22NodeHierarchy}
and the domain specified above.
Writing
\begin{equation}\label{eq:NMBClowNodeCnexp}
    \mathcal N_{22,\mathrm{node}}^{(0)<}
    \simeq
    \frac{
        \mathcal C_{\mathrm{node}}^{(0)<}
    }{
        (k\eta_{s\downarrow})^{
            n_{\mathrm{node}}^{(0)<}
        }
    },
\end{equation}
we identify
\begin{equation}\label{eq:NMBClowNodeCoefficient}
    \mathcal C_{\mathrm{node}}^{(0)<}
    \equiv
    \frac{
        2^{2\nu_0-1}
        \Gamma^2\!\left(\nu_0+\tfrac12\right)
    }{\pi}
    \left(
        \frac{\nu_1-1}{\nu_1-\tfrac12}
    \right)^2
    \left(
        \frac{\eta_{0\downarrow}}
             {\eta_{s\downarrow}}
    \right)^{2\nu_1-2\nu_0},
\end{equation}
and
\begin{equation}\label{eq:NMBClowNodeExponent}
    n_{\mathrm{node}}^{(0)<}
    =
    2\nu_0
    =
    \frac{4}{3w_0+1}.
\end{equation}
No phase averaging is performed for the
node spectra, so no additional factor of
$1/2$ is introduced.

Substituting
Eq.~\eqref{eq:NMBClowNodeCnexp}
into Eq.~\eqref{eq:phfpketarad}
gives the primordial tensor spectrum
\begin{equation}\label{eq:phf1NMBClowNode}
    \mathcal P_{h,\mathrm{node}}^{(0)<}(k)
    \simeq
    \frac{4}{\pi^2}\,
    \frac{k^4}{
        H_0^2\Omega_{\gamma0}m_{\rm pl}^2
    }\,
    \frac{
        \mathcal C_{\mathrm{node}}^{(0)<}
    }{
        (k\eta_{s\downarrow})^{
            n_{\mathrm{node}}^{(0)<}
        }
    }.
\end{equation}
Using Eq.~\eqref{eq:sgwbphfessential},
$k=2\pi f$ for $a_0=1$, and the
bounce-energy relation in
Eq.~\eqref{eq:etasrhos}, we obtain
\begin{align}
    \Omega_{\mathrm{GW,node}}^{(0)<}(f)h^2
    &\simeq
    \frac{h^2}{24}\,
    \frac{8}{\pi}
    \left(
        \frac{f_{H_0}}{f_{m_{\rm pl}}}
    \right)^2
    \frac{
        \mathcal C_{\mathrm{node}}^{(0)<}
    }{
        (2\pi)^{-n_{T,\mathrm{node}}^{(0)<}+1}
    }
    \left(
        \frac{f}{f_{H_0}}
    \right)^{n_{T,\mathrm{node}}^{(0)<}}
    \nonumber\\
    &\quad\times
    \left[
        \frac{\rho_{s\downarrow}^{1/4}}
             {(\rho_{c0}/\Omega_{\gamma0})^{1/4}}
    \right]^{4-n_{T,\mathrm{node}}^{(0)<}}
    \mathcal T_{\rm eq}(f).
    \label{eq:NMBClowNodeSGWB}
\end{align}
When varying the bounce energy, the dependence
of $\mathcal C_{\mathrm{node}}^{(0)<}$ on
$\eta_{0\downarrow}/\eta_{s\downarrow}$
must also be retained.

For fixed background parameters,
$\mathcal C_{\mathrm{node}}^{(0)<}$
is independent of $k$.
Where the hierarchy in
Eq.~\eqref{eq:NMBClowN22NodeHierarchy}
holds, the leading primordial tilt of the
continuous node envelope is therefore
\begin{equation}\label{eq:NMBClowNodeNT}
    n_{T,\mathrm{node}}^{(0)<}
    \equiv
    \frac{
        d\ln\mathcal P_{h,\mathrm{node}}^{(0)<}(k)
    }{d\ln k}
    \simeq
    4-n_{\mathrm{node}}^{(0)<}
    =
    4-2\nu_0
    =
    \frac{4w_0}{w_0+\tfrac13},
    \qquad
    w_0>-\frac13,\quad w_0\neq1.
\end{equation}
This derivative is taken at fixed background
parameters, not along the limiting path
with fixed $k\eta_{0\downarrow}$.
For $0<\nu_0<\tfrac12$, decreasing the
frequency can invalidate the hierarchy,
so this index need not describe the
fixed-ratio asymptotics in
Eq.~\eqref{eq:NMBClowN22NodeLeading}.
The present-day SGWB spectrum additionally
contains the frequency dependence of
$\mathcal T_{\rm eq}(f)$.

\subsubsection{High-frequency phase-node SGWB}

For $k\eta_{0\downarrow}\gg1$,
the Phase~0--I transition leaves the
leading-order spectrum unchanged.
The high-frequency NMBC node kernel
therefore reduces to the MBC node result
in Eq.~\eqref{eq:N221node}.
For $\nu_1>\tfrac12$ and $\nu_1\neq1$,
the high-frequency NMBC phase-node spectrum
reduces at leading order to the MBC node result.
In the regime
$k\eta_{0\downarrow}\gg1$,
$k\eta_{s\downarrow}\ll1$, and
$k\eta_{2\uparrow}\gg1$, evaluated at phase node ($2k\eta_{2\uparrow}=m\pi$), we have
\begin{align}
    \Omega_{\mathrm{GW,node}}^{(0)\ge}(f)h^2
    &\simeq
    \Omega_{\mathrm{GW,node}}^{(1)}(f)h^2
    \nonumber\\
    &\simeq
    \frac{h^2}{24}\,
    \frac{8}{\pi}
    \left(
        \frac{f_{H_0}}{f_{m_{\rm pl}}}
    \right)^2
    \frac{
        \mathcal C_{\mathrm{node}}^{(0)\ge}
    }{
        (2\pi)^{1-n_{T,\mathrm{node}}^{(0)\ge}}
    }
    \left(
        \frac{f}{f_{H_0}}
    \right)^{n_{T,\mathrm{node}}^{(0)\ge}}
    \nonumber\\
    &\quad\times
    \left[
        \frac{\rho_{s\downarrow}^{1/4}}
             {(\rho_{c0}/\Omega_{\gamma0})^{1/4}}
    \right]^{4-n_{T,\mathrm{node}}^{(0)\ge}}
    \mathcal T_{\rm eq}(f),
    \label{eq:NMBChighfreNodeSGWB}
\end{align}
where the coefficient and tensor spectral index are
\begin{align}
    \mathcal C_{\mathrm{node}}^{(0)\ge}
    &=
    \mathcal C_{\mathrm{node}}^{(1)}
    =
    \frac{
        2^{2\nu_1-1}
        \Gamma^2(\nu_1-\tfrac12)
        (\nu_1-1)^2
    }{\pi},
    \label{eq:NMBChighNodeC}
    \\
    n_{T,\mathrm{node}}^{(0)\ge}
    &=
    n_{T,\mathrm{node}}^{(1)}
    =
    4-2\nu_1
    =
    \frac{4w_1}{w_1+\tfrac13}.
    \label{eq:NMBChighNodeNT}
\end{align}
No averaging over the bounce-induced phase is
performed, so no additional factor of $1/2$
is introduced.

For the illustrative plots, we specify the
extrapolated high-frequency pivot amplitude
$\Omega_{\mathrm{GW,node}}^{(0)\ge}(f_\star)h^2$
and write
\begin{equation}\label{eq:NMBChighNodePivotSGWB}
    \Omega_{\mathrm{GW,node}}^{(0)\ge}(f)h^2
    \simeq
    \Omega_{\mathrm{GW,node}}^{(0)\ge}(f_\star)h^2
    \left(\frac{f}{f_\star}\right)^{
        n_{T,\mathrm{node}}^{(0)\ge}
    }
    \frac{
        \mathcal T_{\rm eq}(f)
    }{
        \mathcal T_{\rm eq}(f_\star)
    },
    \qquad
    f_\star\le f\le f_{\rm cut}.
\end{equation}
The plotted curves extend the continuous
node envelope to $f_\star$ and $f_{\rm cut}$.
The asymptotic expression is controlled
in the interior regime
$k\eta_{0\downarrow}\gg1$ and
$k\eta_{s\downarrow}\ll1$, subject to
the nodal matching conditions stated above.

\subsubsection{Two limits connecting the MBC and NMBC
phase-node solutions}
\label{subsubsec:NodeMBCLimits}

The fixed-boundary-ratio kernel in
Eq.~\eqref{eq:NMBClowN22NodeLeading}
admits two reductions to the MBC node solution.
These correspond to identical contracting
exponents and a vanishing duration of Phase~I,
respectively.
Both reductions inherit the asymptotic
matching conditions and Hankel-order
qualifications stated above.

\paragraph{Identical exponents: $\nu_0=\nu_1$.}
For $\nu_0=\nu_1>\tfrac12$, the Phase~0--I
transition becomes trivial.
The second term in
Eq.~\eqref{eq:NMBClowN22NodeInterference}
vanishes, giving
\begin{equation}\label{eq:NodeEqualNuB}
    \left.
        \mathcal B_{\mathrm{node}}
    \right|_{\nu_0=\nu_1}
    =
    \frac{\nu_1-1}{\nu_1-\tfrac12}.
\end{equation}
The factor
$(k\eta_{0\downarrow})^{2\nu_1-2\nu_0}$
also becomes unity.
For $\nu_1\neq1$, the upper branch of
Eq.~\eqref{eq:NMBClowN22NodeLeading}
therefore reduces to
\begin{align}\label{eq:NodeEqualNuReduction}
    \left.
        \mathcal N_{22,\mathrm{node}}^{(0)<}
    \right|_{\nu_0=\nu_1}
    &\simeq
    \frac{
        2^{2\nu_1-1}
        \Gamma^2(\nu_1+\tfrac12)
    }{\pi}
    \left(
        \frac{\nu_1-1}{\nu_1-\tfrac12}
    \right)^2
    \frac{1}{
        (k\eta_{s\downarrow})^{2\nu_1}
    }
    \nonumber\\
    &=
    \frac{
        2^{2\nu_1-1}
        \Gamma^2(\nu_1-\tfrac12)
        (\nu_1-1)^2
    }{
        \pi(k\eta_{s\downarrow})^{2\nu_1}
    }.
\end{align}
Here we used
$\Gamma(\nu_1+\tfrac12)
=(\nu_1-\tfrac12)\Gamma(\nu_1-\tfrac12)$.
With $\tilde\nu_1=\nu_1-\tfrac12$,
the coefficient and exponent are precisely
those in Eqs.~\eqref{eq:C1node}
and \eqref{eq:n1node}.
Thus,
Eq.~\eqref{eq:NodeEqualNuReduction}
recovers the MBC kernel in
Eq.~\eqref{eq:N221node}.
This check concerns the branch
$\nu_1>\tfrac12$, as required by the
NMBC derivation.

\paragraph{Coincident boundaries:
$\eta_{0\downarrow}\to\eta_{s\downarrow}$.}
We now take
$\eta_{0\downarrow}/\eta_{s\downarrow}\to1^+$,
so that Phase~I has zero duration.
For the upper branch, the interference factor
simplifies to
\begin{align}\label{eq:NodeCoincidentB}
    \lim_{\eta_{0\downarrow}\to\eta_{s\downarrow}^{+}}
    \mathcal B_{\mathrm{node}}
    &=
    \frac{\nu_1-1}{\nu_1-\tfrac12}
    -
    \frac{\nu_1-\nu_0}{
        2(\nu_1-\tfrac12)(\nu_0-\tfrac12)
    }
    \nonumber\\
    &=
    \frac{\nu_0-1}{\nu_0-\tfrac12}.
\end{align}
The boundary-argument factors in the two
branches become
\begin{align}
    \frac{
        (k\eta_{0\downarrow})^{2\nu_1-2\nu_0}
    }{
        (k\eta_{s\downarrow})^{2\nu_1}
    }
    &\longrightarrow
    (k\eta_{s\downarrow})^{-2\nu_0},
    \\
    \frac{
        (k\eta_{s\downarrow})^{2\nu_1-2}
    }{
        (k\eta_{0\downarrow})^{2\nu_1-2\nu_0}
    }
    &\longrightarrow
    (k\eta_{s\downarrow})^{-(2-2\nu_0)}.
\end{align}
All dependence on $\nu_1$ therefore disappears.
Using the gamma-function recurrence in the
upper branch and defining
$\tilde\nu_0=|\nu_0-\tfrac12|$, we obtain
\begin{equation}\label{eq:NodeCoincidentReduction}
    \lim_{\eta_{0\downarrow}\to\eta_{s\downarrow}^{+}}
    \mathcal N_{22,\mathrm{node}}^{(0)<}
    \simeq
    \frac{
        2^{2\tilde\nu_0-2}
        \Gamma^2(\tilde\nu_0)
    }{
        \pi(k\eta_{s\downarrow})^{1+2\tilde\nu_0}
    }
    \begin{cases}
        (1-2\tilde\nu_0)^2,
        & \nu_0>\tfrac12,\quad \nu_0\neq1,\\[2pt]
        1,
        & 0<\nu_0<\tfrac12.
    \end{cases}
\end{equation}
This is the MBC node kernel in
Eqs.~\eqref{eq:N221node}--\eqref{eq:n1node},
with its contracting exponent replaced by
$\nu_0$.
Physically, removing Phase~I leaves Phase~0
as the contracting phase preceding the bounce.
Accordingly, $f_\star\to f_{\rm cut}$,
and the separate high-frequency NMBC
asymptotic interval disappears.

Note that, the coincident-boundary reduction requires
the full interference factor in
Eq.~\eqref{eq:NMBClowN22NodeInterference}.
It cannot generally be inferred from the
distinct fixed-argument approximation in
Eq.~\eqref{eq:NMBClowN22NodeFixedArgument}.

\subsection{Phase-node contours and comparison with
integrated constraints}
\label{subsec:DeltaNeffPhaseNode}

\begin{figure}[htbp]
    \centering
    \includegraphics[width=\textwidth]
        {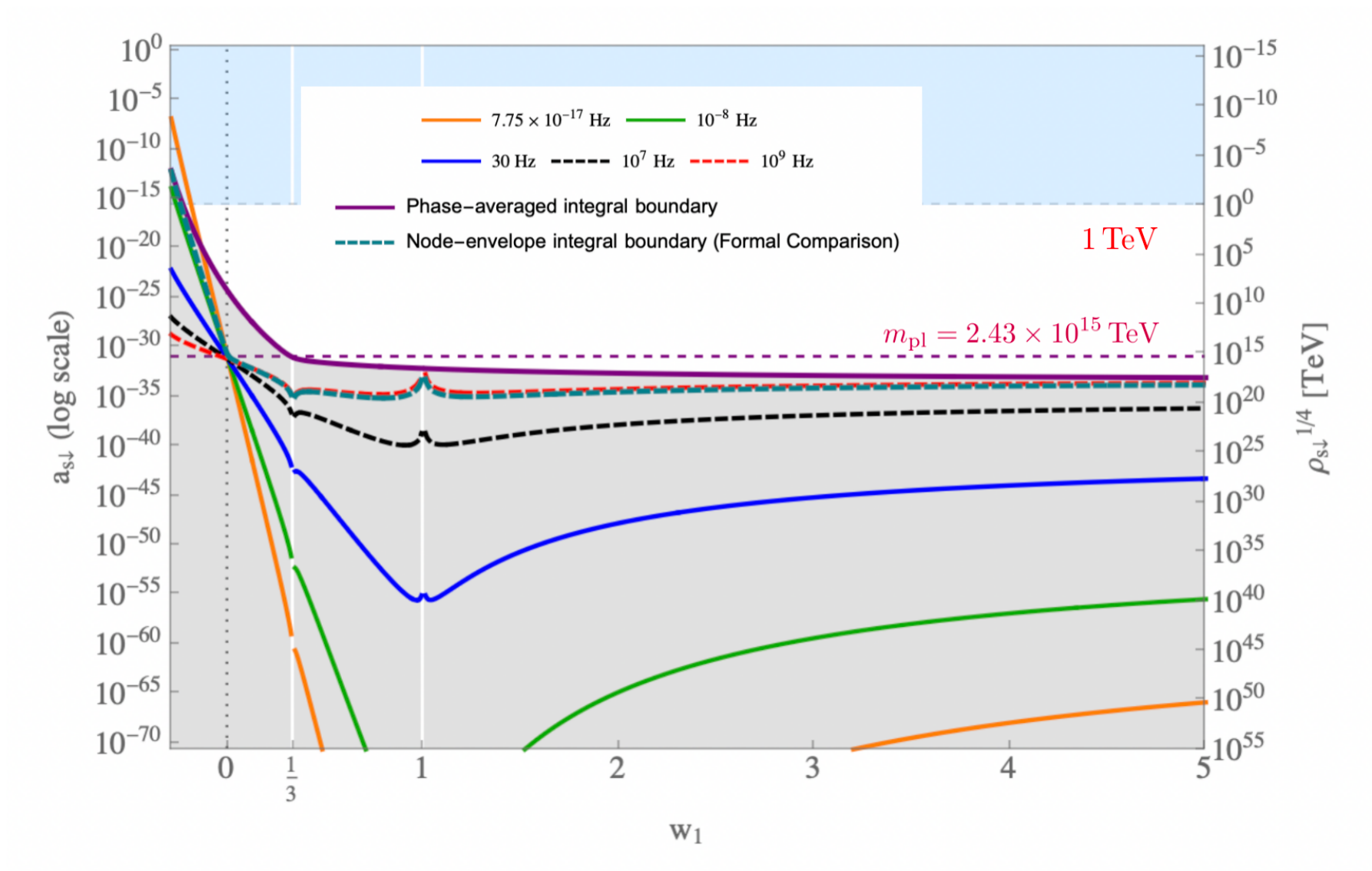}
    \caption{
        Phase-node contours and integrated
        GW-energy-density boundaries in MBC
        under $\tau$-derivative continuity.
        The left vertical axis gives
        $a_{s\downarrow}$ and the right axis gives
        $\rho_{s\downarrow}^{1/4}$ in TeV,
        related by Eq.~\eqref{eq:asdrhosdeq}.
        The solid orange, dark-green, and blue
        curves, and the dashed black and red
        curves, satisfy
        Eq.~\eqref{eq:MBCNodePointwiseContours}
        at $f=7.75\times10^{-17},\,10^{-8},\,30,\,
        10^7$, and $10^9\,\mathrm{Hz}$, respectively.
        Orange and red shading indicates
        pointwise threshold exceedance at the
        lowest and highest reference frequencies.
        The solid purple curve is the
        phase-averaged integral boundary,
        Eq.~\eqref{eq:MBCNodePhaseAverageBoundary};
        gray shading marks exceedance of its budget.
        The dashed dark-cyan curve is the formal
        node-envelope integral boundary,
        Eq.~\eqref{eq:MBCNodeEnvelopeBoundary},
        with no exclusion shading assigned to it.
        Both integrals use
        $10^{-10}\le f\le10^9\,\mathrm{Hz}$
        and the budget $1.7\times10^{-6}$.
        Pale blue marks
        $\rho_{s\downarrow}^{1/4}<1\,\mathrm{TeV}$.
        The bands $|w_1-\tfrac13|<0.01$ and
        $|w_1-1|<0.01$ are omitted because the
        node asymptotics require separate treatment
        near these exceptional values.
        Neither these omitted bands nor the
        pale-blue region represents a
        $\Delta N_{\rm eff}$ exclusion.
    }
    \label{fig:intMBC_Tau_PhaseNode_Contours}
\end{figure}

Figure~\ref{fig:intMBC_Tau_PhaseNode_Contours}
compares pointwise phase-node contours in MBC
with integrated GW-energy-density boundaries.
The pointwise contours satisfy
\begin{equation}
    \Omega_{\rm GW,node}^{(1)}(f)h^2
    =1.7\times10^{-6}
    \label{eq:MBCNodePointwiseContours}
\end{equation}
at the reference frequencies listed in the caption.
Each contour assumes that the selected frequency
coincides with a phase node.
The node normalization is used without an
additional phase-averaging factor of $1/2$.

The solid purple boundary is obtained from
the phase-averaged generic-phase spectrum:
\begin{equation}
    \int_{f_{\min}}^{f_{\max}}
    \overline{\Omega}_{\rm GW}^{(1)}(f)h^2\,
    d\ln f
    =1.7\times10^{-6},
    \qquad
    f_{\min}=10^{-10}\,\mathrm{Hz},
    \quad
    f_{\max}=10^9\,\mathrm{Hz}.
    \label{eq:MBCNodePhaseAverageBoundary}
\end{equation}
The gray region exceeds this integrated budget.
For comparison, the dashed dark-cyan boundary
satisfies
\begin{equation}
    \int_{f_{\min}}^{f_{\max}}
    \Omega_{\rm GW,node}^{(1)}(f)h^2\,
    d\ln f
    =1.7\times10^{-6},
    \label{eq:MBCNodeEnvelopeBoundary}
\end{equation}
with the node envelope formally extended
throughout the same interval.
Since exact nodes of a fixed background are
discrete, this envelope integral does not
replace the broadband $\Delta N_{\rm eff}$
constraint.

The integration limits are illustrative fixed
cutoffs rather than predictions of the bounce model.
The contours use the asymptotic expressions
without an additional $k\eta_{s\downarrow}\ll1$
validity cut.
Their physical interpretation therefore requires
checking the applicable matching regime throughout
the relevant band, together with the validity
of phase averaging for
Eq.~\eqref{eq:MBCNodePhaseAverageBoundary}.

\subsubsection{Scale-invariant phase-node envelope}

At leading order, the high-frequency NMBC
node envelope reduces to the MBC node result.
Its scale-invariant limit satisfies
\begin{equation}
    n_{T,\mathrm{node}}^{(0)\ge}
    =
    n_{T,\mathrm{node}}^{(1)}
    =0
    \quad\Longleftrightarrow\quad
    w_1=0,\qquad \nu_1=2,
\end{equation}
for which $\mathcal C_{\mathrm{node}}^{(1)}=2$.
Using the adopted cosmological constants gives
\begin{equation}
    \begin{aligned}
        \Omega_{\mathrm{GW,node}}^{(0)\ge}(f)h^2
        &\simeq
        \Omega_{\mathrm{GW,node}}^{(1)}(f)h^2\\
        &\simeq
        A_{\mathrm{node}}
        \left(
            \frac{\rho_{s\downarrow}^{1/4}}
                 {1\,\mathrm{TeV}}
        \right)^4
        \mathcal T_{\mathrm{eq}}(f),
        \qquad
        A_{\mathrm{node}}
        \simeq7.95913\times10^{-69},
    \end{aligned}
    \label{eq:NodeFlatSpectrum}
\end{equation}
where
$\mathcal T_{\mathrm{eq}}(f)
=1+\frac{9}{32}(f_{\mathrm{eq}}/f)^2$.
For NMBC, this expression requires
$k\eta_{0\downarrow}\gg1$, together with
$k\eta_{s\downarrow}\ll1$ and
$k\eta_{2\uparrow}\gg1$ at the phase nodes.

\paragraph{Pointwise comparison.}
Applying the adopted threshold in
Eq.~\eqref{eq:MBCNodePointwiseContours}
at a node frequency $f_p$ gives
\begin{equation}
    \rho_{s\downarrow}^{1/4}
    \lesssim
    (1\,\mathrm{TeV})
    \left[
        \frac{1.7\times10^{-6}}
             {A_{\mathrm{node}}
              \mathcal T_{\mathrm{eq}}(f_p)}
    \right]^{1/4}.
    \label{eq:NodeFlatPointwiseBound}
\end{equation}
For $f_p=10^{-6}\,\mathrm{Hz}$,
$\mathcal T_{\mathrm{eq}}(f_p)\simeq1$, yielding
\begin{equation}
    \rho_{s\downarrow}^{1/4}
    \lesssim
    3.82293\times10^{15}\,\mathrm{TeV}.
\end{equation}
This comparison assumes that $f_p$ is a
phase node within the required asymptotic regime.

\paragraph{Formal node-envelope integral.}
Extending Eq.~\eqref{eq:NodeFlatSpectrum}
over a continuous frequency interval gives
\begin{align}
    \mathcal I_{\mathrm{node}}
    \equiv
    \int_{f_{\min}}^{f_{\max}}
    \Omega_{\mathrm{GW,node}}^{(1)}(f)h^2\,
    d\ln f
    \simeq
    A_{\mathrm{node}}
    \left(
        \frac{\rho_{s\downarrow}^{1/4}}
             {1\,\mathrm{TeV}}
    \right)^4 J,
    \end{align}
   with $ J
    =
    \ln\!\left(\frac{f_{\max}}{f_{\min}}\right)
    +\frac{9}{64}f_{\mathrm{eq}}^2
    \left(f_{\min}^{-2}-f_{\max}^{-2}\right)$.
Writing
$\mathcal B=B_\nu\Delta N_{\mathrm{eff}}^{\max}$,
with $B_\nu$ defined in
Eq.~\eqref{eq:NMBC_HF_DeltaNeff},
the condition $\mathcal I_{\mathrm{node}}\le\mathcal B$
gives
\begin{equation}
    \rho_{s\downarrow}^{1/4}
    \lesssim
    (1\,\mathrm{TeV})
    \left[
        \frac{\mathcal B}{A_{\mathrm{node}}J}
    \right]^{1/4}.
    \label{eq:NodeFlatFormalIntegralBound}
\end{equation}
For the separate illustrative band
$f_{\min}=10^{-12}\,\mathrm{Hz}$ to
$f_{\max}=10^8\,\mathrm{Hz}$,
we obtain $J\simeq46.05170$.
Using the same rounded radiation budget
$\mathcal B=1.7\times10^{-6}$ gives
\begin{equation}
    \left.
        \rho_{s\downarrow}^{1/4}
    \right|_{\mathrm{node\ envelope}}
    \lesssim
    1.46752\times10^{15}\,\mathrm{TeV}.
\end{equation}

Although the node envelope is scale invariant
at $w_1=0$, the generic phase-averaged spectrum
has $n_T^{(1)}=-2$.
The formal estimate above therefore cannot
be substituted for the integrated constraint
on the full spectrum.
For an NMBC interpretation, the entire
integration band must additionally lie within
the high-frequency regime.

\subsection{Four illustrative phase-node examples}

\begin{figure}[htbp]
    \centering
    \includegraphics[width=\textwidth]
        {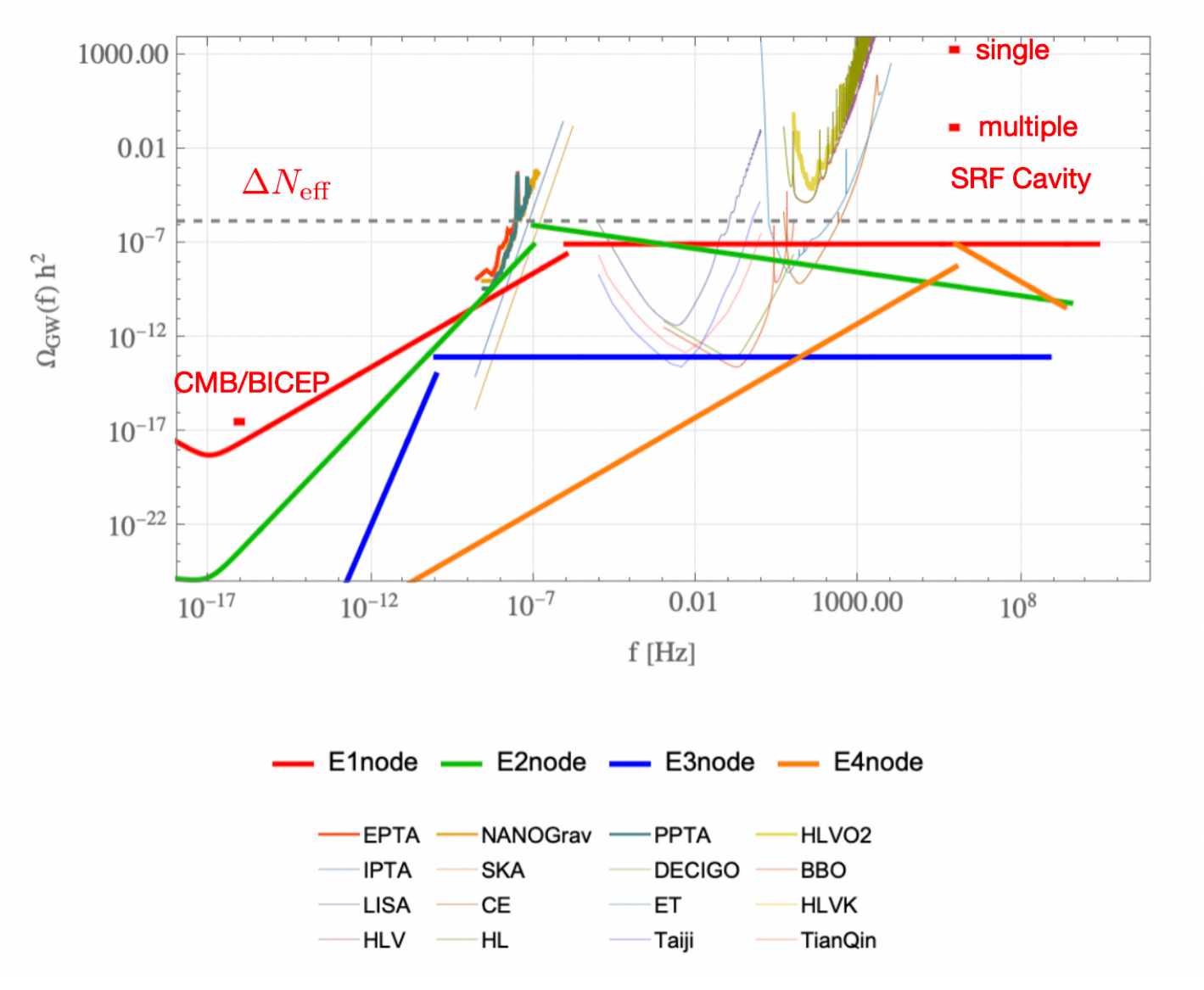}
    \caption{
        Continuous phase-node SGWB envelopes
        for the illustrative NMBC examples
        E1node--E4node.
        The low-frequency branches use the
        fixed-argument approximation in
        Eq.~\eqref{eq:NMBClowNodeSGWB},
        while the high-frequency branches use
        Eq.~\eqref{eq:NMBChighNodePivotSGWB}.
        The colors correspond to the examples
        listed in the text.
        The curves are extended to the break
        and cutoff for illustration; the
        asymptotic expressions apply within
        their stated matching regimes.
        For a fixed background, the physical
        phase nodes sample these envelopes
        only at discrete frequencies.
    }
    \label{fig:FourExamples_Node_FixedArgument}
\end{figure}

Figure~\ref{fig:FourExamples_Node_FixedArgument}
shows four continuous NMBC node envelopes,
E1node--E4node, constructed from
Eqs.~\eqref{eq:NMBClowNodeSGWB}
and \eqref{eq:NMBChighNodePivotSGWB}.
The low-frequency expression uses the
fixed-argument approximation in
Eq.~\eqref{eq:NMBClowN22NodeFixedArgument},
subject to the hierarchy in
Eq.~\eqref{eq:NMBClowN22NodeHierarchy}.

Within the stated domains, the primordial
node-envelope tilts are
\begin{equation}
    n_{T,\mathrm{node}}^{(0)<}=4-2\nu_0,
    \qquad
    n_{T,\mathrm{node}}^{(0)\ge}=4-2\nu_1,
\end{equation}
with $\nu_1>\tfrac12$ and $\nu_1\neq1$.
The corresponding equation-of-state parameters
are
\begin{equation}
    w_0
    =
    \frac{n_{T,\mathrm{node}}^{(0)<}}
         {3\left(4-n_{T,\mathrm{node}}^{(0)<}\right)},
    \qquad
    w_1
    =
    \frac{n_{T,\mathrm{node}}^{(0)\ge}}
         {3\left(4-n_{T,\mathrm{node}}^{(0)\ge}\right)}.
\end{equation}

For each example, the specified amplitude
$\Omega_{\mathrm{GW,node}}^{(0)\ge}(f_\star)h^2$
is the high-frequency node expression
extrapolated to $f=f_\star$.
With $a_0=1$, the transition scales satisfy
\begin{equation}
    \eta_{0\downarrow}
    =\frac{1}{2\pi f_\star},
    \qquad
    \eta_{1\downarrow}
    =\eta_{3\downarrow}
    =\eta_{s\downarrow},
    \qquad
    f_{\rm cut}
    =\frac{1}{2\pi\eta_{s\downarrow}}.
\end{equation}
The remaining equation-of-state parameters are
$w_2,w_3\to-\infty$ and $w_4=\tfrac13$.
The phase scale $\eta_{2\uparrow}$ remains free
and must satisfy $k\eta_{2\uparrow}\gg1$
over the frequency range considered.

The illustrative inputs and reconstructed
parameters are as follows\footnote{Examples~1 and 4 have $\tilde\nu_0=1$,
and Example~4 also has $\tilde\nu_1=2$.
Small deviations from
$n_{T,\mathrm{node}}^{(0)<}=1$ and, for Example~4,
$n_{T,\mathrm{node}}^{(0)\ge}=-1$
give nearby noninteger Hankel orders.
The listed envelopes use the smooth limits
of the leading coefficients as these deviations
vanish.
A direct asymptotic treatment of the nodal
matrices at the exact integer orders requires
the corresponding logarithmic Hankel expansions.}:
\begin{enumerate}
    \item \textbf{E1node (red).}
    \begin{equation}
        \begin{aligned}
            n_{T,\mathrm{node}}^{(0)<}&=1,
            &
            n_{T,\mathrm{node}}^{(0)\ge}&=0,
            \\
            f_\star&=10^{-6}\,\mathrm{Hz},
            &
            \Omega_{\mathrm{GW,node}}^{(0)\ge}
            (f_\star)h^2&=10^{-7}.
        \end{aligned}
    \end{equation}
    These inputs give
    \begin{equation}
        (w_0,w_1)
        =
        \left(\frac19,0\right),
    \end{equation}
    \begin{equation}
        \left(
            \eta_{0\downarrow},\eta_{s\downarrow}
        \right)
        \simeq
        \left(
            1.59\times10^5,\,
            6.95\times10^{-12}
        \right)\,\mathrm{s},
    \end{equation}
    and
    \begin{equation}
        \rho_{s\downarrow}^{1/4}
        \simeq0.774\,m_{\rm pl},
        \qquad
        f_{\rm cut}
        \simeq2.29\times10^{10}\,\mathrm{Hz}.
    \end{equation}

    \item \textbf{E2node (green).}
    \begin{equation}
        \begin{aligned}
            n_{T,\mathrm{node}}^{(0)<}&=1.8,
            &
            n_{T,\mathrm{node}}^{(0)\ge}&=-\frac14,
            \\
            f_\star&=10^{-7}\,\mathrm{Hz},
            &
            \Omega_{\mathrm{GW,node}}^{(0)\ge}
            (f_\star)h^2&=10^{-6}.
        \end{aligned}
    \end{equation}
    These inputs give
    \begin{equation}
        (w_0,w_1)
        =
        \left(\frac3{11},-\frac1{51}\right),
    \end{equation}
    \begin{equation}
        \left(
            \eta_{0\downarrow},\eta_{s\downarrow}
        \right)
        \simeq
        \left(
            1.59\times10^6,\,
            4.70\times10^{-11}
        \right)\,\mathrm{s},
    \end{equation}
    and
    \begin{equation}
        \rho_{s\downarrow}^{1/4}
        \simeq0.114\,m_{\rm pl},
        \qquad
        f_{\rm cut}
        \simeq3.39\times10^9\,\mathrm{Hz}.
    \end{equation}

    \item \textbf{E3node (blue).}
    \begin{equation}
        \begin{aligned}
            n_{T,\mathrm{node}}^{(0)<}&=4-10^{-5},
            &
            n_{T,\mathrm{node}}^{(0)\ge}&=0,
            \\
            f_\star&=10^{-10}\,\mathrm{Hz},
            &
            \Omega_{\mathrm{GW,node}}^{(0)\ge}
            (f_\star)h^2&=10^{-13}.
        \end{aligned}
    \end{equation}
    These inputs give
    \begin{equation}
        (w_0,w_1)
        =
        \left(
            \frac{4\times10^5-1}{3},\,0
        \right)
        =
        (133333,0),
    \end{equation}
    \begin{equation}
        \left(
            \eta_{0\downarrow},\eta_{s\downarrow}
        \right)
        \simeq
        \left(
            1.59\times10^9,\,
            2.20\times10^{-10}
        \right)\,\mathrm{s},
    \end{equation}
    and
    \begin{equation}
        \rho_{s\downarrow}^{1/4}
        \simeq2.45\times10^{-2}\,m_{\rm pl},
        \qquad
        f_{\rm cut}
        \simeq7.24\times10^8\,\mathrm{Hz}.
    \end{equation}

    \item \textbf{E4node (orange).}
    \begin{equation}
        \begin{aligned}
            n_{T,\mathrm{node}}^{(0)<}&=1,
            &
            n_{T,\mathrm{node}}^{(0)\ge}&=-1,
            \\
            f_\star&=10^6\,\mathrm{Hz},
            &
            \Omega_{\mathrm{GW,node}}^{(0)\ge}
            (f_\star)h^2&=10^{-7}.
        \end{aligned}
    \end{equation}
    These inputs give
    \begin{equation}
        (w_0,w_1)
        =
        \left(\frac19,-\frac1{15}\right),
    \end{equation}
    \begin{equation}
        \left(
            \eta_{0\downarrow},\eta_{s\downarrow}
        \right)
        \simeq
        \left(
            1.59\times10^{-7},\,
            7.34\times10^{-11}
        \right)\,\mathrm{s},
    \end{equation}
    and
    \begin{equation}
        \rho_{s\downarrow}^{1/4}
        \simeq7.33\times10^{-2}\,m_{\rm pl},
        \qquad
        f_{\rm cut}
        \simeq2.17\times10^9\,\mathrm{Hz}.
    \end{equation}
\end{enumerate}

~
~.
~
\section{Explicit mismatch factors at $f=f_\star$}
\label{app:mismatch-coefficients}

The low- and high-frequency branches are derived
for $k\eta_{0\downarrow}\ll1$ and
$k\eta_{0\downarrow}\gg1$, respectively.
Extrapolating both to the pivot,
$2\pi f_\star\eta_{0\downarrow}=1$ for $a_0=1$,
generally gives different amplitudes.
The factors below quantify this asymptotic
normalization mismatch, which does not represent
a physical discontinuity.
Resolving the transition requires the full
Phase~0--I matching matrix without either
asymptotic approximation at that boundary.

\subsection{Phase-averaged spectra}
\label{subsec:nmbcfstarmismatch}

For the spectra in
Eqs.~\eqref{eq:NMBCgeoverlineSGWB}
and \eqref{eq:NMBClowoverlineSGWB}, define
\begin{equation}\label{eq:flefgeNMBC}
    \overline{\Omega}_{\rm GW}^{(0)\ge}(f_\star)h^2
    =
    \epsilon_g(\nu_0,\nu_1)\,
    \overline{\Omega}_{\rm GW}^{(0)<}(f_\star)h^2.
\end{equation}
Using their amplitude coefficients and spectral
indices gives
\begin{align}\label{eq:epsilong}
    \epsilon_g(\nu_0,\nu_1)
    &=
    \frac{
        \overline{\mathcal C}^{(0)\ge}
    }{
        \overline{\mathcal C}^{(0)<}
    }
    \left(
        2\pi f_\star\eta_{s\downarrow}
    \right)^{
        n_T^{(0)\ge}-n_T^{(0)<}
    }
    \nonumber\\
    &=
    \frac{
        4^{\nu_1}
        \Gamma^2\!\left(\nu_1+\tfrac12\right)
    }{
        4^{\nu_0}
        \Gamma^2\!\left(\nu_0+\tfrac12\right)
    },
    \qquad
    \nu_0>0,\quad \nu_1>\tfrac12,
\end{align}
within the validity domain of the underlying spectra.
Here
$2\pi f_\star\eta_{0\downarrow}=1$
and
$n_T^{(0)\ge}-n_T^{(0)<}=2\nu_0-2\nu_1$
cancel the dependence on the boundary scales.
The mismatch factor therefore depends only
on the phase exponents.
In particular,
\begin{equation}
    \epsilon_g(\nu_0,\nu_1)=1,
    \qquad \nu_0=\nu_1,
\end{equation}
as expected when the Phase~0--I transition
becomes trivial.

For the main-text examples E1--E4, respectively,
the rounded values are
\begin{equation}\label{eq:epsilongvalue}
    \epsilon_g
    \simeq
    \left\{
        1.48,\,
        1.50,\,
        1.35,\,
        4.02
    \right\}.
\end{equation}

Using Eqs.~\eqref{eq:NMBCgentw1}
and \eqref{eq:NMBClowntw0},
we can also express the factor in terms
of the two spectral indices:
\begin{equation}\label{eq:egntNMBC}
    \epsilon_g\!\left(
        n_T^{(0)<},n_T^{(0)\ge}
    \right)
    =
    2^{\,n_T^{(0)<}-n_T^{(0)\ge}}
    \frac{
        \Gamma^2\!\left(
            \dfrac{3-n_T^{(0)\ge}}{2}
        \right)
    }{
        \Gamma^2\!\left(
            \dfrac{3-n_T^{(0)<}}{2}
        \right)
    },
    \qquad
    n_T^{(0)<}<2,\quad n_T^{(0)\ge}<1.
\end{equation}

\begin{figure}[htbp]
    \centering
    \includegraphics[width=\textwidth]
        {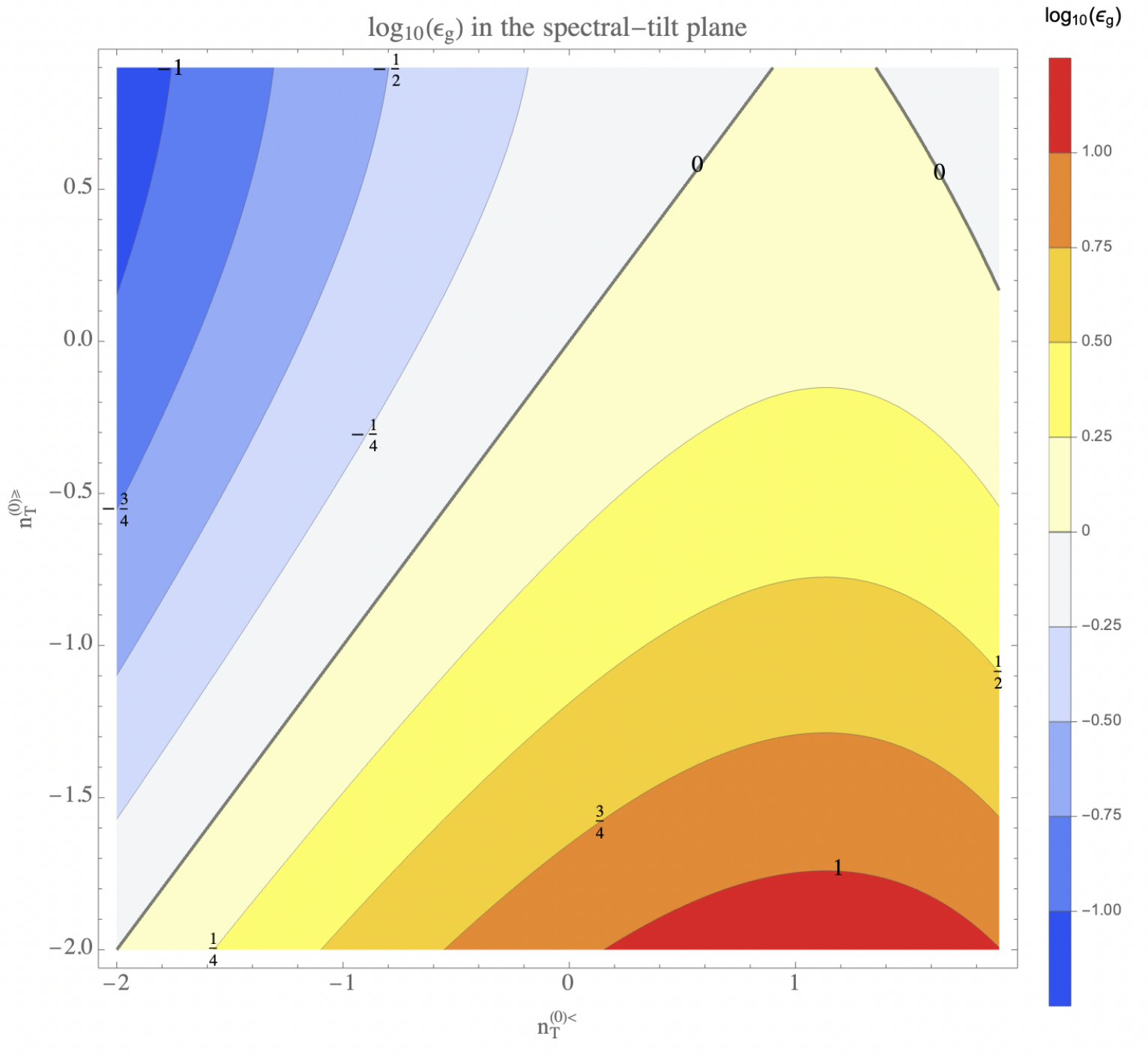}
    \caption{
        Normalization mismatch $\epsilon_g$
        between the phase-averaged asymptotic
        branches, evaluated using
        Eq.~\eqref{eq:egntNMBC}.
        Along the diagonal
        $n_T^{(0)<}=n_T^{(0)\ge}$,
        $\epsilon_g=1$, indicating equal
        amplitudes when the two branches
        are extrapolated to $f=f_\star$.
    }
    \label{fig:Epsilon_g_NMBC_Plots}
\end{figure}

Figure~\ref{fig:Epsilon_g_NMBC_Plots}
shows this dependence.
Over most of the displayed domain,
$-1\lesssim\log_{10}\epsilon_g\lesssim1$,
or equivalently
$10^{-1}\lesssim\epsilon_g\lesssim10$.
Larger departures occur near the upper-left
and lower-right corners.

\subsection{Phase-node spectra}

For the node spectra, we use the fixed-argument
low-frequency coefficient in
Eq.~\eqref{eq:NMBClowNodeCoefficient}
and the high-frequency coefficient in
Eq.~\eqref{eq:NMBChighNodeC}.
The former inherits the hierarchy in
Eq.~\eqref{eq:NMBClowN22NodeHierarchy}.
For $\nu_1>\tfrac12$ and $\nu_1\neq1$,
the gamma-function recurrence
$\Gamma(\nu_1+\tfrac12)
=(\nu_1-\tfrac12)\Gamma(\nu_1-\tfrac12)$
gives
\begin{equation}\label{eq:epsilongnode}
    \epsilon_{g,\mathrm{node}}
    \equiv
    \frac{
        \Omega_{\mathrm{GW,node}}^{(0)\ge}(f_\star)
    }{
        \Omega_{\mathrm{GW,node}}^{(0)<}(f_\star)
    }
    =
    \frac{
        4^{\nu_1}
        \Gamma^2(\nu_1+\tfrac12)
    }{
        4^{\nu_0}
        \Gamma^2(\nu_0+\tfrac12)
    }.
\end{equation}
At $\nu_1=1$, both leading coefficients used
in this ratio vanish, so their cancellation
does not establish the mismatch there.
The earlier qualifications concerning
exceptional Hankel orders also apply.

For E1node--E4node, using the coefficient
continuations specified for those examples,
we obtain
\begin{equation}\label{eq:epsilongnodevalue}
    \epsilon_{g,\mathrm{node}}
    \simeq
    \left\{
        3.5343,\,
        11.0100,\,
        9.00011,\,
        16
    \right\}.
\end{equation}
The low-frequency pivot amplitudes follow
by dividing the specified high-frequency
amplitudes by these factors.

Equations~\eqref{eq:epsilongnode}
and \eqref{eq:epsilong} coincide at fixed
$(\nu_0,\nu_1)$ within their common domain.
Their expressions in terms of spectral
indices differ because the node tilts obey
$n_{T,\mathrm{node}}^{(0)<}=4-2\nu_0$
and
$n_{T,\mathrm{node}}^{(0)\ge}=4-2\nu_1$.
Using Eqs.~\eqref{eq:NMBClowNodeNT}
and \eqref{eq:NMBChighNodeNT} gives
\begin{equation}\label{eq:epsgnodeTilts}
    \epsilon_{g,\mathrm{node}}
    =
    2^{
        n_{T,\mathrm{node}}^{(0)<}
        -n_{T,\mathrm{node}}^{(0)\ge}
    }
    \frac{
        \Gamma^2\!\left(
            \frac{5-n_{T,\mathrm{node}}^{(0)\ge}}{2}
        \right)
    }{
        \Gamma^2\!\left(
            \frac{5-n_{T,\mathrm{node}}^{(0)<}}{2}
        \right)
    }.
\end{equation}

\begin{figure}[htbp]
    \centering
    \includegraphics[width=\textwidth]
        {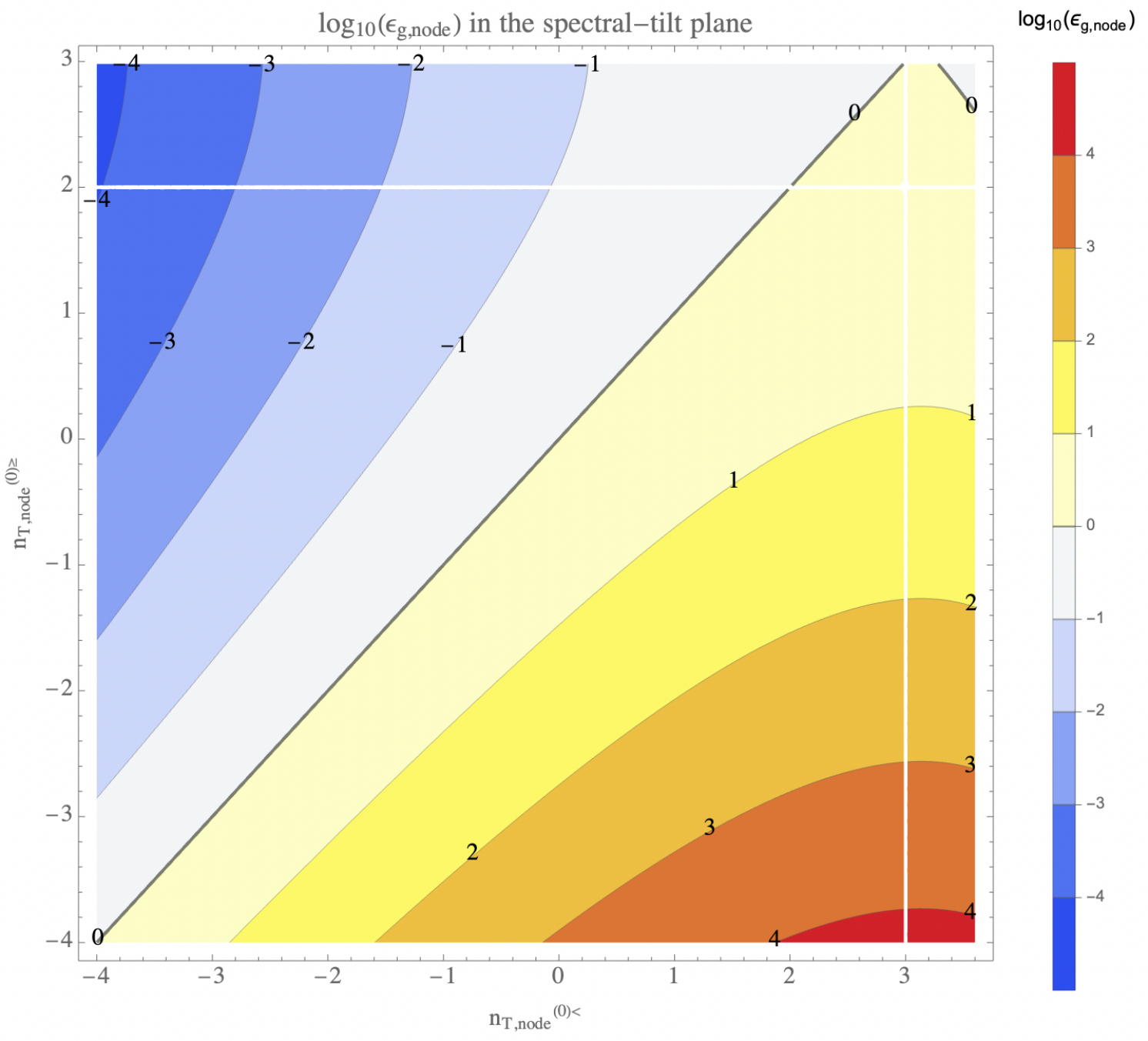}
    \caption{
        Normalization mismatch
        $\epsilon_{g,\mathrm{node}}$
        between the asymptotic node envelopes,
        evaluated using Eq.~\eqref{eq:epsgnodeTilts}
        for the selected fixed-argument approximation.
        Along the diagonal
        $n_{T,\mathrm{node}}^{(0)<}
        =n_{T,\mathrm{node}}^{(0)\ge}$,
        the factor equals unity within the
        stated domain.
        Departures from unity quantify the
        mismatch between the envelopes
        extrapolated to $f=f_\star$.
    }
    \label{fig:Epsilon_g_Node_NMBC_Plots}
\end{figure}

Figure~\ref{fig:Epsilon_g_Node_NMBC_Plots}
illustrates this dependence.
For
$0<n_{T,\mathrm{node}}^{(0)<}<2$
and
$n_{T,\mathrm{node}}^{(0)\ge}\le0$,
the node factor exceeds the generic-phase
factor evaluated at the same numerical
values of the two indices.
Both factors exceed unity in this range,
so the node envelopes exhibit the larger
normalization mismatch.
This comparison fixes the spectral indices,
rather than the background exponents.

~
\bibliographystyle{JHEP}
\bibliography{biblio.bib}

\end{document}